%% file: main.tex
\documentclass[a4paper,11pt]{article}
\usepackage{jheppub} 
\usepackage{lineno}
\usepackage{slashed}
\usepackage{adjustbox}
\usepackage{physics}
\usepackage{tikz}
\usetikzlibrary{decorations.markings}
\usepackage{amsmath,amssymb}
\usepackage{float}
\usepackage{graphicx}
\usepackage{menukeys}
\usepackage{slashed}
\usepackage{bbold}
\usepackage{xr}
\usetikzlibrary{arrows,shapes,patterns,calc}
\usepackage[utf8]{inputenc}
\usepackage{simpler-wick}
\usepackage{comment}
\usepackage{amssymb}
\usepackage{amsmath}
\usepackage{amsfonts}
\usepackage{array}
\usepackage{multirow}
\usepackage{psfrag}
\usepackage{xcolor}
\usepackage{caption}
\usepackage{subcaption}
\usepackage{xfrac}
\newcommand{\ccline}[1]{%
    \cline{#1}%
    \noalign{\vskip-\doublerulesep}
    \cline{#1}%
    \noalign{\vskip\doublerulesep}
}
\usepackage{leftindex}
\usepackage{cleveref}
\usepackage{ifthen}

\usepackage{tabularx}
\usepackage{colortbl}
\definecolor{babypink}{rgb}{0.96, 0.76, 0.76}

\usepackage{lmodern}
\def\ttbf{\ttfamily\fontseries{b}\selectfont}
\def\ttlf{\ttfamily\fontseries{l}\selectfont}

\usepackage{listings}
\DeclareCaptionLabelFormat{boldlisting}{\textbf{Listing~#2}}
\newcommand{\ep}{\varepsilon}
\newcommand{\ii}{\mathrm{i}}
\PassOptionsToPackage{export}{adjustbox}
\chardef\MyArticleWithColor=\pdfcolorstackinit page direct{0 g}

\DeclareMathOperator{\dispersive}{\text{Disp}}
\DeclareMathOperator{\absorptive}{\text{Abs}}

\def\ltd{\mathcal{I}}
\def\renorm{\overline{M}}
\def\bare{M}
\def\remainder{R}
\def\thresholdct{\mathcal{C}}
\def\nfinal{N}
\def\muren{\mu}
\def\mureg{\mu_0}
\def\unitvector{\hat{\mathbf{k}}}
\def\threshold{t}
\def\source{\vec{\mathbf{s}}}
\def\loopmomenta{\vec{\mathbf{k}}}
\def\thresholdset{T}
\def\ddmsbar#1{[\dd[d]{#1}]}
\def\nfcoeff{C}
\def\Nf{N_f}
\def\Nc{N_c}
\def\colouravg{\frac{1}{\Nc}}
\def\virtual{\textrm{V}}
\def\cs{\textrm{CS}}
\def\NLO{\textrm{NLO}}
\def\NfNNLO{{\textrm{NNLO},\Nf}}
\def\NLOoneloop{_\NLO^{\virtual,(1)}}
\def\NLOtree{_\NLO^{\virtual,(0)}}
\def\NfNNLOtwoloop{_\NfNNLO^{\virtual,(2)}}
\def\NfNNLOoneloop{_\NfNNLO^{\virtual,(1)}}
\def\NfNNLOtree{_\NfNNLO^{\virtual,(0)}}
\def\mUV{M}
\def\pmtt{\texttt{+\!\!\!\raisebox{-.48ex}-}}

\DeclareMathOperator{\sgn}{sgn}
\DeclareMathOperator*{\sumint}{%
\int \!\!\!\!\!\!\!{\mbox{\larger$\sum$}}
}

\def\centerarc[#1](#2)(#3:#4:#5)
    { \draw[#1] ($(#2)+({#5*cos(#3)},{#5*sin(#3)})$) arc (#3:#4:#5); }

\title{\boldmath $N_f$-contribution to the virtual correction for electroweak vector boson production at NNLO}

\author[a,b]{Dario Kermanschah,}
\author[c]{Matilde Vicini}
\affiliation[a]{Physics Department, Technical University Munich, James-Franck-Strasse 1, 85748 Garching, Germany}
\affiliation[b]{Rudolf Peierls Centre for Theoretical Physics, Oxford University, Clarendon Laboratory, Parks Road, Oxford OX1 3PU, UK}
\affiliation[c]{Institute for Theoretical Physics, ETH Zurich, Wolfgang-Pauli-Strasse 27, 8093 Z\"urich, Switzerland}

\emailAdd{d.kermanschah@gmail.com, mvicini@phys.ethz.ch}

\abstract{
Multi-loop scattering amplitudes are difficult to evaluate due to singularities of the integrals involved, especially with increasing number of loops, external legs, and mass scales.
For the first time for hadronic collisions at two loops, we enable the combined numerical integration over loop momentum and phase space by tackling infrared, ultraviolet and threshold singularities simultaneously using local subtractions.
We demonstrate the feasibility of our approach by calculating previously unknown perturbative corrections for processes of interest to the Large Hadron Collider, namely the $\Nf$-part of the finite remainder of the phase-space integrated virtual corrections at next-to-next-to-leading order (NNLO) in QCD for the production of up to three massive electroweak vector bosons in proton-proton collisions.
\\~~\\
TUM-HEP-1516/24, OUTP-24-04P
}

\begin{document}
\tikzset{->-/.style={decoration={
  markings,
  mark=at position {#1+0.1} with {\arrow{latex}}},postaction={decorate}}}
\maketitle
\flushbottom
\newpage
\input{sections/introduction}
\newpage
\input{sections/framework}
\newpage
\input{sections/methods_intro}
\input{sections/methods_analytic}
\input{sections/methods_iruvcts}
\input{sections/methods_thresholds}
\newpage
\input{sections/implementation}
\newpage
\input{sections/results}
\newpage
\input{sections/conclusion}

\newpage
\acknowledgments
We are grateful to C.~Anastasiou for his encouragement and valuable discussions throughout this project.
We also thank Z.~Capatti, R.~Haindl, J.~Karlen, A.~Lazopoulos, A.~Pelloni and G.~Sterman for insightful discussions.
Special thanks to F.~Buccioni and V.~Hirschi for their assistance with \textsc{OpenLoops} and \textsc{MadGraph}, respectively. 
We appreciate N.~Kalntis and A.~Schweitzer for their earlier efforts towards the realisation of this project. 
D.K. extends his gratitude to L.~Tancredi and F.~Caola for hosting him at TUM and the University of Oxford, respectively. 
He is also thankful to the Galileo Galilei Institute for their hospitality during the 2023 scientific program.
M.V. is grateful to L.~Tancredi and his group for the opportunity to present part of this project at TUM.
This work was supported by the Swiss National Science Foundation through its Postdoc.Mobility funding scheme (grant number 211062) and project funding scheme (grant number 10001706).

\newpage
\appendix

\input{sections/finite_remainders}
\newpage
\input{sections/phase_space_param}
\newpage
\input{sections/ps_points}
\newpage
\bibliographystyle{JHEP}
\bibliography{biblio}
\end{document}

%% file: sections/introduction.tex
\section{Introduction}
\label{sec:intro}

The calculation of higher-order perturbative corrections within the Standard Model presents a difficult challenge due to the complexity of the loop integrals involved.
These integrals have infrared (IR) and ultraviolet (UV) divergences, which are regularised in $d=4-2\ep$ parametric spacetime dimensions, thus captured as singularities in the dimensional regulator $\ep$.
Additionally, the integrands exhibit threshold singularities when intermediate particles become on-shell.
These singularities are handled by the $\ii\epsilon$ causal prescription, which dictates the analytic continuation of the integral.
\par
The evaluation of loop integrals requires specialised techniques, which quickly reach their limitations when the number of loops, external legs, and involved mass scales are increased.
In particular, resilient numerical integration methods cannot immediately be applied due to the presence of these analytic regulators.
Our paper addresses this challenge for the first time for hadronic collisions at two loops by subtracting IR, UV and threshold singularities with suitable local counterterms.
The resulting hard scattering contributions, which are locally finite in $d=4$ spacetime dimensions, are evaluated using Monte Carlo methods.
As a first application, we target the $\Nf$-part of the finite remainder to the virtual loop corrections at next-to-next-to-leading order (NNLO) in QCD for the production of multiple electroweak vector bosons in proton-proton collisions and present previously unavailable results for up to three massive bosons.
These rare processes are relevant for the Large Hadron Collider (LHC), as they have recently become experimentally accessible to ATLAS \cite{ATLAS:2022xnu,ATLAS:2016jeu} and CMS \cite{CMS:2023rcv,CMS:2020hjs}.
Our calculation demonstrates the feasibility and flexibility of this numerical method, building upon the progress made in recent years.
Below, in section~\ref{sec:overview}, we try to place our method within the broader context of loop calculations.
\par
The rest of the paper is organised as follows.
In section~\ref{sec:framework} we establish the notation for the various components involved in our calculation.
Our method is presented in detail in section~\ref{sec:methods}.
First, in section~\ref{subsec:analytic}, we discuss the singularity structure of the integrand of the loop amplitude.
We then address the treatment of IR and UV, as well as threshold singularities, in sections~\ref{subsec:babis_ct} and \ref{sec:causal_prescription}, respectively. 
The integrated counterterms in dimensional regularisation are supplemented in appendix~\ref{app:finite_remainders}.
The implementation of the method is described in section~\ref{sec:implementation}, with further details on phase space integration in appendix~\ref{app:ps_parametrisation}.
In section~\ref{sec:results}, we present the one-loop and two-loop $\Nf$ contributions to the finite remainder of the virtual corrections for six processes with up to 8 kinematic scales, and provide results for the finite remainders of squared matrix elements at three phase-space points in appendix~\ref{app:fixed_ps_results}.
We conclude with a summary and outlook in section~\ref{sec:conclusion}.

\subsection{Overview}
\label{sec:overview}
The High-Luminosity program at the LHC will collect a substantial amount of new data,
necessitating precise, percent-level theoretical predictions across a wide range of Standard Model processes to uncover new physics \cite{Caola:2022ayt, Boughezal:2022cbl, Andersen:2024czj}.
In response, recent years have witnessed remarkable efforts in perturbative calculations that have led to tremendous progress.
While next-to-leading-order (NLO) calculations have matured and can now be comfortably performed using public automated tools \cite{Giele:2008bc,Berger:2008sj,Badger:2012pg,Czakon:2009ss,Bevilacqua:2011xh,Actis:2012qn,Denner:2017wsf,Hirschi:2011pa,Cascioli:2011va,Buccioni:2019sur}, powerful advances have been made in calculations at NNLO and higher.
NNLO QCD corrections for $2\to2$ processes have become the state of the art, pushing the frontier to $2\to3$ processes, of which exact cutting-edge calculations with photons and/or jets in the final state have recently been performed \cite{Chawdhry:2019bji,Abreu:2020cwb,Chawdhry:2020for,Kallweit:2020gcp,Abreu:2023bdp,Chawdhry:2021hkp,Badger:2021ohm,Badger:2021imn,Badger:2023mgf,Abreu:2018jgq,Abreu:2021oya,Czakon:2021mjy,DeLaurentis:2023nss,Agarwal:2021grm,Agarwal:2021vdh,Agarwal:2023suw,DeLaurentis:2023izi}.
Nonetheless, the availability of multi-scale, multi-loop scattering amplitudes remains a major bottleneck in advancing fixed-order calculations.
The many mass scales, especially in the production of electroweak particles, and the large multiplicity of external legs make the calculation of these amplitudes highly technical and computationally expensive, typically tackled on a case-by-case basis.
A most recent milestone is the computation of all 5-point master integrals with massless internal propagators and one massive external leg  \cite{Chicherin:2021dyp,Abreu:2023rco}.
\par
The latest progress in higher-order calculations was enabled by a better understanding of the function space of loop integrals, as well as the development of new algorithms and optimisations for integral reduction~\cite{Chetyrkin:1981qh,Laporta:2000dsw,Kotikov:1990kg,Henn:2013pwa,vonManteuffel:2014ixa}.
On the other hand, the persistent appearance of new conceptual and computational challenges has given rise to entire programs investigating the mathematical structure of loop integrals and has motivated seeking for alternative approaches to overcome these obstacles.
A recurring theme in this direction has been a shift away from symbolic towards numerical integration.
This idea is viable because a few digits precision of fixed-order cross section calculations are already sufficient.
Moreover, it is particularly promising in overcoming the complexity barrier faced by many methods, as it proves resilient to changes in the number of legs, loops and internal and external masses.
Over the last few decades, significant efforts have been dedicated to Feynman-parametric integrals. 
This progress has been facilitated by sector decomposition within dimensional regularisation \cite{Roth:1996pd,Binoth:2000ps, Heinrich:2008si} combined with contour deformation around threshold singularities \cite{Binoth:2005ff, Nagy:2006xy, Soper:1999xk,Anastasiou:2005pn,Lazopoulos:2007ix,Anastasiou:2007qb,Lazopoulos:2008de,Lazopoulos:2007bv}.
Several public codes are now available that enable the numerical integration of challenging multi-scale, multi-loop integrals \cite{Borowka:2012yc, Borowka:2015mxa, Borowka:2017idc, Bogner:2007cr, Smirnov:2008py, Smirnov:2009pb, Smirnov:2015mct, Smirnov:2021rhf}.
Lately, these techniques were used to calculate the quark-initiated $N_f$-part to $t\bar{t}H$ production at two loops \cite{Agarwal:2024jyq}.
An alternative approach leverages the tropical geometric structure of the parametric integrand to achieve highly efficient Monte Carlo integration of quasi-finite integrals \cite{Borinsky:2023jdv, Borinsky:2020rqs}.
\par
Recently, there has been renewed interest in performing perturbative calculations using direct numerical integration in momentum space.
While naturally aligning with Monte Carlo phase space generators, the factorised denominator structure in momentum space representations has also been instrumental in establishing direct connections to on-shell thresholds and local factorisation.
Moreover, in contrast to Feynman parametrisation, the number of integrations does not increase with the number of propagators.
Additionally, this framework aims to bypass the large systems of integration-by-parts identities involved in integral reduction and the challenges associated with solving master integrals.
Ideally, especially for the sake of automation, the integrand undergoes only minimal manipulations in a largely process-generic manner before being numerically integrated over the physical loop and phase space degrees of freedom.
However, a good understanding of the divergent regions of the integrands of loop amplitudes in four spacetime dimensions becomes essential.
These efforts are therefore connected to the revival of the analytic S-matrix bootstrap program from the 1960s, which explores the analytic structure of scattering amplitudes through the related concepts of cut diagrams, discontinuities and Landau singularities \cite{Bloch:2015efx,Abreu:2014cla,Abreu:2017ptx,Arkani-Hamed:2017ahv,Tomboulis:2017rvd,Collins:2020euz,Bourjaily:2020wvq,Hannesdottir:2021kpd,Klausen:2021yrt,Mizera:2021fap,Correia:2021etg,Bourjaily:2022vti,Flieger:2022xyq,Gambuti:2023eqh,Fevola:2023kaw,Muhlbauer:2023uqh,Hannesdottir:2024cnn,Dlapa:2023cvx,Tourkine:2023xtu,Huber:2022nzs,Gardi:2022khw}.
With subsequent numerical integration in mind, an understanding of the local singularity structure restricted to the physical kinematic region may already be sufficient.
\par
Seminal concepts for NLO calculations in momentum space based on numerical integration were introduced in refs.~\cite{Soper:1998yp,Soper:1998ye, Soper:1999xk, Soper:2001hu, Kramer:2002cd}, which culminated in a computer program \cite{beowulf} that calculates NLO QCD corrections for $e^+ e^- \to 3$ jets.
Among the employed techniques are the analytical integration of loop energies in order to identify and localise singular surfaces, a deformation of the integration contour to regularise threshold singularities according to the $\ii\epsilon$ causal prescription, local UV renormalisation, a treatment of the complications related to self-energies and the construction of specialised importance sampling densities that flatten the integrand for more efficient Monte~Carlo integration.
At the heart of the numerical method lies a local representation of the optical theorem,  expressed as a sum of various cuts of Feynman diagrams that contribute to the forward-scattering amplitude.
This representation exhibits local cancellations of IR singularities due to final-state radiation (FSR) between real and virtual contributions, thereby locally realising the cancellations described by the Kinoshita-Lee-Nauenberg (KLN) theorem \cite{Kinoshita:1962ur,Lee:1964is}.
This paradigm, now referred to as local unitarity \cite{Capatti:2020xjc}, has recently been extended beyond NLO providing automated local renormalisation and treatment of self-energy corrections \cite{Capatti:2022tit}, and has lead to new results for light-by-light scattering \cite{AH:2023kor}.
Ref.~\cite{Capatti:2022tit} also presents a first fully numerical computation of the NNLO corrections to inclusive off-shell photon decay, a calculation where initial-state IR and threshold singularities are absent.
The traction of this approach is further supported by the recent proposal of closely related ideas \cite{Sterman:2023xdj,LTD:2024yrb,Ramirez-Uribe:2024rjg}.
Other considerations aiming at local IR cancellations among real and virtual contributions, based on mappings between loop and phase space, were investigated in refs.~\cite{Buchta:2014dfa,Hernandez-Pinto:2015ysa,Sborlini:2016gbr,Sborlini:2016hat,Runkel:2019zbm,Seth:2016hmv}.
From the outset, remaining open questions were recognised regarding the factorisation associated with initial-state hadrons and practical challenges related to the treatment of threshold singularities.
Both of which may potentially be resolved in combination with the following techniques.
\par
A complementary avenue targets the numerical integration of virtual corrections in momentum space, separately from real contributions.
In this approach, the IR (and UV) divergent regions must first be subtracted using local counterterms before numerical integration.
These counterterms are then analytically integrated in dimensional regularisation and added back.
Note that this strategy has already proven effective for the calculation of real corrections.
First one-loop subtractions were originally formulated graph-by-graph~\cite{Nagy:2003qn} and later, inspired by extensive work on IR factorisation~\cite{Sterman:1995fz,Sterman:2002qn,Sterman:1993hfp,  Erdogan:2014gha,Ma:2019hjq,Akhoury:1978vq,Sen:1982bt,Catani:1996eu,Collins:1989bt,Dixon:2008gr,Becher:2009cu,Feige:2014wja,Sterman:1978bi,Libby:1978qf,Collins:1989gx,Becher:2014oda}, extended to the amplitude level~\cite{Assadsolimani:2009cz,Assadsolimani:2010ka}.
These subtractions, in combination with a suitable contour deformation around the remaining threshold singularities, enabled first NLO calculations with up to 7 jets in electron-positron annihilation \cite{Becker:2011vg,Becker:2012aqa}.
In this context, two variants of contour deformation were explored, one directly in the four-dimensional loop momentum space \cite{Gong:2008ww,Becker:2011vg,Becker:2012aqa,Becker:2012bi} and one where additional auxiliary Feynman parameters \cite{Nagy:2006xy,Becker:2010ng} were introduced.
At two loops, the analysis becomes more complicated due to overlapping singular regions with first systematic subtractions explored in ref.~\cite{Anastasiou:2018rib}.
Recently, a new technique that leverages Ward identities has been developed to achieve local IR factorisation, paying special attention to momentum routings and preserving quadratic propagators.
This method has enabled the formulation of IR (and UV) counterterms for specific classes of two-loop QCD amplitudes, namely for electroweak production in electron-positron \cite{Anastasiou:2020sdt} and quark-antiquark collisions \cite{Anastasiou:2022eym}, as well as in instances of gluon fusion \cite{Anastasiou:2024xvk}.
These counterterms, combined with a novel contour deformation of the spatial loop momenta, were employed in ref.~\cite{Capatti:2019edf} for the numerical integration of one-loop triphoton production amplitudes.
Central to the implementation was the analytic integration over the energy components of the loop momenta, which greatly simplified the local threshold singularity structure.
\par
Representations for loop integrals without explicit loop energy integration originate from old-fashioned perturbation theory of quantum field theory (QFT).
This formulation, also known as time-ordered perturbation theory (TOPT) \cite{Bodwin:1984hc, Collins:1985ue, Sterman:1993hfp, Sterman:1995fz}, breaks manifest Lorentz covariance and expresses a Feynman diagram as a sum of diagrams with time-ordered vertices.
Later, inspired by the Feynman tree theorem, a locally equivalent but algebraically different representation was derived by iteratively integrating out loop energies using the residue theorem.
The resulting identity, called loop-tree duality (LTD) \cite{Catani:2008xa, Aguilera-Verdugo:2019kbz, Aguilera-Verdugo:2020set,JesusAguilera-Verdugo:2020fsn, Ramirez-Uribe:2022sja, Runkel:2019yrs, Capatti:2019ypt}, expresses a graph with loops as a sum of its spanning trees.
TOPT and LTD introduce spurious (non-causal) singularities, which prompted the search for alternative representations, referred to as causal LTD \cite{Aguilera-Verdugo:2020set, Aguilera-Verdugo:2020kzc, JesusAguilera-Verdugo:2020fsn,Ramirez-Uribe:2020hes, Sborlini:2021owe, TorresBobadilla:2021ivx, Bobadilla:2021pvr, Benincasa:2021qcb, Kromin:2022txz, Capatti:2020ytd,Capatti_cLTD_2020}, the cross-free family (CFF) representation \cite{Capatti:2022mly, Capatti:2023shz}, and partially time-ordered perturbation theory (PTOPT) \cite{Sterman:2023xdj}.
While these representations share similarities with TOPT, they explicitly eliminate spurious singular surfaces.
The remaining singular surfaces are precisely the threshold and collinear singularities associated with physical intermediate states.
\par
A deformation of the integration contour of the spatial loop momenta around the threshold singularities was originally proposed in ref.~\cite{Soper:1998yp} and was reconsidered in the context of LTD at one loop \cite{Buchta:2015wna,Kromin:2022txz} and beyond \cite{Capatti:2019edf}.
However, contour deformations are typically difficult to control and lead to an oscillating integrand, which in turn results in inefficient numerical integration, as illustrated in ref.~\cite{Kermanschah:2022mpb}.
Other methods have been explored that set the causal prescription to a small number and use extrapolation \cite{deDoncker:2004bf, Yuasa:2011ff, deDoncker:2017gnb, Baglio:2020ini} or other techniques \cite{Pittau:2021jbs,Pittau:2024ffn}.
An alternative approach based on the subtraction of threshold singularities offers a more robust solution, as it inherently produces a flat integrand.
While this method was first explored in ref.~\cite{Kilian:2009wy}, it did not account for intersecting thresholds.
Recently, a new construction was introduced in ref.~\cite{Kermanschah:2021wbk} that effectively handles complicated threshold singularity structures.
A consequence of this approach is the explicit separation of dispersive and absorptive parts of loop integrals, where resulting expression for the latter constitutes a local representation of the generalised\footnote{
Here, we refer to the optical theorem in a broader context that extends beyond forward-scattering amplitudes, as directly implied by the unitarity of the S-matrix.
} optical theorem \cite{Sterman:1995fz,Sterman:1993hfp}, as explicitly shown at one loop in \cite{Kermanschah:2021wbk}.
\par
In this paper, we will combine the aforementioned techniques to tackle one-loop and the $N_f$-part of two-loop finite remainders to the virtual contributions for six different LHC processes with two and three electroweak vector bosons in the final state.
We do not consider the corresponding real emission contributions, which can be computed with available techniques e.g.~at NLO using FKS subtraction \cite{Frixione:1995ms} automated in \textsc{MadFKS} \cite{Frederix:2009yq}, nor the double-real and real-virtual contributions at NNLO using $q_T$-subtraction \cite{Catani:2007vq} automated in \textsc{Matrix} \cite{Kallweit:2020gcp,Grazzini:2019jkl,Grazzini:2017mhc}.
The missing piece required to complete the full NNLO correction to the cross section is the double-virtual contribution.
In this paper, we provide the finite remainder of the gauge invariant $\Nf$-part, which, since it does not receive any real-virtual corrections, needs to be added only to the corresponding real contributions to obtain a physical result.
Our method relies on using the local IR and UV counterterms of \cite{Anastasiou:2022eym}, integrating out the loop energies in accordance with the CFF representation \cite{Capatti:2022mly} and subtracting the threshold singularities following the construction in \cite{Kermanschah:2021wbk}.
Our calculations not only provide new results for the production of three massive electroweak bosons at the LHC but also demonstrates the applicability of this numerical approach in handling IR, UV and threshold singularities simultaneously in challenging two-loop computations.
This work shows that tackling hadronic collisions through numerical integration over both loop momentum and phase space is practically feasible, which, to the best of our knowledge has not been demonstrated before at the two-loop order.
\par

%% file: sections/framework.tex
\section{Framework}
\label{sec:framework}
We consider the production of $\nfinal$ (possibly distinct) electroweak vector bosons in a proton-proton collision in the Standard Model
\begin{align}
    p + p \to V_1+\dots+V_\nfinal\,.
\end{align}
By $p$ we denote a proton and by $V_i$ any electroweak vector boson, i.e.~either a (virtual) photon $\gamma^{(*)}$ or a $Z$ or $W^\pm$ boson.
We write the hadronic cross section $\sigma_{pp\to \nfinal\times V}$ in terms of the partonic cross sections $\hat\sigma_{ab\to \nfinal\times V}$ using collinear factorisation \cite{Collins:1989gx,Sterman:2022gyf}, which holds up to power corrections,
\begin{align}
\label{eq:factorisation_theorem}
    \dd \sigma_{pp\to \nfinal\times V}(s,\mu_F,\muren)
    = \sum_{a,b}\int \dd x_1 \dd x_2
    f_a(x_1,\mu_F) f_b(x_2,\mu_F)
    \,\dd \hat \sigma_{ab\to \nfinal\times V}(x_1 x_2 s, \muren),
\end{align}
where $s$ is the squared centre of mass energy of the collision, $\mu_F$ and $\muren$ are the factorisation and renormalisation scales and $f_{a,b}(x_{1,2},\mu_F)$ are the parton distribution functions (PDFs) of partons $a,b$ with momentum fraction $x_{1,2}$, respectively.
From now on, we will omit the process specification.
\par
We consider QCD corrections to our process of interest and organise the perturbative expansion of the partonic cross section in terms of the renormalised strong coupling $\alpha_s(\muren)$ as
\begin{align}
\label{eq:expansion_cross_section}
    \dd \hat\sigma
    = 
    \dd \hat\sigma_{\text{LO}}
    + \left(\frac{\alpha_s}{4\pi}\right)\,
    \dd \hat\sigma_\NLO
    + 
    \left(\frac{\alpha_s}{4\pi}\right)^2 \,
    \dd \hat\sigma_{\text{NNLO}}
    + \mathcal{O}\left(\alpha_s^3\right).
\end{align}
For the purposes of this paper, we are not interested in the full NNLO correction
but only in
its $\Nf$-part
\begin{align}
\label{eq:nf_cross_section}
    \dd \sigma_\NfNNLO
\end{align}
defined as the contribution of $\dd \sigma_{\text{NNLO}}$ proportional to the number of massless quarks $\Nf$.
The respective expansion of the hadronic cross section follows analogously to eq.~\eqref{eq:expansion_cross_section}.
Beyond leading order, the cross section receives contributions from real radiation and virtual corrections.
We will refer to the latter by $\dd \sigma^\virtual$. The real radiation contributions will not be considered further in this paper.
\par

The LO, NLO and the NNLO-$\Nf$ contributions to the hadronic cross section are initiated through the channel $q+\bar{q}\to V_1+\dots+V_\nfinal$ by a quark-antiquark pair\footnote{
The quarks can have different flavours if $W$ bosons are produced.
} $q$ and $\bar{q}$.
For example, the relevant Feynman diagrams for diphoton production are given in figures~\ref{fig:1L} and~\ref{fig:BoxNf}.
We can explicitly extract the colour structure and denote the UV-renormalised scattering amplitude with fixed helicities $h$ by $\delta_{ij}\renorm_h$, where $i$ and $j$ are colour indices.
The partonic virtual correction can then be expressed as
\begin{align}
    \dd \hat\sigma^\virtual
    = \colouravg\frac{1}{4}\sum\limits_h \left|\renorm_h\right|^2 \,\frac{\dd \Pi_\nfinal}{F}\,.
\end{align}
Here, $\dd \Pi_\nfinal$ denotes the $\nfinal$-body phase space measure and $F=4\left|\vec{p}^{\phantom{0}}_a p_b^0 - \vec{p}^{\phantom{0}}_b p^0_a\right|=2x_1x_2s$ is the flux factor, where $p_{a,b}$ are the back-to-back momenta of the two colliding quarks, respectively.
The helicities of the final state are summed over and the helicities and colours of the initial state are averaged over.
\par

\input{figures/diags}

The UV renormalisation is performed in the modified minimal subtraction scheme ($\overline{\text{MS}}$) in $d=4-2\ep$ spacetime dimensions. The bare coupling $\alpha_s^0$ satisfies
\begin{align}
    \alpha_s^0\mureg^{2\ep} S_\ep = \alpha_s\muren^{2\ep} Z_{\alpha_s}\,,
    \qquad
    S_{\ep}= (4\pi)^{\ep} e^{-\ep\gamma_E}\,,
\end{align}
where $\mureg$ is the regularisation scale and
\begin{align}
    Z_{\alpha_s} =  1-\frac{2\beta_0}{\ep} \left(\frac{\alpha_s}{4\pi} \right)+\mathcal{O}(\alpha_s^2),
    \qquad
    \text{where}
    \qquad
    \beta_0 = \frac{11 C_A-4\Nf T_F}{6}
\end{align}
is the first coefficient in the QCD beta function.
The perturbative expansion of the renormalised amplitude reads
\begin{align}
    \renorm(\alpha_s) = \renorm^{(0)}
    + \left(\frac{\alpha_s}{4\pi}\right)
    \renorm^{(1)}
    + \left(\frac{\alpha_s}{4\pi}\right)^2
    \renorm^{(2)}
    + \mathcal{O}(\alpha_s)\,.
\end{align}
The renormalised amplitude is expressed in terms of the bare amplitude $\bare$ through $\renorm(\alpha_s) = \bare(\alpha_s^0)$, such that the coefficients in the perturbative expansion are related by 
\begin{align}
    \renorm^{(0)} = \bare^{(0)}, \qquad
     \renorm^{(1)} = \bare^{(1)}, \qquad
     \renorm^{(2)} = \bare^{(2)}
     - \frac{2 \beta_0}{\ep} \bare^{(1)}\,.
\end{align}
The bare amplitude is an integral over $n$ loop momenta
\begin{align}
    M^{(n)} 
    = 
   \int \left(\prod_{i=1}^n \ddmsbar{k_i} \right)\mathcal{M}^{(n)}(\{k_i\})\,,
\end{align}
where the loop measure is defined through
\begin{align}
    \muren^{2\epsilon} S_\ep^{-1} g_s^2\frac{\dd^d k}{(2\pi)^d} = \muren^{2\epsilon} e^{\ep\gamma_E} \left(\frac{\alpha_s}{4\pi}\right) \frac{\dd^d k}{\pi^{d/2}}
    \equiv
    \left(\frac{\alpha_s}{4\pi}\right) \ddmsbar{k}\,.
\end{align}
The two-loop $\Nf$-part we define analogously to eq.~\eqref{eq:nf_cross_section} to be the amplitude contribution proportional to $\Nf$.
The partonic virtual correction has IR singularities, exposed as poles in the dimensional regulator.
To define the IR-finite remainder $R$ of the UV-renormalised amplitude, we subtract the Catani poles \cite{Catani:1998bh,Sterman:2002qn}.
The operators for this type of processes were derived in \cite{Anastasiou:2002zn}.  
The IR divergences of the integrated amplitude factorise as 
$\renorm = \mathbf{Z} \remainder$, which upon expansion in $\alpha_s$ and $\ep$ yields
\begin{align}\label{eq:Z-tree}
\renorm^{(0)}&= \remainder^{(0)}\,,\\
    \label{eq:Z-1L}
    \renorm^{(1)}&= \remainder^{(1)} +\mathbf{Z}^{(1)} \renorm^{(0)}+\mathcal{O}(\ep)\,,
    \\ \label{eq:Z-2L}
    \renorm^{(2)}&= \remainder^{(2)}+\mathbf{Z}^{(1)} \renorm^{(1)} +\mathbf{Z}^{(2)} \renorm^{(0)}+\mathcal{O}(\ep)\,.
\end{align}
In appendix~\ref{app:catani}, we report the relevant operators $\mathbf{Z}^{(1)}$ and $\mathbf{Z}^{(2,\Nf)}$ in the $\overline{\text{MS}}$ factorisation scheme, following the same conventions as ref.~\cite{Chawdhry:2020for}.
We can infer the finite remainders to the partonic virtual correction at higher-orders:
\begin{align}
    \dd \hat \sigma^{\virtual,\cs}_\NLO
    &= 
    \colouravg\frac{1}{4}\sum\limits_h
     2 \Re \left[
         \left(\remainder^{(0)}_h\right)^* 
    \remainder^{(1)}_h
        \right]
        \frac{\dd\Pi_\nfinal}{F}\,,
    \\
    \dd \hat \sigma^{\virtual,\cs}_\NfNNLO
    &= 
    \colouravg\frac{1}{4}\sum\limits_h
     2 \Re \left[
         \left(\remainder^{(0)}_h\right)^*
    \remainder^{(2,\Nf)}_h
    \right]
    \,\frac{\dd\Pi_\nfinal}{F}\,.
\end{align}
The Born contribution is stated for completeness
\begin{align}
    \dd \hat \sigma_{\text{LO}}
    &= 
    \colouravg\frac{1}{4}\sum\limits_h
    \left|\remainder^{(0)}_{h}\right|^2
    \,\frac{\dd\Pi_\nfinal}{F}\,.
\end{align}
The respective contributions to the virtual hadronic cross section are obtained by the convolution with the PDFs in eq.~\eqref{eq:factorisation_theorem}.
\par

%% file: figures/diags.tex
\begin{figure}[t]
    \centering
        \begin{subfigure}{0.24\textwidth}
\includegraphics[width= \textwidth, page=1]{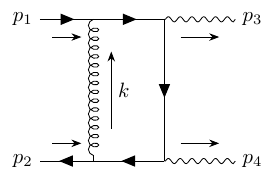}
    \end{subfigure}
        \begin{subfigure}{0.24\textwidth}
\includegraphics[width= \textwidth,page=2]{figures/tikz_TR1loop.pdf}
    \end{subfigure}
        \begin{subfigure}{0.24\textwidth}
\includegraphics[width= \textwidth, page=3]{figures/tikz_TR1loop.pdf}
    \end{subfigure}
        \begin{subfigure}{0.2194\textwidth}
\includegraphics[width= \textwidth, page=4]{figures/tikz_TR1loop.pdf}
    \end{subfigure}
    \caption{The one-loop Feynman diagrams for $q(p_1)\bar q(p_2) \to \gamma(p_3) \gamma (p_4)$ for one fixed ordering in the final state photons.
    The momentum labels consistent with the local IR and UV counterterms reported in section~\ref{subsec:babis_ct}.}
    \label{fig:1L}
\end{figure}
\begin{figure}[b]
    \centering
    \includegraphics[width= 0.24\textwidth, page=6]{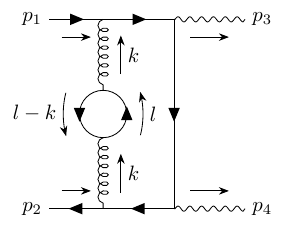}
    \caption{One diagram contributing to the $\Nf$-part of the two-loop amplitude for $q(p_1)\bar q(p_2) \to \gamma(p_3) \gamma (p_4)$.
    All other diagrams can be constructed from the one-loop amplitude by inserting a fermion bubble into the gluon propagator.}
    \label{fig:BoxNf}
\end{figure}
\par

%% file: sections/methods_intro.tex
\section{Method}
\label{sec:methods}
We calculate the one-loop and the two-loop $\Nf$ contributions to the virtual correction of the (differential) cross section for electroweak vector boson production in proton-proton collisions.
Our approach centres around rewriting the finite remainders in terms of integrals that can be efficiently numerically integrated using Monte~Carlo methods.
Since the various contributions involve divergent integrals, we will first perform a variety of analytic manipulations and subtractions at the amplitude level, which expose and locally remove regions in momentum space that produce integral divergences in $d=4$ spacetime dimensions.
In particular, this involves the local subtraction of IR, UV and threshold singularities, as well as the analytic integration over the energy component of the loop momenta.
\par

Our starting points are the one-loop amplitude $\bare^{(1)}$ and the $\Nf$-part of the two-loop amplitude $\bare^{(2,\Nf)}$ in $d=4-2\ep$ spacetime dimensions.
The former is given in terms of Feynman rules and the integrand of the latter we modify according to ref.~\cite{Anastasiou:2020sdt},
\begin{align}
\label{eq:bub-nf}
    \mathcal{M}^{(2,\Nf)}(k,l)= \nfcoeff \frac{1}{l^2+\ii\epsilon}
    \frac{1}{(l+k)^2+\ii\epsilon} \mathcal{M}^{(1)}(k)\,,
    \quad \text{where}
    \quad 
    \nfcoeff = \ii \Nf T_F \frac{2(d-2)}{d-1}\,.
\end{align}
This equivalent integrand representation\footnote{
This representation can be derived by performing a tensor reduction on the fermion bubble, followed by a repeated application of the Ward identity.
Alternatively, it can be viewed as the result of introducing a local counterterm that integrates to zero (the original integrand minus the tensor-reduced one).
} facilitates the construction of suitable local IR and UV counterterms.
The local singularity structure of the one-loop and two-loop $\Nf$-amplitudes are discussed in more detail in section~\ref{subsec:analytic}.
\par

First, as we will show in section~\ref{subsec:babis_ct}, the IR and UV divergences are locally removed from the amplitude $\bare^{(n)}$ at the level of the integrand $\mathcal{M}^{(n)}$ by subtracting the counterterms of refs.~\cite{Anastasiou:2020sdt, Anastasiou:2022eym}.
This achieves a separation of the form
\begin{align}
\label{eq:fin_div_separation}
   \mathcal{M}^{(n)} = \mathcal{M}^{(n)}_\text{finite}
   + \mathcal{M}^{(n)}_\text{singular}\,.
\end{align}
This separation is inherently ambiguous because the singular contribution may have a non-zero finite piece.
However, this ambiguity corresponds to the choice of a UV renormalisation and a IR factorisation scheme.
To facilitate comparisons with existing literature, we convert our finite remainders, $\bare_\text{finite}^{(1)}$ and $ \bare^{(2,\Nf)}_\text{finite}$, in the scheme defined by our chosen counterterms to the finite remainders in $\overline{\text{MS}}$ via the scheme change:
\begin{align}
\label{eq:scheme_change_1l}
    R^{(1)}
    &= \bare_\text{finite}^{(1)} + R^{(1)}_\text{singular} \\
\label{eq:scheme_change_2l}
    R^{(2,\Nf)} &= \bare^{(2,\Nf)}_\text{finite} +
   \bare^{(1)}_\text{finite}  R^{(1,\Nf)}_\text{singular}  + R^{(2,\Nf)}_\text{singular},
\end{align}
where $R^{(1)}_\text{singular}$, $R^{(1,\Nf)}_\text{singular}$ and $R^{(2,\Nf)}_\text{singular}$ are finite in $d=4$ dimensions.
In sections~\ref{subsubsec:one-loop-amplitude} and \ref{subsubsec:two-loop-nf-amplitude}, we explicitly calculate these contributions in dimensional regularisation.
\par
The remaining hard scattering contributions are the loop integrals
\begin{align}
\label{eq:M1finite}
    \bare_\text{finite}^{(1)} &=  \int \frac{\dd[4]{k}}{\pi^2} \mathcal{M}_\text{finite}^{(1)}(k)\,,\\
    \label{eq:M2finite}
    \bare^{(2,\Nf)}_\text{finite} &= \int \frac{\dd[4]{k}}{\pi^2} \frac{\dd[4]{l}}{\pi^2} \mathcal{M}_\text{finite}^{(2,\Nf)}(k,l)\,,
\end{align}
which we aim to evaluate using numerical methods.
Although these integrals are finite in $d=4$ dimensions by construction, their integrands still suffer from threshold singularities.
They are responsible for the amplitudes' absorptive part.
Threshold singularities are regularised by the $\ii \epsilon$ causal prescription, which effectively moves the singularity away from the integration domain.
However, this method is unsuitable for Monte Carlo integration and we have to adopt an alternative approach, which we will describe in section~\ref{subsec:thres_sub}.
\par
First, we analytically integrate out the energy components of the loop momenta,
\begin{align}
    \int\limits \left(\prod_{j=1}^{n} \frac{\dd k_j^0}{\pi^2}\right)
    \mathcal{M}_\text{finite}^{(n)}
    \left(\{k_i\}\right)
    =  
    \ltd_\text{finite}^{(n)}
    \left(\{\vec{k}_i\}\right),
\end{align}
resulting in an expression $\ltd_\text{finite}^{(n)}$ that remains to be integrated over the spatial loop momenta.
This procedure exposes the threshold singularities as poles of $\ltd_\text{finite}^{(n)}$ in the integration domain.
\par
Subsequently, these poles are locally removed by subtracting the threshold counterterms of ref.~\cite{Kermanschah:2021wbk,Kermanschah:2022mpb}.
These counterterms are conveniently constructed such that their dispersive part integrates to zero.
Therefore, a local separation of absorptive and dispersive part is achieved at integrand-level:
\begin{align}
    \ltd_\text{finite}^{(n)} = 
    \dispersive
    \ltd_{\text{ finite}}^{(n)}
    +
    \ii
    \absorptive
    \ltd_{\text{ finite}}^{(n)}.
\end{align}
The real part of the hard scattering amplitude is then given by the integral
\begin{align}
\label{eq:re_summary}
    \Re 
    M_\text{finite}^{(n)}
    = 
    \int 
    \left(
    \prod_{j=1}^n \dd[3]{\vec{k}_j}
    \right)
    \left(
    \Re\dispersive\ltd^{(n)}_{\text{ finite}}
    -
    \Im\absorptive\ltd^{(n)}_{\text{ finite}}
    \right)\,,
\end{align}
where the $\ii\epsilon$ causal prescription can be entirely removed from the integrand prior to integration.
The integrand for the absorptive part contains an additional delta function, and is therefore effectively integrated over the phase space of intermediate on-shell particles, with its explicit expression given in eq.~\eqref{eq:absorptive}.
The absorptive contribution vanishes in the phase-space integrated virtual loop contributions if parity invariant transverse momentum cuts are applied.
However, as we show in appendix~\ref{app:fixed_ps_results} for the process $q\bar{q}\to Z\gamma_1^*\gamma_2^*$, it generally contributes to the finite remainder at a fixed phase space point.
\par
Eventually, we can express the finite remainders of the virtual correction as a sum of terms with different loop multiplicities
\begin{align}
    \label{eq:finite_cross_section_nlo}
    \dd\sigma^{\virtual,\cs}_\NLO
    &= \dd\sigma\NLOoneloop + \dd\sigma\NLOtree\,,\\
      \label{eq:finite_cross_section_nnlo}
    \dd\sigma^{\virtual,\cs}_\NfNNLO
    &= \dd\sigma\NfNNLOtwoloop + \dd\sigma\NfNNLOoneloop + \dd\sigma\NfNNLOtree\,.
\end{align}
Here, $\dd\sigma\NLOoneloop$ and $\dd\sigma\NfNNLOtwoloop$ represent the hard contributions to the virtual corrections, whereas the remaining terms can to be understood as a consequence of a UV renormalisation and IR factorisation scheme change.
They are defined as the integrals
\begin{align}
\label{eq:first_cs_int}
    \dd\sigma\NLOoneloop&= 2\sumint \dd X^{(1)}
     \left(\remainder^{(0)}\right)^*
     \ltd^{(1)}_{\Re,\text{ finite}}\,,\\
     \label{eq:sigmaNLOtree}
     \dd\sigma\NLOtree
     &=2 \Re \sumint \dd X^{(0)}
       \left(\remainder^{(0)}\right)^*
       \remainder^{(1)}_\text{singular}\,,
       \\
    \dd\sigma\NfNNLOtwoloop
    &= 2\sumint \dd X^{(2)}
      \left(\remainder^{(0)}\right)^*
        \ltd^{(2)}_{\Re,\text{ finite}}\,,\\
        \label{eq:sigma1NF}
    \dd\sigma\NfNNLOoneloop
    &=\remainder^{(1,\Nf)}_\text{singular} \dd\sigma\NLOoneloop\,,\\
    \dd\sigma\NfNNLOtree
    &= 2\Re\sumint \dd X^{(0)}
      \left(\remainder^{(0)}\right)^*
     \remainder^{(2,\Nf)}_\text{singular}\,,
\label{eq:last_cs_int}
\end{align}
which can be efficiently numerically integrated using Monte~Carlo methods.
Observables and phase-space cuts can be accommodated by multiplying the integrands with the appropriate measurement functions.
We introduced the short-hand notation
\begin{align}
    \sumint \dd X^{(n)} = 
    \int \dd x_1 \dd x_2
    \sum_{a,b} f_a(x_1,\mu_F) f_b(x_2,\mu_F)
    \frac{\dd \Pi_\nfinal}{F}
    \left(\prod_{j=1}^n\dd[3]{\vec{k}}_j\right)
    \colouravg\frac{1}{4}\sum_h
\end{align}
for the integration over the $2$ Bjorken momentum fractions, the sum over partons, the integration over the $(3\nfinal-4)$--dimensional\footnote{If $\nfinal=1$ the phase space dimension is zero.} $\nfinal$--body phase space and the $n$ $3$--dimensional spatial loop momenta, as well as the sum over helicity configurations.
The helicities and colours of initial state are averaged over.
The dependence of the integrands in~\crefrange{eq:first_cs_int}{eq:last_cs_int} on all of the aforementioned integration and summation variables is left implicit.
We emphasise that the phase space and loop integrations are simultaneously performed in a single Monte~Carlo integration.
For completeness, we also define the Born contribution by
\begin{align}
    \label{eq:cross_section_lo}
    \dd\sigma_\text{LO}
    &= \sumint \dd X^{(0)}
    \left|\remainder^{(0)}\right|^2.
\end{align}
\par
In section~\ref{sec:implementation} we describe how we implemented the above calculations in a computer program.
This program was used to calculate virtual corrections for $2\to2$ and $2\to3$ processes with massless and massive vector bosons in the final state.
Our results are presented in section~\ref{sec:results}.

%% file: sections/methods_analytic.tex
\newpage
\subsection{Scattering kinematics and singular surfaces}
\label{subsec:analytic}
Our process of interest
\begin{align}
    q(p_1) + \bar{q}(p_2) \to V_1(q_1) + \dots +  V_\nfinal(q_\nfinal)
\end{align}
imposes the on-shell conditions $p_1^2 = p_2^2 = 0$ and $q_i^2 = m_i^2 \geq 0$.
Moreover, we define 
\begin{align}
    p_S = \sum_{i\in S} q_i
\end{align}
for all non-empty, $S\neq \emptyset$, proper subsets $S\subsetneq \{1,\dots,\nfinal\}$, to distinguish between positive and negative invariants $p_S^2 \geq 0$ and $(p_1-p_S)^2\leq 0$, respectively.
\par

\begin{figure}[b]
    \centering
    \begin{subfigure}[b]{0.45\textwidth}
        \centering
        \begin{tikzpicture}
            \coordinate (a) at (1,-1);
            \draw[thick,->-=0.5] (0,0) node[left]{$k$} -- (a);
            \draw[thick,->-=0.5] (a) --  (0,-2) node[left]{$k-p$};
            \draw[thick,->-=0.5] (a) -- (2,-1) node[right]{$p$};
            \draw[->] (0.5,-2.5) -- node[above]{$\text{time}$} (1.5,-2.5);  
        \end{tikzpicture}
        \caption{Causality imposes \\\phantom{(a) }$\sgn k^0 = -\sgn (k-p)^0$.}
        \label{fig:one_loop_momentum_conservation}
    \end{subfigure}
    \hspace{15pt}%
    \begin{subfigure}[b]{0.45\textwidth}
    \centering
    \begin{tikzpicture}
        \coordinate (a) at (1,-1);
        \draw[thick,->-=0.5] (0,0) node[left]{$k+l$} -- (a);
        \draw[thick,->-=0.5] (a) --  (0,-1) node[left]{$l$};
        \draw[thick,->-=0.5] (a) --  (0,-2) node[left]{$k-p$};
        \draw[thick,->-=0.5] (a) -- (2,-1) node[right]{$p$};
        \draw[->] (0.5,-2.5) -- node[above]{$\text{time}$} (1.5,-2.5);  
    \end{tikzpicture}
    \caption{Causality imposes \\\phantom{(b) }$\sgn (k+l)^0 = -\sgn l^0 = -\sgn (k-p)^0$.}
    \label{fig:two_loop_momentum_conservation}
    \end{subfigure}
    \caption{Diagrammatic representation of momentum conservation in one- and two-loop diagrams.
On-shell momenta must satisfy the respective conditions in eqs.~\eqref{eq:one-loop-threshold} and \eqref{eq:two-loop-threshold}, for the specified (or equivalently reversed) time direction.}
\end{figure}

The singularities of loop integrands are dictated by unitarity and causality and can easily be inferred from the Cutkosky cuts of the amplitude\footnote{
The singular surfaces can be deduced from the denominator structure of the contributing Feynman diagrams.
In particular, they become manifest after the energy component of the loop momenta has been integrated out, such as in the (causal) LTD, CFF, or TOPT representations.
However, the considerations presented in this section do not rely on a particular representation.
}.
In the one-loop amplitude the singularities lie on surfaces of the form
\begin{align}
\label{eq:one-loop-threshold}
    \threshold_1\equiv|\vec{k}|+|\vec{k}-\vec{p}|-|p^0|=0\,,
\end{align}
where $p$ can represent various linear combinations of external momenta, as we will discuss below.
In the $\Nf$-part of the two-loop amplitude we additionally find the surface
\begin{align}
\label{eq:two-loop-threshold}
    \threshold_2\equiv|\vec{l}| + |\vec{l}+\vec{k}|+|\vec{k}-\vec{p}|-|p^0| = 0\,.
\end{align}
The conditions $\threshold_1$ and $\threshold_2$ are a consequence of energy conservation of two and three on-shell loop momenta, respectively, along with a total momentum $p$.
Causality aligns the energy flow and thereby fixes the signs of the on-shell energies of the (massless) momenta propagating in the loop.
The conservation of the total spatial momentum is automatically satisfied by the loop momentum routing.
Therefore, eqs.~\eqref{eq:one-loop-threshold} and \eqref{eq:two-loop-threshold} are characteristic of the two- and three-body phase space and graphically depicted in figure~\ref{fig:one_loop_momentum_conservation} and \ref{fig:two_loop_momentum_conservation}, respectively.
\par
The solutions of $\threshold\in\{\threshold_1,\threshold_2\}$ is referred to as a pinch surface if $p^2= 0$ and as a threshold singularity if $p^2> 0$.
While threshold singularities are regularised by shifting poles away from the integration contour using the $\ii \epsilon$ causal prescription, pinch surfaces are not.
This is because, as $\epsilon\to0$, two poles pinch the contour, forcing it to pass through a non-integrable singularity.
If $p^2 < 0$ the equation $\threshold=0$ does not have solutions for real momenta and therefore does not describe singularities in the integration domain.
\par
For our purposes, we differentiate between three situations, which we discuss below.
\par
\subsubsection{Negative invariants $\left(p=p_1-p_S\right)$}
If $p^2 < 0$, the surfaces $\threshold_1$ and $\threshold_2$ do not have solutions.
They become pinch surfaces when $p^2 = 0$, requiring all $q_i$ for $i \in S$ to be lightlike and collinear to $p_1$.
However, transverse momentum and/or rapidity cuts within the phase space exclude these configurations, ensuring that the pinch surface is never reached.
\par

\subsubsection{$p=p_1$ or $p=p_2$}
In this case $p^2=0$ and the equations $\threshold_1$ and $\threshold_2$ describe pinch surfaces.
\par
The surface $\threshold_1$ is parameterised by the lightlike collinear (and soft) loop configurations $k=x p$, where $x\in[0,1]$ is a momentum fraction of $p$.
By writing $\vec{k} \to x \vec{p} + \delta \vec{k}_\perp$, we find that in the collinear limit the one-dimensional surface vanishes quadratically in the two transverse directions $\vec{k}_\perp$, such that the integrand of the one-loop amplitude scales as
\begin{align}
    \frac{\dd[3]{\vec{k}}}{\threshold_1} \sim \frac{\delta^2}{\delta^2}=\delta^0
\end{align}
and therefore develops a logarithmic divergence.
\par
The surface $\threshold_2$ is parameterised by lightlike (anti-)collinear loop momenta, denoted as $k = x_1 p$ and $l = -x_2 p$, where $x_{1,2}\in[0,1]$ are positive momentum fractions that satisfy $x_1 \geq x_2$.
In the six-dimensional integration domain of the two-loop amplitude, $\threshold_2$ is therefore a two-dimensional slice within the larger four-dimensional one-loop pinch surface $\threshold_1$, which only constrains $k$ to be a momentum fraction of $p$, while $l$ can be any on-shell vector.
In the (double) collinear limit, $\vec{k} \to x_1 \vec{p} + \delta \vec{k}_\perp$, $\vec{l}\to -x_2\vec{p}+\delta \vec{l}_\perp$, both $\threshold_2$ and $\threshold_1$ vanish quadratically in the four transverse directions:
\begin{align}
    \frac{\dd[3]{\vec{k}}\dd[3]{\vec{l}}}{\threshold_1\threshold_2} \sim \frac{\delta^4}{\delta^2\delta^2}=\delta^0\,,
\end{align}
leading to a logarithmic divergence.
\par
To obtain the correct scaling at the soft endpoints, one must account for the overlaps of pinch surfaces, which can vanish along with on-shell energies.
By employing the power counting procedure of refs.~\cite{Sterman:1978bi,Libby:1978qf,Collins:1989gx,Anastasiou:2018rib,Anastasiou:2020sdt,Anastasiou:2022eym}, we can identify that the soft configurations $k=0$ and $l=0$ also lead to logarithmic divergences at one loop and in our representation of the two-loop $\Nf$-part.
Both the collinear and soft divergences will be locally removed using the IR counterterms specified in section~\ref{subsec:babis_ct}.
\par

\subsubsection{Positive invariants $\left(p=p_S \right)$}
If $p^2>0$\,, the surface $\threshold_1$ describes a threshold singularity and is regularised by the $\ii \epsilon$ causal prescription.
To eventually enable numerical integration, we will instead subtract the singularity with the threshold counterterms described in section~\ref{subsec:thres_sub}.
\par
If $p^2=0$, all $q_i$ for $i\in S$ must be collinear to a common lightlike direction $k$.
In this case, $\threshold_1$ describes a pinch surface.
The loop momenta $l$ and $l-p$ then have to be collinear and anti-collinear to the same direction $k$, respectively.
They can be parameterised as $l=xk$, $l-p=-yk$ and $q_i=x_ik$ for some positive momentum fractions $x,y,x_i>0$ that satisfy $\sum_{i\in S} x_i = x+y$.
However, for this collinear configuration, the numerator contains
\begin{align}
    \slashed{k}\slashed{\epsilon}_i^*\slashed{k} = -\slashed{\epsilon}_i^*k^2 + 2 (\epsilon_i^* \cdot k) \slashed{k}\,
\end{align}
which, in the limit $k$ parallel to $q_i$, vanishes for physical (transverse) polarisation vectors $\epsilon_i^* \equiv \epsilon_i^*(q_i)$,
\begin{align}
    \epsilon_i^*\cdot q_i = \epsilon_i^* \cdot k = 0\,,
\end{align}
thereby preventing the amplitude from exhibiting an IR divergence.
\par
Note that the momentum $p$ in the surface $\threshold_2$ cannot be timelike.
Consequently, the threshold singularities of the two-loop $\Nf$-contribution are the same as those already encountered at one loop and are described by $\threshold_1$.

%% file: sections/methods_iruvcts.tex
\newpage
\subsection{Construction of IR- and UV- finite amplitudes}
\label{subsec:babis_ct}
In this section, we follow the procedure of refs.~\cite{Anastasiou:2020sdt,Anastasiou:2022eym} to locally remove the IR and UV singularities from the one-loop and the two-loop $\Nf$ amplitude.
The approach achieves a separation of the amplitude at the level of the integrand into a finite and a divergent part in $d=4$ spacetime dimensions
\begin{align}
\label{eq:ampl_finite_singular_decomposition}
    \mathcal{M}^{(n)} =  \mathcal{M}^{(n)}_\text{finite} + \mathcal{M}^{(n)}_\text{singular}.
\end{align}
For completeness we report the necessary details in sections~\ref{subsubsec:one-loop-amplitude} and \ref{subsubsec:two-loop-nf-amplitude}, respectively.

\subsubsection{The one-loop amplitude}\label{subsubsec:one-loop-amplitude}
The IR and UV divergences of the one-loop amplitude are locally removed by introducing suitable counterterms for the respective singularities.
More precisely, the full counterterm can be expressed as
\begin{align}
\label{eq:full_ct}
     \mathcal{M}^{(1)}_{\rm singular} = 
    \mathcal{R}_{\rm IR} \mathcal{M}^{(1)} +\mathcal{R}_{\rm UV} \mathcal{M}^{(1)} -\mathcal{R}_{\rm UV}\mathcal{R}_{\rm IR} \mathcal{M}^{(1)},
\end{align}
where $\mathcal{R}_{\rm IR} \mathcal{M}^{(1)}$ and $\mathcal{R}_{\rm UV} \mathcal{M}^{(1)}$ have the same local IR and UV behaviour as $\mathcal{M}^{(1)}$, respectively, and $\mathcal{R}_{\rm UV}\mathcal{R}_{\rm IR} \mathcal{M}^{(1)}$ removes the UV divergences introduced by the IR counterterm $\mathcal{R}_{\rm IR} \mathcal{M}^{(1)}$.
\par
The infrared counterterm corresponds to a form factor triangle diagram and reads
\begin{equation}\label{eq:form-factor-IR}
    \mathcal{R}_{\rm IR} \mathcal{M}^{(1)} =
    \mathcal{F}^{(1)} [\mathbf{P}_1 \widetilde{\mathcal{M}}^{(0)}\mathbf{P}_1 ]=
  {-\ii  C_F} \frac{\bar{v}(p_2)\gamma^{\mu} (\slashed{k}-\slashed{p}_2) \left[
    \mathbf{P}_1 \widetilde{\mathcal{M}}^{(0)}\mathbf{P}_1\right] (\slashed{p}_1 +\slashed{k})\gamma_{\mu} u(p_1)
    }{k^2 (k-p_2)^2  (k+p_1)^2} \, ,
\end{equation}
where 
\begin{equation}\label{eq:projector-P1}
\mathbf{P}_1=\frac{\slashed{p}_1\slashed{p}_2}{2 p_1\cdot p_2}\,
\end{equation}
is a projector
and $\widetilde{\mathcal{M}}^{(0)}$ is defined through the tree-level amplitude via
\begin{equation}
\mathcal{M}^{(0)} = \bar{v}(p_2) \widetilde{\mathcal{M}}^{(0)} u(p_1)\,.
\end{equation}
Note the above IR counterterm has an $s$-channel threshold singularity for $s=2p_1\cdot p_2>0$, which will have to be addressed before numerical integration.
\par
Its corresponding UV counterterm is 
\begin{equation}\label{eq:form-factor-IR-UV}
   \mathcal{R}_{\rm UV} \mathcal{R}_{\rm IR} \mathcal{M}^{(1)}(k) =
    {-\ii  C_F} \frac{\bar{v}(p_2)\gamma^{\mu} \slashed{k}\left[
    \mathbf{P}_1 \widetilde{\mathcal{M}}^{(0)}\mathbf{P}_1\right] \slashed{k}\gamma_{\mu} u(p_1)
    }{(k^2-\mUV^2)^3} \, ,
\end{equation}
where an extra mass parameter $M$ is introduced to avoid originating additional IR singularities while retaining the leading $k\to\infty$ behaviour of eq.~\eqref{eq:form-factor-IR}.
The UV mass $M$ can be considered a renormalisation scale, whose dependence will be replaced by $\muren$ once combined with the integrated counterterms -- a change in the UV-renormalisation scheme.
\par
The UV divergences of the one-loop amplitude either arise from the subamplitudes with a vertex $\mathcal{M}^{(1)}_\text{vertex}$ or a quark self-energy correction $\mathcal{M}^{(1)}_\text{self-energy}$:
\begin{align}
\label{eq:uv_ct}
    \mathcal{R}_\text{UV} \mathcal{M}^{(1)}
    = \mathcal{R}_\text{UV}\mathcal{M}^{(1)}_\text{vertex} + \mathcal{R}_\text{UV}\mathcal{M}^{(1)}_\text{self-energy}\,.
\end{align}
\par
The counterterm $\mathcal{R}_\text{UV}\mathcal{M}^{(1)}_\text{vertex}$ is constructed by replacing the vertex correction
\begin{align}
g_0\Gamma^{(1),\nu}(p,q,k)=
&
\parbox{20mm}{\includegraphics[width=0.125\textwidth,page=2]{figures/template_diagrams.pdf}}
={-g_0 g_{0,s}^2C_F } \frac{\gamma^{\mu}(\slashed{p}- \slashed{q}+\slashed{k})V^{\nu}(\slashed{p}+\slashed{k})\gamma_{\mu}}{(p-q+k)^2(p+k)^2k^2}\,,
\end{align}
in each Feynman diagram with
\begin{equation}
    \mathcal{R}_{\rm UV}[g\Gamma^{(1),\nu}(p,q,k)]= {-g_0 g_{0,s}^2C_F } \frac{\gamma^{\mu}\slashed{k}V^{\nu}\slashed{k}\gamma_{\mu}}{(k^2-\mUV^2)^3}\,,
\end{equation}
where $g_{0}V^{\nu}=Q e_{0} \gamma^{\nu}$ or  $g_{0}V^{\nu}= e_{0} (v \gamma^{\nu}-a\gamma^{\nu} \gamma^5)$ depending on the external gauge boson and the quark flavour.
Similarly, in order to construct $\mathcal{R}_\text{UV}\mathcal{M}^{(1)}_\text{self-energy}$, we replace the self-energy correction
\begin{align}
\Pi^{(1)}(p,k)&=\parbox{20mm}{\includegraphics[width=0.125\textwidth, page=1]{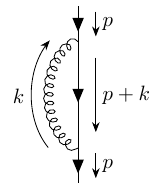}}
 = {-  g_{0,s}^2 C_F} \frac{\gamma^{\mu}(\slashed{p}+\slashed{k})\gamma_{\mu}}{(p+k)^2k^2}\,,
\end{align}
with its corresponding UV counterterm from ref.~\cite{Anastasiou:2022eym}
\begin{equation}
    \mathcal{R}_{\rm UV}[\Pi^{(1)}(p,k)]= -{ g_{0,s}^2 C_F}\left( \frac{\gamma^{\mu}(\slashed{p}+\slashed{k})\gamma_{\mu}}{(k^2-\mUV^2)^2}
    -2k\cdot p  \frac{\gamma^{\mu}\slashed{k}\gamma_{\mu}}{(k^2-\mUV^2)^3}
    \right)\,.
\end{equation}
Note that the above UV counterterms do not introduce additional threshold singularities.
\par

Finally, the finite remainder of the one-loop amplitude reads
\begin{align}
\label{eq:1_loop_finite_remainder}
    R^{(1)}
    &= \bare_\text{finite}^{(1)} + R^{(1)}_\text{singular}\,,
\end{align}
where
\begin{align}
\label{eq:r_1_singular}
    R^{(1)}_\text{singular}
    &= \bare^{(1)}_\text{singular} - \mathbf{Z}^{(1)} \bare^{(0)}
    + \mathcal{O}(\ep)
\end{align}
is given explicitly in appendix~\ref{app:finite_remainders}.

\subsubsection{The $\Nf$-contribution to the two-loop amplitude}
\label{subsubsec:two-loop-nf-amplitude}
The local IR and UV counterterms for the two-loop $\Nf$-amplitude of eq.~\eqref{eq:bub-nf} were devised in ref.~\cite{Anastasiou:2020sdt}.
A separation of the two-loop $\Nf$ integrand into finite and singular parts, as in eq.~\eqref{eq:ampl_finite_singular_decomposition}, is then achieved by
\begin{equation}
    \label{eq:M2-Nf-finite}
    \mathcal{M}^{(2,\Nf)}_{\rm finite} 
(k,l)= \nfcoeff \left( \frac{1}{l^2(l+k)^2} 
-\frac{1}{(l^2-\mUV^2)^2} \right)\mathcal{M}^{(1)}_{\rm finite}(k)\,,
\end{equation}
and
\begin{equation}
    \label{eq:M2-Nf-singular}
    \mathcal{M}^{(2,\Nf)}_{\rm singular} 
(k,l)= \nfcoeff \left( \frac{
\mathcal{M}^{(1)}_{\rm singular}(k)
}{l^2(l+k)^2
} 
+\frac{\mathcal{M}^{(1)}_{\rm finite}(k)}{(l^2-\mUV^2)^2} \right)\,.
\end{equation}

To evaluate the Catani finite remainder, we follow the same procedure as was used in ref.~\cite{Haindl:2022hcb} for the corresponding QED correction.
The calculation is more subtle than at one loop, as it mixes one- and two-loop contributions.
The finite remainder reads
\begin{align}
\label{eq:R-2-nf}
    R^{(2,\Nf)} &= \bare^{(2,\Nf)}
    - \frac{2\beta_0^{(\Nf)}}{\ep} \bare^{(1)}
    - \mathbf{Z}^{(2,\Nf)} \bare^{(0)} +\mathcal{O}(\ep) \\
    &= \bare^{(2,\Nf)}_\text{finite} + \bare^{(2,\Nf)}_\text{singular}
    - \frac{2\beta_0^{(\Nf)}}{\ep}\left(\bare^{(1)}_\text{finite} + \bare^{(1)}_\text{singular}\right)
    - \mathbf{Z}^{(2,\Nf)} \bare^{(0)}+\mathcal{O}(\ep) \,,
\end{align}
where we used that $\mathbf{Z}^{(1)}$ is independent of $\Nf$ and $\beta_0^{(\Nf)}$ is the $\Nf$-part of $\beta_0$,
\begin{equation}
    \beta_0^{(\Nf)} = -\frac{2}{3}\Nf T_F\,.
\end{equation}
Using the counterterm in eq.~\eqref{eq:M2-Nf-singular}, the finite remainder of the $\Nf$ part of the two-loop amplitude can be expressed as
\begin{align}
\label{eq:fin_rem_2_nf}
        R^{(2,\Nf)} = \bare^{(2,\Nf)}_\text{finite} +
   \bare^{(1)}_\text{finite}  R^{(1,\Nf)}_\text{singular}  + R^{(2,\Nf)}_\text{singular}\,,
\end{align}
where the coefficients are computed analytically in $d=4-2\ep$ dimensions and read
\begin{align}
    \label{eq:fin_rem_2l_nf_coeff_m1}
     R^{(1,\Nf)}_\text{singular}  &=
     -\frac{2\beta_0^{(\Nf)}}{\ep}
    +
    \nfcoeff
    \int 
    [\dd^d l]
    \frac{1}{(l^2-\mUV^2)^2}+\mathcal{O}(\ep)
    = \frac{4}{9} \Nf T_F \left(
    1 - 3\log\left(\frac{\mu^2}{\mUV^2}\right)
    \right)\,,
    \\
    \label{eq:fin_rem_2_nf_explicit}
    R^{(2,\Nf)}_\text{singular} &=
    -\frac{2\beta_0^{(\Nf)}}{\ep}
    \bare^{(1)}_\text{singular}
    +
    \nfcoeff
    \int
    [\dd^d k]
    [\dd^d l]
    \frac{
        \mathcal{M}^{(1)}_\text{singular}(k)
    }{l^2(l+k)^2}
    - \mathbf{Z}^{(2,\Nf)} \bare^{(0)}+\mathcal{O}(\ep)\,.
\end{align}
The integrated expression for $R^{(2,\Nf)}_\text{singular}$ is given explicitly in appendix~\ref{app:finite_remainders}.
We emphasise that contributions of order $O(\ep)$ of $M^{(1)}_{\rm finite}$,
whose calculation would be tedious and cumbersome especially if $\gamma^5$ is present, are not needed in eq.~\eqref{eq:fin_rem_2_nf}, since the corresponding coefficient $R^{(1,\Nf)}_\text{singular}$ is finite.

%% file: sections/methods_thresholds.tex
\newpage
\subsection{Treatment of thresholds}
\label{sec:causal_prescription}
Although the amplitude $\bare_\text{finite}^{(n)}$ is UV- and IR-finite by construction, it is still challenging to evaluate using Monte Carlo numerical integration.
This is because it relies on the $\ii\epsilon$ causal prescription of the propagators to be well-defined.
Without it, the on-shell poles of the integrand $\mathcal{M}_\text{finite}^{(n)}$ fall within the integration domain.
While the $\ii\epsilon$ causal prescription formally regularises the integral, it is not suitable for numerical integration.
In this section, we explain how we construct an integrand representation that is appropriate for Monte Carlo integration.
\par

\subsubsection{Integration of loop energies}
\label{subsec:ltd}
First, we focus on the two poles implied by each (inverse) propagator
\begin{align}
    D_i = l_i^2 - m_i^2 + \ii \epsilon = (l_i^0 - E_i)(l_i^0 + E_i)\,,
\end{align}
which occur at positive and negative on-shell energies.
According to the residue theorem, we can explicitly get rid of them by analytically integrating out the energy component of the loop momenta:
\begin{align}
\label{eq:ltd}
    \int\limits \left(\prod_{j=1}^{n} \frac{\dd k_j^0}{\pi^2}\right)
    \mathcal{M}_\text{finite}^{(n)}
    \left(\{k_i\}\right)
    = 
    \ltd^{(n)}
    \left(\{\vec{k}_i\}\right)\,.
\end{align}
The resulting integrand over the spatial loop momenta $\ltd^{(n)}$ is a rational function of the on-shell energies and can be written in various algebraic forms.
These different, yet locally identical, expressions include the LTD~\cite{Catani:2008xa,Capatti:2019ypt}, CFF~\cite{Capatti:2022mly}, and TOPT \cite{Bodwin:1984hc, Collins:1985ue, Sterman:1993hfp, Sterman:1995fz} representations.
Below, we address their respective advantages.
\par
The LTD representation is the most compact of the three.
Diagrammatically, it relates a loop graph to the sum of all its spanning trees, where the momenta flowing through the remaining (cut) edges are on-shell.
However, the expression has spurious singularities, which lead to large local cancellations among the summands and consequently impair the stability of the numerical evaluation.
Moreover, the construction of the LTD expression in the presence of raised propagators becomes cumbersome as it involves taking derivatives.
Nonetheless, for our purposes, the LTD representation serves as a convenient intermediate expression.
More explicitly, the phase space residues, introduced in section~\ref{subsec:thres_sub}, are in a more compact form when derived from LTD rather than CFF or TOPT.
Moreover, their interpretation as interferences of tree diagrams becomes manifest.

\par
The CFF and TOPT expressions share many similarities.
In particular, their construction is unaffected by the presence of raised propagators and the numerators of the individual terms take the same form.
The CFF expression distinguishes itself from TOPT and LTD through its absence of spurious singularities, making its evaluation more numerically stable.
This advantage of CFF, despite typically involving more terms, makes it our preferred representation for numerical evaluation.

\subsubsection{Subtraction of threshold singularities}
\label{subsec:thres_sub}
After analytic integration over the loop energies, the finite amplitude is expressed as an integral over spatial loop momenta.
The integrand $\ltd^{(n)}$, defined in eq.~\eqref{eq:ltd}, still has poles for loop momentum configurations, where $n+1$ propagators go on-shell.
Those poles are precisely the threshold singularities that are responsible for the absorptive part of the amplitude.
Their existence depends on the external momentum configuration and internal masses.
In section~\ref{subsec:analytic} we already discussed the thresholds of the one-loop and the two-loop $\Nf$-contributions.
Following the procedure outlined in ref.~\cite{Kermanschah:2021wbk}, we will locally eliminate these threshold singularities and thereby separate dispersive and absorptive part of the amplitude into two distinct integral representations.
We can safely remove the $\ii\epsilon$ causal prescription from both integrals and efficiently calculate them using Monte~Carlo numerical integration.
We stress that in the phase-space integrated virtual corrections presented in section~\ref{sec:results}, the absorptive part of the amplitude could be ignored as it cancels in the result.
However, as shown in section~\ref{app:fixed_ps_results} for $Z\gamma_1^*\gamma_2^*$ production at fixed phase-space points, the absorptive part is generally non-zero.
\par

We intend to locally subtract the threshold singularities of the amplitude using suitable counterterms.
We first express the integrand $\ltd^{(n)}$ in hyperspherical coordinates $(\vec{k}_1,\dots,\vec{k}_n)=\loopmomenta=r \unitvector$ and expand it in the vicinity of a single threshold singularity~$\threshold$
\begin{align}
\label{eq:threshold_expansion}
    \ltd(r \unitvector) = \frac{1}{r^{3n-1}}
    \frac{\Res\left[r^{3n-1} \ltd,r=r_\threshold(\unitvector)\right]}{r-r_\threshold(\unitvector)} + \mathcal{O}
    \left(\left(r-r_\threshold(\unitvector)\right)^0\right)\,,
\end{align}
where $r_\threshold(\unitvector)$ parameterises the threshold singularity for any unit direction $\unitvector$, defined implicitly through the on-shell condition
\begin{align}
\label{eq:threshold_param}
    \threshold(r_\threshold(\unitvector)\unitvector) = 0\,.
\end{align}
Any valid threshold counterterm has to subtract the leading term in the above expansion.
\par
Our parametrisation is characterised by the origin of the spherical coordinates system, which we call the source $\source$.
It plays a crucial role for two main reasons:
\begin{enumerate}
    \item We demand the source to lie within the threshold, $\threshold(\source)<0$.
    If it does not, the residue in eq.~\eqref{eq:threshold_expansion} will have undesirable integrable singularities at the tangents of the surface $\threshold(\vec{\mathbf{k}})=0$ through the source.
    \item The source is also constrained by the intersections of thresholds.
    At these intersections, the expansion in eq.~\eqref{eq:threshold_expansion} generally\footnote{
    \label{footnote:poles}
At an intersection $\loopmomenta_*=r_*\unitvector_*$ of two thresholds $\threshold_1$ and $\threshold_2$, along some direction $\unitvector_*$ where $\threshold_1(\loopmomenta_*)=\threshold_2(\loopmomenta_*)=0$, the integrand has higher-order poles at $r=r_*$ if $\ltd\propto (\threshold_1\threshold_2)^{-1}$ but not if $\mathcal{I}\propto \threshold_1^{-1}+\threshold_2^{-1}$.
    } has to be modified to include higher-order poles.
    To correctly subtract these poles, the intersecting thresholds must be parameterised with respect to the same source.
\end{enumerate}
Consequently, the source must lie inside all intersecting thresholds simultaneously\footnote{
This requirement is related to the construction of the contour deformation of ref.~\cite{Capatti:2019edf}, where the overlaps of all ellipsoids are identified algorithmically.
}.
If this is not possible, one may resort to a multi-channelling procedure, where the integrand is partitioned into channels, as
\begin{align}
\label{eq:multi-channelling}
    \ltd = \sum_{c \in C} m_c \ltd\,, \qquad \text{where} \qquad \sum_{c\in C}m_c=1\,.
\end{align}
The channels must be constructed in such a way that a single source $\source_c$ can be identified for each channel $m_c\ltd$. 
In the following, we will investigate the thresholds and their overlap structure for our process of interest.
\par
\input{figures/thresholds}
First, we split the finite amplitude into partial amplitudes $\bare^{(n,\sigma)}_\text{finite}$, whose external bosons are attached in a fixed order $\sigma$, as
\begin{align}
    \bare^{(n)}_\text{finite} = \sum_{\sigma\in S_\nfinal} \bare^{(n,\sigma)}_\text{finite}\,,
\end{align}
where $S_\nfinal$ is the set of all permutations of $\nfinal$ external bosons.
Note that according to the counterterms in section~\ref{subsec:babis_ct}, the partial amplitudes $\bare^{(n,\sigma)}_\text{finite}$ are individually IR- and UV-finite.
We can therefore consider them separately.
\par
\newcommand{\coloredthreshold}[1]{
    \ifthenelse{\equal{#1}{12}}{\textcolor{red}{\threshold_{s_{12}}}}{
    \ifthenelse{\equal{#1}{13}}{\textcolor{green}{\threshold_{s_{13}}}}{
    \ifthenelse{\equal{#1}{23}}{\textcolor{cyan}{\threshold_{s_{23}}}}{
    \ifthenelse{\equal{#1}{1}}{\textcolor{blue}{\threshold_{s_1}}}{
    \ifthenelse{\equal{#1}{3}}{\textcolor{orange}{\threshold_{s_3}}}{
    \ifthenelse{\equal{#1}{2}}{\textcolor{violet}{\threshold_{s_2}}}{
    err}}}}}}
}
The partial amplitudes $\bare^{(1,\sigma)}_\text{finite}$ and $\bare^{(2,\Nf,\sigma)}_\text{finite}$ at one loop and for the $\Nf$-part at two loops, respectively, have
\begin{align}
    \begin{pmatrix}
    \nfinal+1 \\ 2
    \end{pmatrix} -\nfinal_0
\end{align}
thresholds, where $\nfinal_0$ external vector bosons are massless and $\nfinal-\nfinal_0$ are massive.
In table~\ref{tab:thresholds}, we explicitly list the thresholds that appear in their respective integrands $\ltd^{(1,\sigma)}$ and $\ltd^{(2,\Nf,\sigma)}$ with two and three external vector bosons.
The corresponding Cutkosky cuts for the $2\to3$ process are illustrated in figure~\ref{subfig:cutkosky}. 
The threshold singularities are defined as
\begin{align}
\label{eq:threshold_first}
    \coloredthreshold{12} &= E_1 + E_2 - (p_1^0+p_2^0)\,, \\
    \coloredthreshold{13} &= E_1 + E_4 - (q_1^0+q_3^0)\,, \\
    \coloredthreshold{23} &= E_2 + E_3 - (q_2^0+q_3^0)\,, \\
    \coloredthreshold{1} &= E_1 + E_3 - q_1^0 \,,\\
    \coloredthreshold{3} &= E_3 + E_4 - q_3^0 \,,\\
    \label{eq:threshold_last}
    \coloredthreshold{2} &= E_2 + E_4 - q_2^0\,,
\end{align}
where we choose the momentum routing such that the on-shell energies are
\begin{align}
    E_1 = |\vec{k}+\vec{p}_1|, \quad
    E_2 = |\vec{k}-\vec{p}_2|, \quad
    E_3 = |\vec{k}+\vec{p}_1-\vec{q}_1|, \quad
    E_4 = |\vec{k}-\vec{p}_2+\vec{q}_2|\,.
\end{align}
While in the $2\to3$ process with massive external bosons all of the above thresholds exist, the surfaces involving $q_3$ drop out in the $2\to2$ process.
Moreover, the surface $\threshold_{s_i}$ only exists if $q_i^2=m_i^2>0$.
\par

\input{tables/thresholds_tab}

The threshold singularities $\threshold(\vec{k})=0$ describe rotational ellipsoids in spatial loop momentum space of $\vec{k}$.
For a single ellipsoid, a natural choice for the source $\vec{s}$ may be one of the two focal points, i.e. $\vec{k}=0$ and $\vec{k}=-\vec{p}$ for the ellipsoid $|\vec{k}+\vec{p}|+|\vec{k}|-p^0 = 0$.
\par
Moreover, the source must lie within the overlaps of intersecting ellipsoids.
Note that thresholds change their shape and their intersections can generally move and can possibly disappear under Lorentz transformations.
However, the presence of those intersections that result in higher-order poles (see footnote~\ref{footnote:poles}) is a Lorentz invariant property.
These intersections are constrained by the scattering kinematics.
A hierarchy becomes apparent if we arrange them in a diagram
\begin{align}
\label{eq:threshold_hierarchy}
    \begin{tikzpicture}
        \node[draw] at (0, 0) (a) {$\coloredthreshold{12}$};
        \node[draw] at (-1, -1) (b1) {$\coloredthreshold{13}$}; 
        \node[draw] at (1, -1) (b2) {$\coloredthreshold{23}$}; 
        \node[draw] at (-2, -2) (c1) {$\coloredthreshold{1}$};
        \node[draw] at (0, -2) (c2) {$\coloredthreshold{3}$};
        \node[draw] at (2, -2) (c3) {$\coloredthreshold{2}$};
        \draw (a) -- (b1);
        \draw (a) -- (b2);
        \draw (b1) -- (c1);
        \draw (b1) -- (c2);
        \draw (b2) -- (c2);
        \draw (b2) -- (c3);
    \end{tikzpicture}
\end{align}
where the parent and child nodes $P$ and $C$ satisfy $P\leq C$, as a consequence of the triangle inequality.
Therefore, the ellipsoid $C=0$ is contained within the ellipsoid $P=0$.
Due to the this hierarchy, it is sufficient to understand the overlaps of the smallest ellipsoids, which we will address next.
\par
In the centre-of-mass (COM) frame, where $p_1+p_2$ is at rest, the two focal points at $\vec{k}=-\vec{p}_1$ and $\vec{k}=\vec{p}_2$ coincide and automatically lie inside all ellipsoids except $\coloredthreshold{3}$.
Therefore, for the $2\to2$ process with massless and massive final momenta as well as the $2\to3$ with massless final momenta, where $\coloredthreshold{3}$ does not exist, we can choose the source at $\vec{s}=-\vec{p}_1=\vec{p}_2$.
For the $2\to3$ process with massive final momenta, the threshold $\coloredthreshold{3}$ exists and a single point in the interior of all ellipsoids can in general not be found.
A plot of the singular surfaces for particular phase space configuration is illustrated in figure~\ref{subfig:ellipsoids}.
In the rest frame of $p_1 + p_2$ there are three pairwise overlapping regions among the threshold singularities $\coloredthreshold{1}, \coloredthreshold{3}$ and $\coloredthreshold{2}$.
Therefore, we will resort to the multi-channelling procedure of eq.~\eqref{eq:multi-channelling} to construct three channels, as summarised in table~\ref{tab:thresholds}.
\par
Strictly speaking, one can actually avoid multi-channelling by boosting the external momenta to the frame where $q_3$ is at rest.
In this Lorentz frame, the surface $\coloredthreshold{3}=0$ is a sphere, whose centre $\vec{s}=-\vec{p}_1+\vec{q}_1=\vec{p}_2-\vec{q}_2$ simultaneously lies inside all ellipsoids.
In practice, this means that each subamplitude with fixed final state ordering has to be evaluated in a different boosted frame.
\par
Since the thresholds of the two-loop $\Nf$-part and the one-loop amplitude are the same, the above considerations apply for both.
Note that also at two loops a parametrisation of the thresholds in the magnitude $|\vec{k}|$ would be sufficient.
However, in future two-loop applications, such a projection to a one-loop problem is generally not possible anymore.
Therefore, we chose to parameterise the thresholds in the magnitude $r$ of the combined loop momenta $(\vec{k},\vec{l})=r \unitvector$.
Note that in this case the source reads $\source=(\vec{s}_1,\vec{s}_2)$, where we picked $\vec{s}_2=0$, although in principle any other vector would be fine.
\par
Eventually, we can express the dispersive part of the amplitude as an integral over $n$ spatial loop momenta,
\begin{align}
\label{eq:dispersive}
    \dispersive I &= 
    \sum_{c\in C}
    \int \dd[3n]{\loopmomenta}
    \left(
    \ltd_c(\loopmomenta)
    - \sum_{\threshold \in\thresholdset_c} \thresholdct_\threshold[\ltd_c](\loopmomenta)
    \right),
\end{align}
where $\ltd_c(\loopmomenta) = m_c\ltd(\loopmomenta-\source_c)$
and the absorptive part as a phase space integral,
\begin{align}
    \label{eq:absorptive}
    \absorptive I &=
    \pi\sum_{c\in C}
    \int \dd[3n-1]{\unitvector}
    \sum_{\threshold \in \thresholdset_c} 
    R_t[\ltd_c](\unitvector)\,,
\end{align}
where $r_\threshold$ is the solution $r=r_\threshold>0$ of the on-shell condition 
\begin{align}
\label{eq:os_radius}
    \threshold(r\unitvector-\source)=0\,.
\end{align}
For our purposes, the threshold condition reduces to a quadratic equation, whose solution\footnote{
Since ref.~\cite{Kermanschah:2021wbk} assumed $\hat{k}$ to be a unit vector, we need to modify its eq.~(3.18) and replace $1$ by $\hat{k}^2$.
}
was explicitly worked out in eq.~(3.21) of ref.~\cite{Kermanschah:2021wbk}.
We emphasise that the expressions in eqs.~\eqref{eq:dispersive} and \eqref{eq:absorptive} are locally integrable, allowing for the safe removal of the $\ii\epsilon$ causal prescription.
This makes them well-suited for direct numerical integration using Monte Carlo methods.
\par
The threshold counterterm we define as
\begin{align}
    \thresholdct_\threshold
    \left[
    f(r\unitvector)
    \right]
    &=
    \frac{R_\threshold[f]}{r^{3n-1}}
    \left(
    \frac{
        \chi(r-r_\threshold)
    }{
        r-r_\threshold -\ii\epsilon
        }
    +
    \frac{
    \chi(-r-r_\threshold)
    }{
        -r-r_\threshold -\ii\epsilon
        }
    \right)\,,
\end{align}
where $\chi$ is any symmetric function around the origin, which satisfies $\chi(0)=1$ and $\chi(x)\to 0$ for $x\to \pm\infty$.
For our results we opted for a Gaussian with a cutoff,
\begin{align}
    \chi(x) = \exp\left[-\left(\frac{x}{E_\text{CM}}\right)^2\right]
    \Theta\left[(3 E_\text{CM})^2-x^2\right]\,.
\end{align}
The integrated counterterm only contributes to the absorptive part of the integral since the Cauchy principal value vanishes by construction, i.e.
\begin{align}
   \int_0^\infty \dd{r} r^{3n-1} \thresholdct_\threshold\left[f(r\unitvector)\right] = \ii \pi R_\threshold[f](\unitvector)\,.
\end{align}
The residue reads
\begin{align}
\label{eq:ps_residue}
    R_\threshold[f(\loopmomenta)] = \Res[r^{3n-1}f(r\unitvector),r=r_\threshold(\unitvector)]
    &= 
    \left. 
    \frac{r^{3n-1}}{\frac{\dd \threshold}{\dd r}}
    \lim_{\threshold\to 0} \left(\threshold  f\right)
    \right\vert_{r=r_\threshold(\unitvector)}\,.
\end{align}
At one loop, it is precisely the (helicity summed) interference of the two trees separated by the Cutkosky cut, including the phase space measure.
Specifically, the denominator is an artifact of the parametrisation.
Therefore, the expression for the absorptive part in eq.~\eqref{eq:absorptive} constitutes a local representation of the generalised optical theorem.
In the two-loop $\Nf$-part, the residue is the one-loop interference multiplied by a scalar bubble after integration over the loop energy.
\par
The multi-channelling is determined by the groups of thresholds $\thresholdset_c$ in table~\ref{tab:thresholds}.
Naturally, they must satisfy $\bigcup_{c\in C} \thresholdset_c = \thresholdset$.
We define the multi-channelling coefficients as
\begin{align}
\label{eq:multi_channel_factor}
    m_c = 
    \frac{
      \left(
            \prod_{\threshold\in\thresholdset\setminus\thresholdset_c} \frac{\threshold}{E_\text{CM}}
            \right)^2
    }{
        \sum_{c\in C}\left(
            \prod_{\threshold\in\thresholdset\setminus\thresholdset_c} \frac{\threshold}{E_\text{CM}}
            \right)^2
    }\,,
\end{align}
where $C$ denotes the set of channels and $\sum_{c \in C} m_c =1$.
Note that by design the denominator cannot be zero, and the numerator cancels all thresholds in $\ltd_c=m_c\ltd$ except for the desired ones in $\thresholdset_c$.
For numerical stability, it is important to algebraically perform the cancellations between the numerator of $m_c$ and the denominator of $\ltd$.
Additionally, since the number of thresholds per group is generally different, we normalise by $E_\text{CM}$ so that all terms are dimensionless.
This normalisation is essential for maintaining numerical stability.
\par
Note that the residue in eq.~\eqref{eq:ps_residue} is generally more compact if derived from the LTD expression than from CFF, since in the former $\threshold$ generally appears in fewer summands.
Moreover, we will exploit that the cross section only depends on the dispersive part of the amplitude.
Therefore, the absorptive part in eq.~\eqref{eq:absorptive} is not needed for the results presented in section~\ref{sec:results}.

%% file: figures/thresholds.tex
\begin{figure}[b]
    \centering
    \begin{subfigure}{0.55\textwidth}
    \centering
    \resizebox{!}{3.8cm}{
    \begin{tikzpicture}[scale=3]
    \draw [thick, pattern=north west lines] (2,1) rectangle (1,0) node[pos=.5,text width=3cm,align=center] {one loop or \\ two-loop-$\Nf$}; 
        \coordinate (p2a) at (0.5,0);
        \coordinate (p1a) at (0.5,1);
        \coordinate (p2b) at (1,0);
        \coordinate (p1b) at (1,1);
        \coordinate (q1a) at (2,1);
        \coordinate (q1b) at (2.5,1);
        \coordinate (q2a) at (2,0);
         \coordinate (q2b) at (2.5,0);
           \coordinate (q3a) at (2,0.5);
         \coordinate (q3b) at (2.5,0.5);
      \draw[thick,->-=0.5] (p1a) node[left] {$p_1$}  -- (p1b);
        \draw[thick,->-=0.5] (q1a) --  (q1b) node[right] {$q_1$};
        \draw[thick,->-=0.5] (p2a) node[left] {$p_2$}  --  (p2b);
        \draw[thick,->-=0.5] (q2a) -- (q2b) node[right] {$q_2$};
        \draw[thick,->-=0.5] (q3a) -- (q3b) node[right] {$q_3$};
        \draw [line width=2.5pt,red,opacity=0.6] (1.5,1.125) -- (1.5,-0.125);
        \draw [line width=2.5pt,blue,opacity=0.6] (1.75,1.125) -- (2.125,0.75);
        \draw [line width=2.5pt,green,opacity=0.6] (1.625,1.125) --  (2.1,0.125);
        \draw [line width=2.5pt,violet,opacity=0.6] (1.75,-0.125) -- (2.125,0.25);
        \draw [line width=2.5pt,cyan,opacity=0.6] (1.625,-0.125) --  (2.1,0.875);
        \draw [line width=2.5pt,orange,opacity=0.6] (2.125,0.25) arc (-85:-275:0.25) ;
  \end{tikzpicture}
  }
  \caption{Cutkosky cuts.\\~\\~}
  \label{subfig:cutkosky}
  \end{subfigure}%
  \hfill
  \begin{subfigure}{0.45\textwidth}
    \centering
      \includegraphics[height=3.8cm]{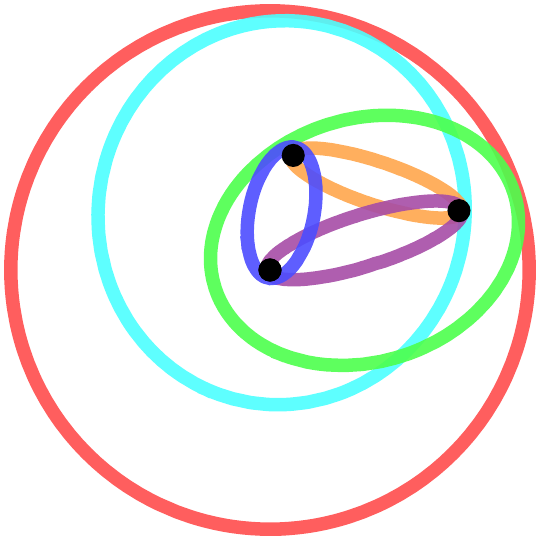}
      \vfill
      \caption{Threshold singularities  on a slice of spatial one-loop momentum space. They lie on ellipsoids, whose focal points are marked in black.}
      \label{subfig:ellipsoids}
  \end{subfigure}
    \caption{Thresholds of the one-loop and the $\Nf$-part of the two-loop partial amplitude for $q(p_1)\bar q(p_2)\to V(q_1)V(q_3)V(q_2)$, where the massive electroweak vector bosons $V$ are attached in the depicted order.
    With matching colours we highlight in (\protect\subref{subfig:cutkosky}) the Cutkosky cuts, in (\protect\subref{subfig:ellipsoids}) the singular surfaces for one particular configuration in phase space in the COM frame with example external boson masses
    $m_{\{1,2,3\}}=\hat{E}_\text{CM}/\{8,10,12\}$, and in \crefrange{eq:threshold_first}{eq:threshold_last} their respective definitions.
    }
    \label{fig:thresholds}
\end{figure}

%% file: tables/thresholds_tab.tex
\begin{table}[t]
    \centering
    \def\arraystretch{1.3}%
    \begin{tabular}{|l|l|c|c|}
    \hline
    \multirow{2}{*}{Process (fixed ordering cf.~\ref{subfig:cutkosky})} & \multirow{2}{*}{
        Thresholds $\thresholdset$
    } & \multicolumn{2}{c|}{Channels $C$ in COM frame $(\vec{p}_1=-\vec{p}_2)$} \\
    \cline{3-4}
     &   & Groups $\thresholdset_c$ & Source $\vec{s}_c$ \\
     \hline
        $q(p_1) \bar q(p_2) \to \gamma(q_1) \gamma(q_2)$ 
        &$\{\coloredthreshold{12}\}$ 
        &$\thresholdset$
        & $\vec{p}_2$\\
        \hline
        $q(p_1) \bar q(p_2) \to V(q_1) V(q_2)$
        & $\{\coloredthreshold{12},\coloredthreshold{1},\coloredthreshold{2}\}$
        & $\thresholdset$
        & $\vec{p}_2$\\
        \hline
        $q(p_1) \bar q(p_2) \to \gamma(q_1) \gamma(q_3) \gamma(q_2)$ 
        & $\{\coloredthreshold{12},\coloredthreshold{13},\coloredthreshold{23}\}$ 
        & $\thresholdset$
        & $\vec{p}_2$\\
        \hline
        \multirow{3}{*}{$q(p_1) \bar q(p_2) \to V(q_1) V(q_3) V(q_2)$}
        & \multirow{3}{*}{\shortstack[c]{$\{\coloredthreshold{12},\coloredthreshold{13},\coloredthreshold{23},$\\
        $\coloredthreshold{1},\coloredthreshold{3},\coloredthreshold{2}\}$}}
        & $\{\coloredthreshold{1},\coloredthreshold{3}\}$
        & $-\vec{p}_1+\vec{q}_1$\\
        \cline{3-4}
        & & $\{\coloredthreshold{3},\coloredthreshold{2}\}$
        & $\vec{p}_2-\vec{q}_2$\\
        \cline{3-4}
        & & $\{\coloredthreshold{12},\coloredthreshold{13},\coloredthreshold{23},\coloredthreshold{1},\coloredthreshold{2}\}$
        & $\vec{p}_2$\\
        \hline
    \end{tabular}
    \caption{The channels and their respective parametrisations according to the thresholds present in the one-loop and the two-loop-$\Nf$ contributions to the $2\to2$ and $2\to3$ processes.
    By $V$ we denote any massive vector boson.}
    \label{tab:thresholds}
\end{table}

%% file: sections/implementation.tex
\section{Implementation}
\label{sec:implementation}
In this section we describe the numerical implementation we used to produce the results in section~\ref{sec:results}.
Some aspects of our pipeline are generic for arbitrary processes, while others are specifically tailored to our current application.
Our pipeline is separated into two parts: integrand generation and Monte~Carlo integration.
In the following sections we discuss these two parts in detail.

\subsection{Integrand generation}
\label{subsec:integrand_generation}
Our starting point is the list of Feynman diagrams that contribute to the amplitude with final state photons, generated by \textsc{QGRAF} \cite{Nogueira:1991ex}.
The Feynman rules are substituted in and the IR and UV counterterms of section~\ref{subsec:babis_ct}
are supplied as extra diagrams.
Each diagram is of the form
\begin{align}
\label{eq:diagram_generation}
   \frac{\mathcal{N}(l_1,\dots,l_m)}{\prod_{i=1}^m D_i}
\end{align}
with a numerator $\mathcal{N}$ that depends on the propagator momenta $l_i$ and quadratic Feynman propagators
\begin{align}
    D_i = l_i^2 - m_i^2 + \ii \epsilon\,.
\end{align}
The propagator momenta $l_i$ are a linear combination of loop momenta $k_j$ and incoming and outgoing momenta $p_j$ and $q_j$, respectively.
For each diagram, we generate both the CFF representation $\ltd_\text{CFF}$ of ref.~\cite{Capatti:2022mly} and the LTD representation $\ltd_\text{LTD}$ of ref.~\cite{Capatti:2019ypt}.
Recall that the two representations are locally identical, $\ltd_\text{CFF} =\ltd_\text{LTD}$. 
However, due to the presence of spurious singularities, the LTD representation is generally more numerically unstable than the CFF expression.
Therefore, we will eventually not use LTD for direct evaluation but merely for the generation of factorised threshold counterterms.
Moreover, the CFF expression remains unaffected by the presence of raised propagators, which appear in the UV counterterms.
\par
We generate the CFF expression $\ltd_\text{CFF}$ according to the algorithm of ref.~\cite{Capatti:2022mly}, which we implemented in \textsc{Python}.
The numerators multiplying each CFF summands, we denote compactly
\begin{align}
    \mathcal{N}_\text{CFF}(\sigma_1,\dots,\sigma_m)
    =
    \mathcal{N}(l_1,\dots,l_m)\vert_{l_j^0=\sigma_j E_j}
\end{align}
where $\sigma_j \in \{+1,-1\}$ and $E_j = \sqrt{{\vec{l}_j}^{\ 2}+m_j^2-\ii \epsilon}$ is the positive on-shell energy.
If the numerator is independent of $l_j$ we set $\sigma_j=0$ in order to later collect identical expressions for optimisation purposes.
\par
The LTD expression $\ltd_\text{LTD}$ we derive explicitly in \textsc{Form}~\cite{Vermaseren:2000nd,Ruijl:2017dtg,Ueda:2020wqk} under the assumptions that raised propagators are absent.
Since we will use LTD solely as an intermediate step to construct the threshold counterterms, this assumption will be sufficient for our purposes, as our expressions include only raised UV propagators and not raised thresholds.
\par
First, we iteratively identify the poles in $k_j^0$ and sum those residues, whose poles lie within the integration contour.
By default we choose all integration contours to close in the lower complex half-plane, such that only poles at $k_j^0 = k_j^*$ contribute, where $k_j^*$ is a sum of positive on-shell energies and some shift given as a linear combination of the energy components of the external momenta as well as the loop energies $k_r^0$ with $r>j$, which are integrated out in subsequent iterations.
We keep track of the numerators of the LTD residues as
\begin{align}
    \mathcal{N}_\text{LTD}(k_1^*,\dots,k_n^*)
    =
    \mathcal{N}(l_1,\dots,l_m)\vert_{k_j^0=k_j^*}.
\end{align}
\par
We use \textsc{Form}'s \texttt{extrasymbols} to collect the on-shell energies $E_i$ and all singular surfaces $S_i$ arising in the CFF and LTD expressions into separate variables and write them into a \textsc{Yaml} dictionary.
\par
For our processes of interest, the threshold singularities as well as the channels are supplied manually, as identified in table~\ref{tab:thresholds}.
The groups of thresholds, that define the corresponding channels, are supplied to a \textsc{Form} routine. The latter constructs the multi-channelling factors in eq.~\eqref{eq:multi_channel_factor}, the channels $C_i\ltd$ for both $\ltd \in \{\ltd_\text{CFF},\ltd_\text{LTD}\}$ and carries out algebraic simplification between the numerator of the multi-channel factor and the denominator of $\ltd$.
These channels are then exported in a \textsc{Yaml} dictionary.
\par
To construct the threshold counterterms, we first need to derive the corresponding phase space residues at a given threshold $\threshold$.
In the LTD representation, the denominator factor $\threshold$ typically appears in fewer terms compared to the CFF representation.
Therefore, we choose to derive the residue from LTD, as it is generally more compact.
However, we express the UV-tadpoles that multiply the residues for the $\Nf$-part in their CFF representation rather than in the LTD one, in order to more easily accommodate equal UV masses (which lead to raised propagators).
We use \textsc{Form} to algebraically remove the threshold $\threshold$ as in eq.~\eqref{eq:ps_residue} and export the symbolic expression in a \textsc{Yaml} dictionary.
Note that we perform numerically at runtime the construction of the counterterm and the evaluation of the residue at the on-shell radius.
\par
The CFF and LTD expressions and the phase space residues are expressed symbolically as functions of on-shell energies, energy components of external momenta and a numerator.
While the CFF and LTD expressions can be evaluated for arbitrary spatial loop momenta, the phase space residue is only meaningful if evaluated for on-shell configurations, defined by some threshold $\threshold=0$.
Note that in both the CFF and in the LTD representation the numerator retains its functional form and it is only evaluated at different on-shell configurations.
It is therefore sufficient to implement a single numerator per diagram as a function of loop four-momenta.
The spin chains in the numerator are numerically evaluated using the Weyl representation of gamma matrices.
Repeated indices are summed over recursively.
To accommodate external vector bosons other than photons, we replace the corresponding polarisations $\slashed{\epsilon}_i^*$ with $\slashed{\epsilon}_i^*V_i$, where $V_i= v\mathbb{1}-a\gamma^5$.
Here, $v$ and $a$ are respectively the vector and axial couplings to the $Z$ or $W^{\pm}$ bosons.
\par
The integrand generation is steered by a \textsc{Python} script and the final expressions in the \textsc{Yaml} files are exported in a \textsc{Rust} dynamic library.
The integrated counterterms proportional to tree-level amplitudes  in eqs.~\eqref{eq:r_1_singular} and \eqref{eq:fin_rem_2_nf_explicit} are automatically generated and also exported in a \textsc{Rust} dynamic library.
\par

\subsection{Parametrisation and Monte Carlo integration}
\label{subsec:mc_integration}

The dynamic libraries generated as discussed in the previous section are imported into a \textsc{Rust} code, where the integration domain is parameterised and the Monte Carlo integration is performed.
The code evaluates the individual diagrams, the threshold counterterms, sums permutations of final momentum configurations, computes the interferences, convolutes with the PDFs and applies phase space cuts.
\par
We use the Vegas adaptive Monte-Carlo algorithm  \cite{Lepage:1977sw,Lepage:2020tgj} of the \textsc{Cuba} library \cite{Hahn:2004fe} for multidimensional numerical integration.
For each Monte Carlo sample on the hypercube we generate the Bjorken variables and the phase space and spatial loop momenta.
The phase space parametrisation is discussed in appendix~\ref{app:ps_parametrisation}.
The spinors of the initial quarks as well as the polarisation vectors of the final bosons are generated according to the \textsc{HELAS} conventions \cite{Murayama:1992gi}.
We generate the loop momenta in hyperspherical coordinates, as $\loopmomenta = r \unitvector$, where at one loop $\loopmomenta=\vec{k}$ and at two loops $\loopmomenta=(\vec{k}, \vec{l})$.
We sample the magnitude from the unit hypercube variable $x\in[0,1]$ as
\begin{align}
    r = E_\text{CM} \frac{x}{1-x}\,, \qquad
    \frac{\dd r}{\dd x} = \frac{(E_\text{CM} + r)^2}{E_\text{CM}}\,.
\end{align}
The unit vector is parameterised in the standard hyperspherical coordinates with $3n-1$ angles $\phi_i$.
The angles are sampled linearly from the unit hypercube.
The on-shell radii $r_\threshold(\unitvector)$ on which the threshold counterterms (and residues) depend are calculated according to ref.~\cite{Kermanschah:2021wbk}.
Internal parameters, such as the UV mass and couplings are supplied as constants.
If the kinematic configuration is allowed by any phase space cut, the loop and conjugate tree amplitudes are evaluated for all permutations of final kinematics and multiplied before they are summed over all (non-vanishing) helicity configurations.
The interference is multiplied by the PDFs taken from \textsc{LHAPDF6} \cite{Buckley:2014ana}.
\par
The stability of the numerical evaluation can optionally be checked at runtime using a rotation test, as performed in ref.~\cite{Capatti:2019ypt}:
if the integrand evaluation at the original and at a rotated kinematic configuration have a specified minimum number of digits in common, the evaluation is considered stable. 
The rotation test can optionally be repeated in quadruple precision, which potentially rescues an unstable sample.
If the evaluation remains unstable it is set to zero.
The number of unstable and rescued samples is monitored and the former should not surpass some upper bound in order not to distort the Monte Carlo estimate.
\par
The integration is steered by a designated settings file in $\textsc{Yaml}$ format.
In particular, it is possible to individually enable or disable the PDFs, the phase space or loop integrations, the helicity sum and the permutation of final state kinematics.
The phase (dispersive or absorptive part) of the amplitude(s) can also be specified.
For our calculation of the phase-space integrated virtual corrections to the  cross section only the dispersive part is needed.
However, at fixed phase-space points, the absorptive part is generally non-zero, as shown explicitly in appendix~\ref{app:fixed_ps_results}.
Moreover, when integrating over a phase space of $n$ indistinguishable final state particles, we exploit this symmetry to save a factor of $n!$ in evaluation time by only permuting the final state of the (computationally much lighter) tree amplitude.
Note that the aforementioned settings allow for the calculation of intermediate results, such as helicity amplitudes and interferences at fixed phase space configurations.
This enabled us to conduct early comparisons of matrix elements with \textsc{OpenLoops} \cite{Cascioli:2011va,Kallweit:2014xda,Buccioni:2017yxi,Buccioni:2019sur}, \textsc{MadGraph} \cite{Hirschi:2011pa,Alwall:2014hca} and own analytic calculations documented in ref.~\cite{Anastasiou:2023koq}.
In appendix~\ref{app:fixed_ps_results}, we present numerical results for the finite remainders of the the one-loop and two-loop $\Nf$ contributions to squared matrix elements for six processes at three fixed phase-space points, with comparisons to reference results where available.
\par
We also supply other physical and technical parameters though the settings file.
Our physical parameters are the centre-of-mass energy $E_\text{CM}$, minimum transverse momentum $p_{T,\text{min}}$ for the final state bosons and the factorisation and renormalisation scales.
Among the technical parameters there are Vegas and stability related settings.

%% file: sections/results.tex
\section{Results}
\label{sec:results}

First, we highlight the new results\footnote{As of the time of publication, the corresponding two-loop amplitudes cannot be analytically computed, since the master integrals are not known.} calculated with our method, namely the $\Nf$-part of the virtual corrections contributing at NNLO for the processes $pp\to\gamma^*\gamma^*\gamma^*$ and $p_dp_d\to Z\gamma_1^*\gamma_2^*$, where for the latter we only present the $d\bar{d}$-initiated contribution.

\def\arraystretch{1.3}%
\begin{table}[h]
    \centering
    \begin{tabular}{l c c}
        \hline\hline
         Process
         & $\bigl(\frac{\alpha_s}{4\pi}\bigr)^2\sigma_\NfNNLO^{\virtual,\cs}$ [{\ttlf$\texttt{10}^{\texttt{-5}}$} pb]
         & $\Delta\ [\%]$ \\\hline\hline
         $pp\to\gamma^*\gamma^*\gamma^*$
         & {\ttlf-1.4212 \pmtt~0.0140}
         &{\ttlf0.988}\\\hline
        $p_dp_d\to Z\gamma_1^*\gamma_2^*$
        & {\ttlf-2.4301 \pmtt~0.0229}
        & {\ttlf0.941} \\\hline
    \end{tabular}
    \caption{
    $\Nf$-part of the virtual loop-corrections contributing at NNLO for the production of three massive vector bosons.
    The parameters are specified in table~\ref{tab:parameters}.
    }
    \label{tab:new_results}
\end{table}

A summary of all our results for the LO, NLO, and the $\Nf$-part of the NNLO virtual corrections, as defined in eqs.~\eqref{eq:cross_section_lo}, \eqref{eq:finite_cross_section_nlo}, and \eqref{eq:finite_cross_section_nnlo}, respectively, are displayed in table~\ref{tab:0} for $2\to2$ processes, and in tables~\ref{tab:1} and \ref{tab:2} for $2\to3$ processes.
The tables also show the various summands contributing to the higher-order corrections, according to \crefrange{eq:first_cs_int}{eq:last_cs_int}, each evaluated using separate Monte Carlo integrations.
For all processes considered, we find that the $\Nf$-part of the NNLO virtual corrections is approximately an order of magnitude smaller than the NLO virtual corrections.
This behaviour suggests that the perturbative expansion is well-behaved.
However, we note that the virtual correction alone is not directly comparable to an observable quantity.
\par

\input{tables/stability}

Moreover, table~\ref{tab:1} shows two alternative computations of the one-loop and the $\Nf$-part of the two-loop virtual corrections for $pp\to\gamma^*\gamma^*\gamma^*$.
The one labelled \emph{stabilised} applied the numerical stability check described in section~\ref{sec:implementation}.
Statistics on the stability of the integrand evaluations are shown in table~\ref{tab:stability}.
We can consider our integral estimate reliable as we encounter numerically unstable points in only $\mathcal{O}(1\text{\textperthousand})$ or fewer evaluations.
The other computation used a different UV mass $M=1\,\text{GeV}$, upon which the local UV counterterms depend.
Since the virtual corrections must be independent of $M$\footnote{
The dependence of the hard scattering contribution on the UV mass $M$ is replaced by a dependence on the renormalistation scale $\mu$ once combined with the integrated counterterms (cf.~eqs.~\eqref{eq:scheme_change_1l} and \eqref{eq:scheme_change_2l}).
}, this result serves a strong internal cross-check for our implementation.
Note that while the two contributions at NLO are constant in $M$ after integration, the three contributions at $\Nf$-NNLO individually depend on $M$.
Their sum, however, must be independent of $M$, as confirmed by our results.
\par
We compare our results for the finite remainders of the virtual corrections with available reference results and found agreement within $2\sigma$ in all cases.
For the LO cross section and the virtual correction at NLO, we provide reference benchmarks generated using \textsc{MadGraph}~\cite{Hirschi:2011pa,Alwall:2014hca}.
Therefore, for comparisons at NLO, we adopt, instead of the $\overline{\text{MS}}$ factorisation scheme, the BLHA normalisation convention \cite{Binoth:2010xt,Alioli:2013nda} that \textsc{MadGraph} implements.
This affects only the numerical result for $\sigma\NLOtree$, defined in eq.~\eqref{eq:sigmaNLOtree}, and amounts to a replacement of the $\overline{\text{MS}}$ remainder $R^{(1)}_\text{singular}$ in eq.~\eqref{eq:M-singular-integrated} with
\begin{align}
\label{eq:blha}
   R^{(1),(\text{BLHA})}_\text{singular}
    = R^{(1)}_\text{singular}
    -\frac{\pi^2}{6} C_F \bare^{(0)}.
\end{align}
Notably, the hard contribution $\sigma\NLOoneloop$ is unaffected by the factorisation scheme and can therefore be reused for $\sigma\NfNNLOoneloop$ as per eq.~\eqref{eq:sigma1NF}.
\par
The reference results for the two-loop $\Nf$ parts were obtained using the analytic expressions of ref.~\cite{Anastasiou:2002zn} for $\gamma\gamma$, ref.~\cite{Gehrmann:2015ora} for $\gamma^*\gamma^*$ and $ZZ$, and the \texttt{FivePointAmplitudes-cpp} library of ref.~\cite{Abreu:2023bdp} for $\gamma\gamma\gamma$ production.
In all cases, the results were converted to the $\overline{\rm MS}$ factorisation scheme and integrated with our own phase space generator.
To the best of our knowledge, there are no benchmark results available for the two-loop $\Nf$-contributions to $\gamma^*\gamma^*\gamma^*$ and $Z\gamma_1^*\gamma_2^*$ production.
\par

\input{tables/parameters}

The parameters used for the runs are listed in table~\ref{tab:parameters}.
We computed the finite remainder of the virtual corrections at the LHC's collision energy of 13\,TeV.
Transverse momentum cuts are imposed for both massless and massive vector bosons, but they are only needed to make the hard part of the virtual corrections involving massless bosons finite.
We used the \texttt{CT10} NLO PDF set \cite{Lai:2010vv} from \textsc{LHAPDF6} \cite{Buckley:2014ana}.
Five quark flavours were used for all virtual corrections except those involving a $Z$ boson in the final state, for which only the PDFs of the $d$ and $\bar{d}$ quarks were used.
Thus, in the table, we denote the initial state by $p_dp_d$ for these processes.
The renormalisation scale $\muren$, factorisation scale $\mu_F$, and UV mass $M$ are all fixed to the $Z$ boson mass $M_Z$ unless explicitly specified in table~\ref{tab:1}.
The invariant masses of all virtual photons are also set to $M_Z$, except for the process $p_d p_d \to Z \gamma_1^* \gamma_2^*$, where all invariant masses are different.
The strong coupling $\alpha_s(M_Z)$ is specified by the PDF set, and the electroweak coupling $\alpha$, as well as the axial and vector couplings $a_{Zd}$ and $v_{Zd}$ of the $Z$ boson to the $d$ quark were taken from \textsc{MadGraph}.
The remaining constants are $\Nc=3$, $\Nf=5$, $T_F=\frac{1}{2}$, and $C_F=\frac{4}{3}$.
\par

In the table headers, $N_p$ refers to the number of Monte Carlo evaluations, $\sfrac{t}{p}$ to the average time per evaluation, $\text{Exp.}$ to the power of ten that multiplies the result, and $\Delta$ to the relative uncertainty of the result.
Our calculations were performed on a standard computer with an Intel Xeon Gold 6136 CPU @ 3.00GHz in two sockets with 12 cores each.
The average evaluation time was determined by dividing the total runtime by the number of Monte Carlo samples.
The number of Monte Carlo evaluations and the evaluation time are not specified for the MadGraph reference results, nor for $\sigma\NfNNLOoneloop$, since, according to eq.~\eqref{eq:sigma1NF}, it is recycled from $\sigma\NLOoneloop$.
They are also not specified for the higher-order virtual corrections, as they are determined by sums of terms.
\par

The Vegas parameters \texttt{nstart} and \texttt{nincrease} are set to $10^5$.
The integrand evaluations were all performed in double precision and without numerical stability checks unless explicitly noted in table~\ref{tab:1}.
For our calculations, we aimed for an uncertainty of $1$\% or less.
This level of precision should be adequate for phenomenological applications, especially considering that NNLO scale variations are typically of a similar magnitude.
For the NNLO $N_f$-contribution with the unphysical choice for the UV mass $M=1\,\text{GeV}$ shown in table~\ref{tab:2}, we observe large cancellations among the contributions, making it more difficult to achieve a small uncertainty.
\par
The $\Nf$-part to the NNLO virtual correction of $p_dp_d\to Z \gamma_1^*\gamma_2^*$, the most complex of the performed calculations, ran for almost a month to achieve 1\% accuracy on a standard computer.
This runtime was approximately $3!$ times longer than that for the corresponding part of $pp\to \gamma^*\gamma^*\gamma^*$, where the symmetry of indistinguishable final-state particles could be leveraged.
Additionally, in the triphoton case, the couplings of the quark flavours factorised, unlike for the case with the $Z$-boson, which couples to both vector and axial currents.
Given the long runtime in our current setup, we decided to postpone the computation of the contributions from the other quarks to future work, such as when virtual contributions are combined with real emissions.
However, we do not expect any further difficulties and anticipate to accelerate the full calculation by implementing suitable reweighting of vector and axial currents at runtime together with the optimisations elaborated on below.
Nonetheless, the results presented for the $d$-quark contribution, along with the summed contribution over five quark flavours in off-shell triphoton production, serve as a robust benchmark for future calculations.
\par
Generally, we anticipate significant runtime optimisations through improved Dirac chain evaluations, removal of integrable singularities at vanishing on-shell energies, and employing importance sampling to flatten phase-space resonances.
Furthermore, thanks to the parallelisability of Monte Carlo integrations, substantial speed improvements are anticipated on high-performance computer clusters.
\raggedbottom
\def\arraystretch{1.3}%
\input{tables/results_tab}

%% file: tables/stability.tex
\def\arraystretch{1.3}%
\begin{table}[b]
    \centering
    \begin{tabular}{lrrr}\hline\hline
Part & Unstable [\textperthousand]& Rescued [\%] & Discarded [\textperthousand]\\\hline\hline
$\sigma\NLOtree$
& \ttlf\texttt{ 0.41 }
&\ttlf\texttt{ 53.3 }
&\ttlf\texttt{ 0.19 }
\\\hline
$\sigma\NLOoneloop$
& \ttlf\texttt{ 0.26 }
&\ttlf\texttt{ 82.3 }
&\ttlf\texttt{ 0.05 }
\\\hline
$\sigma\NfNNLOtree$
& \ttlf\texttt{ 0.36 }
&\ttlf\texttt{ 51.8 }
&\ttlf\texttt{ 0.17 }
\\\hline
$\sigma\NfNNLOtwoloop$
& \ttlf\texttt{ 2.56 }
&\ttlf\texttt{ 87.9 }
&\ttlf\texttt{ 0.31 }
\\\hline
    \end{tabular}
    \caption{\label{tab:stability}
    Stability analysis for the process $pp \to \gamma^* \gamma^* \gamma^*$ based on the integrations labelled \emph{stabilised} in table~\ref{tab:1}.
    Stability checks were performed as detailed in section~\ref{sec:implementation}, testing for 7 numerically stable digits.
    The table presents the relative number of unstable evaluations in double precision, the percentage of those recovered by repeating the evaluation in quadruple precision and the discarded evaluations that remained unstable even in quadruple precision, which were consequently set to zero.}
\end{table}

%% file: tables/parameters.tex
\begin{table}[b]
    \centering
    \begin{tabularx}{\textwidth}{l|X}
    \hline
    centre-of-mass energy & $E_\text{CM}\equiv\sqrt{s}=13$\,TeV \\
    \hline
    phase-space cuts & $p_T^\text{min} = 
    \begin{cases}
        50\,\text{GeV} & \text{for massless bosons}\\
        25\,\text{GeV} & \text{for massive bosons}
    \end{cases}$ \\
    \hline
    PDFs & \texttt{CT10} NLO PDF, \texttt{lhapdf\_id}: 11000 \cite{Lai:2010vv}\newline
     flavours = $\begin{cases}
        d,\bar{d} & \text{if $Z$ boson in final state}\\
        u,\bar{u},d,\bar{d},c,\bar{c},s,\bar{s},b,\bar{b} & \text{otherwise}
    \end{cases}$\\
    \hline
    masses / scales & $M_Z=91.1876\,\text{GeV}$, $M_{\gamma_1^*}=20\,\text{GeV}$, $M_{\gamma_2^*}=50\,\text{GeV}$\newline
     $\muren=\mu_F=M=M_{\gamma^*}=M_Z$\\
    \hline
    couplings at $M_Z$ & $\alpha_s = 0.118001$, $\alpha=(132.5070)^{-1}$, \newline  $a_{Zd}=\frac{1}{4\cos \theta_W\sin \theta_W}$, $v_{Zd}=\frac{3-4\sin^2\theta_W}{12\cos \theta_W\sin \theta_W}$\,, $\sin^2\theta_W=0.22225$ \\[0.6cm]
    \hline
    other constants & $\Nc=3$, $\Nf=5$, $T_F=\frac{1}{2}$, $C_F=\frac{4}{3}$
    \\\hline
    \end{tabularx}
    \caption{\label{tab:parameters}
    Process specification parameters.}
\end{table}

%% file: tables/results_tab.tex
\begin{table}[H]
\centering
\resizebox{\columnwidth}{!}{%
\begin{tabular}{lllrrrcr}
\hline\hline
Process & Order & Part & $N_p \ [10^6]$ & $\sfrac{t}{p} \ [\mu \text{s}]$ & Exp. & Result $[\text{pb}]$ & $\Delta \ [\%]$ \\
\hline\hline\multirow{10}{*}{$pp\to\gamma\gamma$}
& \multirow{2}{*}{$\textrm{LO}$}
& \boldmath{\bf$\sigma_\textrm{LO}$}
& {\ttbf\texttt{10}}
& {\ttbf\texttt{0.1}}
&
& {\ttbf\texttt{~6.0995~$\pmtt$~0.0021}}
& {\ttbf\texttt{0.035}}
\\
[\doublerulesep]\ccline{3-3}\ccline{4-4}\ccline{5-5}\ccline{6-6}\ccline{7-7}\ccline{8-8}& & \bf reference\textbf{}
&
&
&
& {\ttbf\texttt{~6.0890~$\pmtt$~0.0038}}
& {\ttbf\texttt{0.062}}
\\
[\doublerulesep]\ccline{2-2}\ccline{3-3}\ccline{4-4}\ccline{5-5}\ccline{6-6}\ccline{7-7}\ccline{8-8}& \multirow{4}{*}{$\textrm{NLO}$}
& $\bigl(\frac{\alpha_s}{4\pi}\bigr)$$\sigma\NLOtree$
& {\ttlf\texttt{10}}
& {\ttlf\texttt{0.2}}
& {\ttlf$\texttt{10}^{\texttt{-1}}$}
& {\ttlf\texttt{~3.6477~$\pmtt$~0.0014}}
& {\ttlf\texttt{0.037}}
\\
\cline{3-3}\cline{4-4}\cline{5-5}\cline{6-6}\cline{7-7}\cline{8-8}& & $\bigl(\frac{\alpha_s}{4\pi}\bigr)$$\sigma\NLOoneloop$
& {\ttlf\texttt{366}}
& {\ttlf\texttt{2\phantom{.0}}}
& {\ttlf$\texttt{10}^{\texttt{-1}}$}
& {\ttlf\texttt{~1.6374~$\pmtt$~0.0164}}
& {\ttlf\texttt{1.000}}
\\
[\doublerulesep]\ccline{3-3}\ccline{4-4}\ccline{5-5}\ccline{6-6}\ccline{7-7}\ccline{8-8}& & \boldmath{\bf$\bigl(\frac{\alpha_s}{4\pi}\bigr)$$\sigma_\textrm{NLO}^{\virtual,\cs}$}\textbf{}
&
&
& {\ttbf$\texttt{10}^{\texttt{-1}}$}
& {\ttbf\texttt{~5.2851~$\pmtt$~0.0164}}
& {\ttbf\texttt{0.311}}
\\
[\doublerulesep]\ccline{3-3}\ccline{4-4}\ccline{5-5}\ccline{6-6}\ccline{7-7}\ccline{8-8}& & \bf reference\textbf{}
&
&
& {\ttbf$\texttt{10}^{\texttt{-1}}$}
& {\ttbf\texttt{~5.2560~$\pmtt$~0.0036}}
& {\ttbf\texttt{0.068}}
\\
[\doublerulesep]\ccline{2-2}\ccline{3-3}\ccline{4-4}\ccline{5-5}\ccline{6-6}\ccline{7-7}\ccline{8-8}& \multirow{4}{*}{$\textrm{NNLO},\Nf$}
& $\bigl(\frac{\alpha_s}{4\pi}\bigr)^2$$\sigma\NfNNLOtree$
& {\ttlf\texttt{10}}
& {\ttlf\texttt{0.3}}
& {\ttlf$\texttt{10}^{\texttt{-2}}$}
& {\ttlf\texttt{-2.8241~$\pmtt$~0.0010}}
& {\ttlf\texttt{0.036}}
\\
\cline{3-3}\cline{4-4}\cline{5-5}\cline{6-6}\cline{7-7}\cline{8-8}& & $\bigl(\frac{\alpha_s}{4\pi}\bigr)^2$$\sigma\NfNNLOoneloop$
&
&
& {\ttlf$\texttt{10}^{\texttt{-3}}$}
& {\ttlf\texttt{~1.7084~$\pmtt$~0.0171}}
& {\ttlf\texttt{1.000}}
\\
\cline{3-3}\cline{4-4}\cline{5-5}\cline{6-6}\cline{7-7}\cline{8-8}& & $\bigl(\frac{\alpha_s}{4\pi}\bigr)^2$$\sigma\NfNNLOtwoloop$
& {\ttlf\texttt{1720}}
& {\ttlf\texttt{3\phantom{.0}}}
& {\ttlf$\texttt{10}^{\texttt{-2}}$}
& {\ttlf\texttt{-3.4942~$\pmtt$~0.0348}}
& {\ttlf\texttt{0.997}}
\\
[\doublerulesep]\ccline{3-3}\ccline{4-4}\ccline{5-5}\ccline{6-6}\ccline{7-7}\ccline{8-8}& & \boldmath{\bf$\bigl(\frac{\alpha_s}{4\pi}\bigr)^2$$\sigma_\NfNNLO^{\virtual,\cs}$}\textbf{}
&
&
& {\ttbf$\texttt{10}^{\texttt{-2}}$}
& {\ttbf\texttt{-6.1475~$\pmtt$~0.0349}}
& {\ttbf\texttt{0.568}}
\\
[\doublerulesep]\ccline{3-3}\ccline{4-4}\ccline{5-5}\ccline{6-6}\ccline{7-7}\ccline{8-8}& & \bf reference\textbf{}
& {\ttlf\texttt{11}}
&
& {\ttbf$\texttt{10}^{\texttt{-2}}$}
& {\ttbf\texttt{-6.1483~$\pmtt$~0.0025}}
& {\ttbf\texttt{0.040}}
\\
\hline\hline\multirow{10}{*}{$pp\to\gamma^*\gamma^*$}
& \multirow{2}{*}{$\textrm{LO}$}
& \boldmath{\bf$\sigma_\textrm{LO}$}
& {\ttbf\texttt{11}}
& {\ttbf\texttt{0.3}}
&
& {\ttbf\texttt{~2.5022~$\pmtt$~0.0005}}
& {\ttbf\texttt{0.020}}
\\
[\doublerulesep]\ccline{3-3}\ccline{4-4}\ccline{5-5}\ccline{6-6}\ccline{7-7}\ccline{8-8}& & \bf reference\textbf{}
&
&
&
& {\ttbf\texttt{~2.4970~$\pmtt$~0.0016}}
& {\ttbf\texttt{0.064}}
\\
[\doublerulesep]\ccline{2-2}\ccline{3-3}\ccline{4-4}\ccline{5-5}\ccline{6-6}\ccline{7-7}\ccline{8-8}& \multirow{4}{*}{$\textrm{NLO}$}
& $\bigl(\frac{\alpha_s}{4\pi}\bigr)$$\sigma\NLOtree$
& {\ttlf\texttt{11}}
& {\ttlf\texttt{1\phantom{.0}}}
& {\ttlf$\texttt{10}^{\texttt{-1}}$}
& {\ttlf\texttt{~1.3053~$\pmtt$~0.0005}}
& {\ttlf\texttt{0.038}}
\\
\cline{3-3}\cline{4-4}\cline{5-5}\cline{6-6}\cline{7-7}\cline{8-8}& & $\bigl(\frac{\alpha_s}{4\pi}\bigr)$$\sigma\NLOoneloop$
& {\ttlf\texttt{11}}
& {\ttlf\texttt{12\phantom{.0}}}
& {\ttlf$\texttt{10}^{\texttt{-1}}$}
& {\ttlf\texttt{~3.0119~$\pmtt$~0.0089}}
& {\ttlf\texttt{0.295}}
\\
[\doublerulesep]\ccline{3-3}\ccline{4-4}\ccline{5-5}\ccline{6-6}\ccline{7-7}\ccline{8-8}& & \boldmath{\bf$\bigl(\frac{\alpha_s}{4\pi}\bigr)$$\sigma_\textrm{NLO}^{\virtual,\cs}$}\textbf{}
&
&
& {\ttbf$\texttt{10}^{\texttt{-1}}$}
& {\ttbf\texttt{~4.3172~$\pmtt$~0.0089}}
& {\ttbf\texttt{0.206}}
\\
[\doublerulesep]\ccline{3-3}\ccline{4-4}\ccline{5-5}\ccline{6-6}\ccline{7-7}\ccline{8-8}& & \bf reference\textbf{}
&
&
& {\ttbf$\texttt{10}^{\texttt{-1}}$}
& {\ttbf\texttt{~4.3160~$\pmtt$~0.0028}}
& {\ttbf\texttt{0.065}}
\\
[\doublerulesep]\ccline{2-2}\ccline{3-3}\ccline{4-4}\ccline{5-5}\ccline{6-6}\ccline{7-7}\ccline{8-8}& \multirow{4}{*}{$\textrm{NNLO},\Nf$}
& $\bigl(\frac{\alpha_s}{4\pi}\bigr)^2$$\sigma\NfNNLOtree$
& {\ttlf\texttt{11}}
& {\ttlf\texttt{1\phantom{.0}}}
& {\ttlf$\texttt{10}^{\texttt{-3}}$}
& {\ttlf\texttt{-7.4003~$\pmtt$~0.0025}}
& {\ttlf\texttt{0.033}}
\\
\cline{3-3}\cline{4-4}\cline{5-5}\cline{6-6}\cline{7-7}\cline{8-8}& & $\bigl(\frac{\alpha_s}{4\pi}\bigr)^2$$\sigma\NfNNLOoneloop$
&
&
& {\ttlf$\texttt{10}^{\texttt{-3}}$}
& {\ttlf\texttt{~3.1425~$\pmtt$~0.0093}}
& {\ttlf\texttt{0.295}}
\\
\cline{3-3}\cline{4-4}\cline{5-5}\cline{6-6}\cline{7-7}\cline{8-8}& & $\bigl(\frac{\alpha_s}{4\pi}\bigr)^2$$\sigma\NfNNLOtwoloop$
& {\ttlf\texttt{99}}
& {\ttlf\texttt{26\phantom{.0}}}
& {\ttlf$\texttt{10}^{\texttt{-2}}$}
& {\ttlf\texttt{-3.2685~$\pmtt$~0.0322}}
& {\ttlf\texttt{0.986}}
\\
[\doublerulesep]\ccline{3-3}\ccline{4-4}\ccline{5-5}\ccline{6-6}\ccline{7-7}\ccline{8-8}& & \boldmath{\bf$\bigl(\frac{\alpha_s}{4\pi}\bigr)^2$$\sigma_\NfNNLO^{\virtual,\cs}$}\textbf{}
&
&
& {\ttbf$\texttt{10}^{\texttt{-2}}$}
& {\ttbf\texttt{-3.6943~$\pmtt$~0.0322}}
& {\ttbf\texttt{0.873}}
\\
[\doublerulesep]\ccline{3-3}\ccline{4-4}\ccline{5-5}\ccline{6-6}\ccline{7-7}\ccline{8-8}& & \bf reference\textbf{}
& {\ttlf\texttt{11}}
&
& {\ttbf$\texttt{10}^{\texttt{-2}}$}
& {\ttbf\texttt{-3.6767~$\pmtt$~0.0016}}
& {\ttbf\texttt{0.044}}
\\
\hline\hline\multirow{10}{*}{$p_dp_d\to ZZ$}
& \multirow{2}{*}{$\textrm{LO}$}
& \boldmath{\bf$\sigma_\textrm{LO}$}
& {\ttbf\texttt{11}}
& {\ttbf\texttt{1\phantom{.0}}}
&
& {\ttbf\texttt{~4.0516~$\pmtt$~0.0008}}
& {\ttbf\texttt{0.020}}
\\
[\doublerulesep]\ccline{3-3}\ccline{4-4}\ccline{5-5}\ccline{6-6}\ccline{7-7}\ccline{8-8}& & \bf reference\textbf{}
&
&
&
& {\ttbf\texttt{~4.0490~$\pmtt$~0.0028}}
& {\ttbf\texttt{0.069}}
\\
[\doublerulesep]\ccline{2-2}\ccline{3-3}\ccline{4-4}\ccline{5-5}\ccline{6-6}\ccline{7-7}\ccline{8-8}& \multirow{4}{*}{$\textrm{NLO}$}
& $\bigl(\frac{\alpha_s}{4\pi}\bigr)$$\sigma\NLOtree$
& {\ttlf\texttt{11}}
& {\ttlf\texttt{2\phantom{.0}}}
& {\ttlf$\texttt{10}^{\texttt{-1}}$}
& {\ttlf\texttt{~2.1312~$\pmtt$~0.0008}}
& {\ttlf\texttt{0.037}}
\\
\cline{3-3}\cline{4-4}\cline{5-5}\cline{6-6}\cline{7-7}\cline{8-8}& & $\bigl(\frac{\alpha_s}{4\pi}\bigr)$$\sigma\NLOoneloop$
& {\ttlf\texttt{11}}
& {\ttlf\texttt{21\phantom{.0}}}
& {\ttlf$\texttt{10}^{\texttt{-1}}$}
& {\ttlf\texttt{~4.8755~$\pmtt$~0.0158}}
& {\ttlf\texttt{0.325}}
\\
[\doublerulesep]\ccline{3-3}\ccline{4-4}\ccline{5-5}\ccline{6-6}\ccline{7-7}\ccline{8-8}& & \boldmath{\bf$\bigl(\frac{\alpha_s}{4\pi}\bigr)$$\sigma_\textrm{NLO}^{\virtual,\cs}$}\textbf{}
&
&
& {\ttbf$\texttt{10}^{\texttt{-1}}$}
& {\ttbf\texttt{~7.0067~$\pmtt$~0.0159}}
& {\ttbf\texttt{0.226}}
\\
[\doublerulesep]\ccline{3-3}\ccline{4-4}\ccline{5-5}\ccline{6-6}\ccline{7-7}\ccline{8-8}& & \bf reference\textbf{}
&
&
& {\ttbf$\texttt{10}^{\texttt{-1}}$}
& {\ttbf\texttt{~7.0020~$\pmtt$~0.0046}}
& {\ttbf\texttt{0.066}}
\\
[\doublerulesep]\ccline{2-2}\ccline{3-3}\ccline{4-4}\ccline{5-5}\ccline{6-6}\ccline{7-7}\ccline{8-8}& \multirow{4}{*}{$\textrm{NNLO},\Nf$}
& $\bigl(\frac{\alpha_s}{4\pi}\bigr)^2$$\sigma\NfNNLOtree$
& {\ttlf\texttt{11}}
& {\ttlf\texttt{2\phantom{.0}}}
& {\ttlf$\texttt{10}^{\texttt{-2}}$}
& {\ttlf\texttt{-1.2060~$\pmtt$~0.0004}}
& {\ttlf\texttt{0.033}}
\\
\cline{3-3}\cline{4-4}\cline{5-5}\cline{6-6}\cline{7-7}\cline{8-8}& & $\bigl(\frac{\alpha_s}{4\pi}\bigr)^2$$\sigma\NfNNLOoneloop$
&
&
& {\ttlf$\texttt{10}^{\texttt{-3}}$}
& {\ttlf\texttt{~5.0869~$\pmtt$~0.0165}}
& {\ttlf\texttt{0.325}}
\\
\cline{3-3}\cline{4-4}\cline{5-5}\cline{6-6}\cline{7-7}\cline{8-8}& & $\bigl(\frac{\alpha_s}{4\pi}\bigr)^2$$\sigma\NfNNLOtwoloop$
& {\ttlf\texttt{113}}
& {\ttlf\texttt{44\phantom{.0}}}
& {\ttlf$\texttt{10}^{\texttt{-2}}$}
& {\ttlf\texttt{-5.2390~$\pmtt$~0.0520}}
& {\ttlf\texttt{0.992}}
\\
[\doublerulesep]\ccline{3-3}\ccline{4-4}\ccline{5-5}\ccline{6-6}\ccline{7-7}\ccline{8-8}& & \boldmath{\bf$\bigl(\frac{\alpha_s}{4\pi}\bigr)^2$$\sigma_\NfNNLO^{\virtual,\cs}$}\textbf{}
&
&
& {\ttbf$\texttt{10}^{\texttt{-2}}$}
& {\ttbf\texttt{-5.9363~$\pmtt$~0.0520}}
& {\ttbf\texttt{0.876}}
\\
[\doublerulesep]\ccline{3-3}\ccline{4-4}\ccline{5-5}\ccline{6-6}\ccline{7-7}\ccline{8-8}& & \bf reference\textbf{}
& {\ttlf\texttt{11}}
&
& {\ttbf$\texttt{10}^{\texttt{-2}}$}
& {\ttbf\texttt{-5.9495~$\pmtt$~0.0026}}
& {\ttbf\texttt{0.044}}
\\
\cline{3-3}\cline{4-4}\cline{5-5}\cline{6-6}\cline{7-7}\cline{8-8}\hline\hline
\end{tabular}%
}%
\caption{\label{tab:0}Virtual contributions to $2\to2$ cross sections.
The respective contributions are defined in eqs.~\eqref{eq:finite_cross_section_nlo} to \eqref{eq:cross_section_lo}.
}%
\end{table}

\begin{table}[H]
\centering
\resizebox{\columnwidth}{!}{%
\begin{tabular}{lllrrrcr}
\hline\hline
Process & Order & Part & $N_p \ [10^6]$ & $\sfrac{t}{p} \ [\mu \text{s}]$ & Exp. & Result $[\text{pb}]$ & $\Delta \ [\%]$ \\
\hline\hline\multirow{24}{*}{$pp\to\gamma^*\gamma^*\gamma^*$}
& \multirow{2}{*}{$\textrm{LO}$}
& \boldmath{\bf$\sigma_\textrm{LO}$}
& {\ttbf\texttt{11}}
& {\ttbf\texttt{2\phantom{.0}}}
& {\ttbf$\texttt{10}^{\texttt{-3}}$}
& {\ttbf\texttt{~1.0876~$\pmtt$~0.0005}}
& {\ttbf\texttt{0.042}}
\\
[\doublerulesep]\ccline{3-3}\ccline{4-4}\ccline{5-5}\ccline{6-6}\ccline{7-7}\ccline{8-8}& & \bf reference\textbf{}
&
&
& {\ttbf$\texttt{10}^{\texttt{-3}}$}
& {\ttbf\texttt{~1.0890~$\pmtt$~0.0008}}
& {\ttbf\texttt{0.073}}
\\
[\doublerulesep]\ccline{2-2}\ccline{3-3}\ccline{4-4}\ccline{5-5}\ccline{6-6}\ccline{7-7}\ccline{8-8}& \multirow{10}{*}{$\textrm{NLO}$}
& $\bigl(\frac{\alpha_s}{4\pi}\bigr)$$\sigma\NLOtree$
& {\ttlf\texttt{11}}
& {\ttlf\texttt{4\phantom{.0}}}
& {\ttlf$\texttt{10}^{\texttt{-5}}$}
& {\ttlf\texttt{-9.9080~$\pmtt$~0.0052}}
& {\ttlf\texttt{0.052}}
\\
\cline{3-3}\cline{4-4}\cline{5-5}\cline{6-6}\cline{7-7}\cline{8-8}& & $\bigl(\frac{\alpha_s}{4\pi}\bigr)$$\sigma\NLOoneloop$
& {\ttlf\texttt{78}}
& {\ttlf\texttt{127\phantom{.0}}}
& {\ttlf$\texttt{10}^{\texttt{-4}}$}
& {\ttlf\texttt{~2.4600~$\pmtt$~0.0144}}
& {\ttlf\texttt{0.584}}
\\
[\doublerulesep]\ccline{3-3}\ccline{4-4}\ccline{5-5}\ccline{6-6}\ccline{7-7}\ccline{8-8}& & \boldmath{\bf$\bigl(\frac{\alpha_s}{4\pi}\bigr)$$\sigma_\textrm{NLO}^{\virtual,\cs}$}\textbf{}
&
&
& {\ttbf$\texttt{10}^{\texttt{-4}}$}
& {\ttbf\texttt{~1.4692~$\pmtt$~0.0144}}
& {\ttbf\texttt{0.979}}
\\
[\doublerulesep]\ccline{3-3}\ccline{4-4}\ccline{5-5}\ccline{6-6}\ccline{7-7}\ccline{8-8}& & $\bigl(\frac{\alpha_s}{4\pi}\bigr)$$\sigma\NLOtree$ $(\textrm{stabilised})$
& {\ttlf\texttt{11}}
& {\ttlf\texttt{8\phantom{.0}}}
& {\ttlf$\texttt{10}^{\texttt{-5}}$}
& {\ttlf\texttt{-9.9080~$\pmtt$~0.0052}}
& {\ttlf\texttt{0.052}}
\\
\cline{3-3}\cline{4-4}\cline{5-5}\cline{6-6}\cline{7-7}\cline{8-8}& & $\bigl(\frac{\alpha_s}{4\pi}\bigr)$$\sigma\NLOoneloop$ $(\textrm{stabilised})$
& {\ttlf\texttt{78}}
& {\ttlf\texttt{305\phantom{.0}}}
& {\ttlf$\texttt{10}^{\texttt{-4}}$}
& {\ttlf\texttt{~2.4572~$\pmtt$~0.0141}}
& {\ttlf\texttt{0.575}}
\\
[\doublerulesep]\ccline{3-3}\ccline{4-4}\ccline{5-5}\ccline{6-6}\ccline{7-7}\ccline{8-8}& & \boldmath{\bf$\bigl(\frac{\alpha_s}{4\pi}\bigr)$$\sigma_\textrm{NLO}^{\virtual,\cs}$}\textbf{ $(\textrm{stabilised})$}
&
&
& {\ttbf$\texttt{10}^{\texttt{-4}}$}
& {\ttbf\texttt{~1.4664~$\pmtt$~0.0141}}
& {\ttbf\texttt{0.964}}
\\
[\doublerulesep]\ccline{3-3}\ccline{4-4}\ccline{5-5}\ccline{6-6}\ccline{7-7}\ccline{8-8}& & $\bigl(\frac{\alpha_s}{4\pi}\bigr)$$\sigma\NLOtree$ $(M=1$\,GeV$)$
& {\ttlf\texttt{11}}
& {\ttlf\texttt{4\phantom{.0}}}
& {\ttlf$\texttt{10}^{\texttt{-5}}$}
& {\ttlf\texttt{-9.9080~$\pmtt$~0.0052}}
& {\ttlf\texttt{0.052}}
\\
\cline{3-3}\cline{4-4}\cline{5-5}\cline{6-6}\cline{7-7}\cline{8-8}& & $\bigl(\frac{\alpha_s}{4\pi}\bigr)$$\sigma\NLOoneloop$ $(M=1$\,GeV$)$
& {\ttlf\texttt{118}}
& {\ttlf\texttt{125\phantom{.0}}}
& {\ttlf$\texttt{10}^{\texttt{-4}}$}
& {\ttlf\texttt{~2.4487~$\pmtt$~0.0143}}
& {\ttlf\texttt{0.585}}
\\
[\doublerulesep]\ccline{3-3}\ccline{4-4}\ccline{5-5}\ccline{6-6}\ccline{7-7}\ccline{8-8}& & \boldmath{\bf$\bigl(\frac{\alpha_s}{4\pi}\bigr)$$\sigma_\textrm{NLO}^{\virtual,\cs}$}\textbf{ $(M=1$\,GeV$)$}
&
&
& {\ttbf$\texttt{10}^{\texttt{-4}}$}
& {\ttbf\texttt{~1.4579~$\pmtt$~0.0143}}
& {\ttbf\texttt{0.982}}
\\
[\doublerulesep]\ccline{3-3}\ccline{4-4}\ccline{5-5}\ccline{6-6}\ccline{7-7}\ccline{8-8}& & \bf reference\textbf{}
&
&
& {\ttbf$\texttt{10}^{\texttt{-4}}$}
& {\ttbf\texttt{~1.4490~$\pmtt$~0.0008}}
& {\ttbf\texttt{0.055}}
\\
[\doublerulesep]\ccline{2-2}\ccline{3-3}\ccline{4-4}\ccline{5-5}\ccline{6-6}\ccline{7-7}\ccline{8-8}& \multirow{12}{*}{$\textrm{NNLO},\Nf$}
& $\bigl(\frac{\alpha_s}{4\pi}\bigr)^2$$\sigma\NfNNLOtree$
& {\ttlf\texttt{11}}
& {\ttlf\texttt{4\phantom{.0}}}
& {\ttlf$\texttt{10}^{\texttt{-6}}$}
& {\ttlf\texttt{~2.7263~$\pmtt$~0.0015}}
& {\ttlf\texttt{0.053}}
\\
\cline{3-3}\cline{4-4}\cline{5-5}\cline{6-6}\cline{7-7}\cline{8-8}& & $\bigl(\frac{\alpha_s}{4\pi}\bigr)^2$$\sigma\NfNNLOoneloop$
&
&
& {\ttlf$\texttt{10}^{\texttt{-6}}$}
& {\ttlf\texttt{~2.5667~$\pmtt$~0.0150}}
& {\ttlf\texttt{0.584}}
\\
\cline{3-3}\cline{4-4}\cline{5-5}\cline{6-6}\cline{7-7}\cline{8-8}& & $\bigl(\frac{\alpha_s}{4\pi}\bigr)^2$$\sigma\NfNNLOtwoloop$
& {\ttlf\texttt{1252}}
& {\ttlf\texttt{269\phantom{.0}}}
& {\ttlf$\texttt{10}^{\texttt{-5}}$}
& {\ttlf\texttt{-1.9594~$\pmtt$~0.0136}}
& {\ttlf\texttt{0.696}}
\\
[\doublerulesep]\ccline{3-3}\ccline{4-4}\ccline{5-5}\ccline{6-6}\ccline{7-7}\ccline{8-8}& & \boldmath{\bf$\bigl(\frac{\alpha_s}{4\pi}\bigr)^2$$\sigma_\NfNNLO^{\virtual,\cs}$}\textbf{}
&
&
& {\ttbf$\texttt{10}^{\texttt{-5}}$}
& {\ttbf\texttt{-1.4301~$\pmtt$~0.0137}}
& {\ttbf\texttt{0.959}}
\\
[\doublerulesep]\ccline{3-3}\ccline{4-4}\ccline{5-5}\ccline{6-6}\ccline{7-7}\ccline{8-8}& & $\bigl(\frac{\alpha_s}{4\pi}\bigr)^2$$\sigma\NfNNLOtree$ $(\textrm{stabilised})$
& {\ttlf\texttt{11}}
& {\ttlf\texttt{7\phantom{.0}}}
& {\ttlf$\texttt{10}^{\texttt{-6}}$}
& {\ttlf\texttt{~2.7264~$\pmtt$~0.0015}}
& {\ttlf\texttt{0.053}}
\\
\cline{3-3}\cline{4-4}\cline{5-5}\cline{6-6}\cline{7-7}\cline{8-8}& & $\bigl(\frac{\alpha_s}{4\pi}\bigr)^2$$\sigma\NfNNLOoneloop$ $(\textrm{stabilised})$
&
&
& {\ttlf$\texttt{10}^{\texttt{-6}}$}
& {\ttlf\texttt{~2.5637~$\pmtt$~0.0147}}
& {\ttlf\texttt{0.575}}
\\
\cline{3-3}\cline{4-4}\cline{5-5}\cline{6-6}\cline{7-7}\cline{8-8}& & $\bigl(\frac{\alpha_s}{4\pi}\bigr)^2$$\sigma\NfNNLOtwoloop$ $(\textrm{stabilised})$
& {\ttlf\texttt{1174}}
& {\ttlf\texttt{1040\phantom{.0}}}
& {\ttlf$\texttt{10}^{\texttt{-5}}$}
& {\ttlf\texttt{-1.9502~$\pmtt$~0.0140}}
& {\ttlf\texttt{0.716}}
\\
[\doublerulesep]\ccline{3-3}\ccline{4-4}\ccline{5-5}\ccline{6-6}\ccline{7-7}\ccline{8-8}& & \boldmath{\bf$\bigl(\frac{\alpha_s}{4\pi}\bigr)^2$$\sigma_\NfNNLO^{\virtual,\cs}$}\textbf{ $(\textrm{stabilised})$}
&
&
& {\ttbf$\texttt{10}^{\texttt{-5}}$}
& {\ttbf\texttt{-1.4212~$\pmtt$~0.0140}}
& {\ttbf\texttt{0.988}}
\\
[\doublerulesep]\ccline{3-3}\ccline{4-4}\ccline{5-5}\ccline{6-6}\ccline{7-7}\ccline{8-8}& & $\bigl(\frac{\alpha_s}{4\pi}\bigr)^2$$\sigma\NfNNLOtree$ $(M=1$\,GeV$)$
& {\ttlf\texttt{11}}
& {\ttlf\texttt{4\phantom{.0}}}
& {\ttlf$\texttt{10}^{\texttt{-5}}$}
& {\ttlf\texttt{~1.8117~$\pmtt$~0.0008}}
& {\ttlf\texttt{0.042}}
\\
\cline{3-3}\cline{4-4}\cline{5-5}\cline{6-6}\cline{7-7}\cline{8-8}& & $\bigl(\frac{\alpha_s}{4\pi}\bigr)^2$$\sigma\NfNNLOoneloop$ $(M=1$\,GeV$)$
&
&
& {\ttlf$\texttt{10}^{\texttt{-5}}$}
& {\ttlf\texttt{-6.6625~$\pmtt$~0.0389}}
& {\ttlf\texttt{0.585}}
\\
\cline{3-3}\cline{4-4}\cline{5-5}\cline{6-6}\cline{7-7}\cline{8-8}& & $\bigl(\frac{\alpha_s}{4\pi}\bigr)^2$$\sigma\NfNNLOtwoloop$ $(M=1$\,GeV$)$
& {\ttlf\texttt{914}}
& {\ttlf\texttt{272\phantom{.0}}}
& {\ttlf$\texttt{10}^{\texttt{-5}}$}
& {\ttlf\texttt{~3.3998~$\pmtt$~0.0320}}
& {\ttlf\texttt{0.943}}
\\
[\doublerulesep]\ccline{3-3}\ccline{4-4}\ccline{5-5}\ccline{6-6}\ccline{7-7}\ccline{8-8}& & \boldmath{\bf$\bigl(\frac{\alpha_s}{4\pi}\bigr)^2$$\sigma_\NfNNLO^{\virtual,\cs}$}\textbf{ $(M=1$\,GeV$)$}
&
&
& {\ttbf$\texttt{10}^{\texttt{-5}}$}
& {\ttbf\texttt{-1.4509~$\pmtt$~0.0504}}
& {\ttbf\texttt{3.476}}
\\
\cline{3-3}\cline{4-4}\cline{5-5}\cline{6-6}\cline{7-7}\cline{8-8}\hline\hline
\end{tabular}%
}%
\caption{\label{tab:1}Virtual contributions to the $pp\to\gamma^*\gamma^*\gamma^*$ cross section. We compare with two alternative setups using numerical stability checks and a different UV mass $M$.
The respective contributions are defined in eqs.~\eqref{eq:finite_cross_section_nlo} to \eqref{eq:cross_section_lo}.
}%
\end{table}

\begin{table}[H]
\centering
\resizebox{\columnwidth}{!}{%
\begin{tabular}{lllrrrcr}
\hline\hline
Process & Order & Part & $N_p \ [10^6]$ & $\sfrac{t}{p} \ [\mu \text{s}]$ & Exp. & Result $[\text{pb}]$ & $\Delta \ [\%]$ \\
\hline\hline\multirow{11}{*}{$pp\to\gamma\gamma\gamma$}
& \multirow{2}{*}{$\textrm{LO}$}
& \boldmath{\bf$\sigma_\textrm{LO}$}
& {\ttbf\texttt{10}}
& {\ttbf\texttt{0.3}}
& {\ttbf$\texttt{10}^{\texttt{-3}}$}
& {\ttbf\texttt{~2.5999~$\pmtt$~0.0023}}
& {\ttbf\texttt{0.089}}
\\
[\doublerulesep]\ccline{3-3}\ccline{4-4}\ccline{5-5}\ccline{6-6}\ccline{7-7}\ccline{8-8}& & \bf reference\textbf{}
&
&
& {\ttbf$\texttt{10}^{\texttt{-3}}$}
& {\ttbf\texttt{~2.5980~$\pmtt$~0.0018}}
& {\ttbf\texttt{0.069}}
\\
[\doublerulesep]\ccline{2-2}\ccline{3-3}\ccline{4-4}\ccline{5-5}\ccline{6-6}\ccline{7-7}\ccline{8-8}& \multirow{4}{*}{$\textrm{NLO}$}
& $\bigl(\frac{\alpha_s}{4\pi}\bigr)$$\sigma\NLOtree$
& {\ttlf\texttt{10}}
& {\ttlf\texttt{1\phantom{.0}}}
& {\ttlf$\texttt{10}^{\texttt{-4}}$}
& {\ttlf\texttt{-1.0386~$\pmtt$~0.0017}}
& {\ttlf\texttt{0.165}}
\\
\cline{3-3}\cline{4-4}\cline{5-5}\cline{6-6}\cline{7-7}\cline{8-8}& & $\bigl(\frac{\alpha_s}{4\pi}\bigr)$$\sigma\NLOoneloop$
& {\ttlf\texttt{1420}}
& {\ttlf\texttt{6\phantom{.0}}}
& {\ttlf$\texttt{10}^{\texttt{-4}}$}
& {\ttlf\texttt{~2.5260~$\pmtt$~0.0139}}
& {\ttlf\texttt{0.549}}
\\
[\doublerulesep]\ccline{3-3}\ccline{4-4}\ccline{5-5}\ccline{6-6}\ccline{7-7}\ccline{8-8}& & \boldmath{\bf$\bigl(\frac{\alpha_s}{4\pi}\bigr)$$\sigma_\textrm{NLO}^{\virtual,\cs}$}\textbf{}
&
&
& {\ttbf$\texttt{10}^{\texttt{-4}}$}
& {\ttbf\texttt{~1.4874~$\pmtt$~0.0140}}
& {\ttbf\texttt{0.940}}
\\
[\doublerulesep]\ccline{3-3}\ccline{4-4}\ccline{5-5}\ccline{6-6}\ccline{7-7}\ccline{8-8}& & \bf reference\textbf{}
&
&
& {\ttbf$\texttt{10}^{\texttt{-4}}$}
& {\ttbf\texttt{~1.5090~$\pmtt$~0.0010}}
& {\ttbf\texttt{0.066}}
\\
[\doublerulesep]\ccline{2-2}\ccline{3-3}\ccline{4-4}\ccline{5-5}\ccline{6-6}\ccline{7-7}\ccline{8-8}& \multirow{5}{*}{$\textrm{NNLO},\Nf$}
& $\bigl(\frac{\alpha_s}{4\pi}\bigr)^2$$\sigma\NfNNLOtree$
& {\ttlf\texttt{10}}
& {\ttlf\texttt{1\phantom{.0}}}
& {\ttlf$\texttt{10}^{\texttt{-6}}$}
& {\ttlf\texttt{~1.4781~$\pmtt$~0.0074}}
& {\ttlf\texttt{0.499}}
\\
\cline{3-3}\cline{4-4}\cline{5-5}\cline{6-6}\cline{7-7}\cline{8-8}& & $\bigl(\frac{\alpha_s}{4\pi}\bigr)^2$$\sigma\NfNNLOoneloop$
&
&
& {\ttlf$\texttt{10}^{\texttt{-6}}$}
& {\ttlf\texttt{~2.6355~$\pmtt$~0.0145}}
& {\ttlf\texttt{0.549}}
\\
\cline{3-3}\cline{4-4}\cline{5-5}\cline{6-6}\cline{7-7}\cline{8-8}& & $\bigl(\frac{\alpha_s}{4\pi}\bigr)^2$$\sigma\NfNNLOtwoloop$
& {\ttlf\texttt{7703}}
& {\ttlf\texttt{12\phantom{.0}}}
& {\ttlf$\texttt{10}^{\texttt{-5}}$}
& {\ttlf\texttt{-2.9574~$\pmtt$~0.0236}}
& {\ttlf\texttt{0.798}}
\\
[\doublerulesep]\ccline{3-3}\ccline{4-4}\ccline{5-5}\ccline{6-6}\ccline{7-7}\ccline{8-8}& & \boldmath{\bf$\bigl(\frac{\alpha_s}{4\pi}\bigr)^2$$\sigma_\NfNNLO^{\virtual,\cs}$}\textbf{}
&
&
& {\ttbf$\texttt{10}^{\texttt{-5}}$}
& {\ttbf\texttt{-2.5460~$\pmtt$~0.0237}}
& {\ttbf\texttt{0.929}}
\\
[\doublerulesep]\ccline{3-3}\ccline{4-4}\ccline{5-5}\ccline{6-6}\ccline{7-7}\ccline{8-8}& & \bf reference\textbf{}
& {\ttbf\texttt{177}}
&
& {\ttbf$\texttt{10}^{\texttt{-5}}$}
& {\ttbf\texttt{-2.5732~$\pmtt$~0.0006}}
& {\ttbf\texttt{0.023}}
\\
\hline\hline\multirow{10}{*}{$p_dp_d\to Z\gamma_1^*\gamma_2^*$}
& \multirow{2}{*}{$\textrm{LO}$}
& \boldmath{\bf$\sigma_\textrm{LO}$}
& {\ttbf\texttt{11}}
& {\ttbf\texttt{5\phantom{.0}}}
& {\ttbf$\texttt{10}^{\texttt{-3}}$}
& {\ttbf\texttt{~1.3831~$\pmtt$~0.0010}}
& {\ttbf\texttt{0.072}}
\\
[\doublerulesep]\ccline{3-3}\ccline{4-4}\ccline{5-5}\ccline{6-6}\ccline{7-7}\ccline{8-8}& & \bf reference\textbf{}
&
&
& {\ttbf$\texttt{10}^{\texttt{-3}}$}
& {\ttbf\texttt{~1.3830~$\pmtt$~0.0009}}
& {\ttbf\texttt{0.066}}
\\
[\doublerulesep]\ccline{2-2}\ccline{3-3}\ccline{4-4}\ccline{5-5}\ccline{6-6}\ccline{7-7}\ccline{8-8}& \multirow{4}{*}{$\textrm{NLO}$}
& $\bigl(\frac{\alpha_s}{4\pi}\bigr)$$\sigma\NLOtree$
& {\ttlf\texttt{11}}
& {\ttlf\texttt{21\phantom{.0}}}
& {\ttlf$\texttt{10}^{\texttt{-5}}$}
& {\ttlf\texttt{-2.6582~$\pmtt$~0.0077}}
& {\ttlf\texttt{0.289}}
\\
\cline{3-3}\cline{4-4}\cline{5-5}\cline{6-6}\cline{7-7}\cline{8-8}& & $\bigl(\frac{\alpha_s}{4\pi}\bigr)$$\sigma\NLOoneloop$
& {\ttlf\texttt{108}}
& {\ttlf\texttt{1056\phantom{.0}}}
& {\ttlf$\texttt{10}^{\texttt{-4}}$}
& {\ttlf\texttt{~2.7258~$\pmtt$~0.0210}}
& {\ttlf\texttt{0.771}}
\\
[\doublerulesep]\ccline{3-3}\ccline{4-4}\ccline{5-5}\ccline{6-6}\ccline{7-7}\ccline{8-8}& & \boldmath{\bf$\bigl(\frac{\alpha_s}{4\pi}\bigr)$$\sigma_\textrm{NLO}^{\virtual,\cs}$}\textbf{}
&
&
& {\ttbf$\texttt{10}^{\texttt{-4}}$}
& {\ttbf\texttt{~2.4600~$\pmtt$~0.0210}}
& {\ttbf\texttt{0.855}}
\\
[\doublerulesep]\ccline{3-3}\ccline{4-4}\ccline{5-5}\ccline{6-6}\ccline{7-7}\ccline{8-8}& & \bf reference\textbf{}
&
&
& {\ttbf$\texttt{10}^{\texttt{-4}}$}
& {\ttbf\texttt{~2.4440~$\pmtt$~0.0016}}
& {\ttbf\texttt{0.065}}
\\
[\doublerulesep]\ccline{2-2}\ccline{3-3}\ccline{4-4}\ccline{5-5}\ccline{6-6}\ccline{7-7}\ccline{8-8}& \multirow{4}{*}{$\textrm{NNLO},\Nf$}
& $\bigl(\frac{\alpha_s}{4\pi}\bigr)^2$$\sigma\NfNNLOtree$
& {\ttlf\texttt{19}}
& {\ttlf\texttt{20\phantom{.0}}}
& {\ttlf$\texttt{10}^{\texttt{-7}}$}
& {\ttlf\texttt{-2.1367~$\pmtt$~0.0204}}
& {\ttlf\texttt{0.954}}
\\
\cline{3-3}\cline{4-4}\cline{5-5}\cline{6-6}\cline{7-7}\cline{8-8}& & $\bigl(\frac{\alpha_s}{4\pi}\bigr)^2$$\sigma\NfNNLOoneloop$
&
&
& {\ttlf$\texttt{10}^{\texttt{-6}}$}
& {\ttlf\texttt{~2.8440~$\pmtt$~0.0219}}
& {\ttlf\texttt{0.771}}
\\
\cline{3-3}\cline{4-4}\cline{5-5}\cline{6-6}\cline{7-7}\cline{8-8}& & $\bigl(\frac{\alpha_s}{4\pi}\bigr)^2$$\sigma\NfNNLOtwoloop$
& {\ttlf\texttt{927}}
& {\ttlf\texttt{2403\phantom{.0}}}
& {\ttlf$\texttt{10}^{\texttt{-5}}$}
& {\ttlf\texttt{-2.6932~$\pmtt$~0.0228}}
& {\ttlf\texttt{0.845}}
\\
[\doublerulesep]\ccline{3-3}\ccline{4-4}\ccline{5-5}\ccline{6-6}\ccline{7-7}\ccline{8-8}& & \boldmath{\bf$\bigl(\frac{\alpha_s}{4\pi}\bigr)^2$$\sigma_\NfNNLO^{\virtual,\cs}$}\textbf{}
&
&
& {\ttbf$\texttt{10}^{\texttt{-5}}$}
& {\ttbf\texttt{-2.4301~$\pmtt$~0.0229}}
& {\ttbf\texttt{0.941}}
\\
\cline{3-3}\cline{4-4}\cline{5-5}\cline{6-6}\cline{7-7}\cline{8-8}\hline\hline
\end{tabular}%
}%
\caption{\label{tab:2}Virtual contributions to $2\to3$ cross sections.
The respective contributions are defined in eqs.~\eqref{eq:finite_cross_section_nlo} to \eqref{eq:cross_section_lo}.
}%
\end{table}

%% file: sections/conclusion.tex
\section{Conclusion}
\label{sec:conclusion}

We introduced an efficient numerical method to evaluate virtual corrections for the production of multiple electroweak vector bosons in proton-proton collisions.
Our approach offers a potential solution to overcome the complexity barrier encountered by established methods when calculating two-loop $2\to3$ scattering amplitudes involving multiple massive particles in the final state. 
The feasibility of this method was demonstrated by computing previously inaccessible $\Nf$-contributions to the virtual corrections at NNLO, involving up to three equal-mass off-shell photons, and the $d\bar{d}$-initiated contributions to the production of two different-mass off-shell photons alongside a $Z$ boson.
Additionally, we showcased its flexibility by calculating these and other lower-order contributions using the same computational pipeline.
\par
The method relies on the local subtraction of UV and IR divergences using counterterms, based on the local factorisation paradigm of refs.~\cite{Anastasiou:2020sdt,Anastasiou:2022eym}, to construct locally finite amplitudes that capture the hard scattering contributions.
The integration of these counterterms and their combination with the IR and UV poles is analytically performed within dimensional regularisation, which allows to choose any preferred IR factorisation and UV renormalisation schemes.
The integrated counterterm expressions apply universally to the production of any number of external electroweak vector bosons, without the need to calculate additional integrals nor carry out Dirac algebra away from four spacetime dimensions.
\par
The hard scattering amplitude is evaluated by first analytically integrating out the loop energies according to the loop-tree duality and the cross-free family expression of the respective refs.~\cite{Capatti:2019ypt,Capatti:2022mly}.
Subsequently, we subtract the threshold singularities following the method of ref.~\cite{Kermanschah:2021wbk}, before numerically performing the interference with the tree amplitude.
Finally, the expression is numerically integrated over the (spatial) loop momenta and the phase space using Monte~Carlo methods.
\par
Looking forward, this approach can readily accommodate four or more external electroweak bosons, necessitating only considerations of additional thresholds.
Other future directions include the combination of the virtual contributions with the real ones.
Our next goal is to extend this method beyond the $\Nf$-part and encompass the complete NNLO virtual contribution.

%% file: sections/finite_remainders.tex
\section{Catani finite remainders and integrated counterterms}
\label{app:finite_remainders}

The Catani finite remainders of the one-loop and the $\Nf$-part of the two-loop amplitudes are given in eq.~\eqref{eq:1_loop_finite_remainder} and eq.~\eqref{eq:fin_rem_2_nf}, respectively.
In this section we calculate the two contributions in $d=4-2\ep$ spacetime dimensions,
\begin{align}
    R^{(1)}_\text{singular}
    &= \mathcal{R}_{\rm IR} \bare^{(1)} +\mathcal{R}_{\rm UV} \bare^{(1)} -\mathcal{R}_{\rm UV}\mathcal{R}_{\rm IR} \bare^{(1)} - \mathbf{Z}^{(1)} \bare^{(0)} +\mathcal{O}(\ep)\,,\\
    R^{(2,\Nf)}_\text{singular} &=
    \left(
    -\frac{2\beta_0^{(\Nf)}}{\ep}
    \bare^{(1)}_\text{singular}
    +
    \nfcoeff
    \left(
    Y_\text{IR} - Y_\text{UV\&IR} + Y_\text{UV}
    \right)
    \right)
    - \mathbf{Z}^{(2,\Nf)} \bare^{(0)} +\mathcal{O}(\ep)\,,
\end{align}
where $\beta_0^{(\Nf)}$ is the part of $\beta_0$ proportional to $\Nf$:
\begin{equation}
    \beta_0^{(\Nf)} = -\frac{2}{3}\Nf T_F\,. \label{eq:beta0-nf-part}
    \end{equation}
The coefficients $\mathbf{Z}^{(1)}$ and $\mathbf{Z}^{(2,\Nf)}$ can be found in section~\ref{app:catani}.
The integrated one-loop counterterms evaluate to
\begin{align}
    \label{eq:form-factor-integrated}
   \mathcal{R}_{\rm IR}\bare^{(1)}
    &= -\ii \left[ - \mathcal{T}_1 \left( \frac{2}{\ep} +\frac{\ep}{1-\ep} \right) + \mathcal{T}_2 \frac{1}{4}\frac{1}{1-\ep}  \right] 
    {\rm Bub}(s_{12})\,,\\
   \mathcal{R}_{\rm UV}\mathcal{R}_{\rm IR} \bare^{(1)}
    &=  -\ii \mathcal{T}_2  \frac{1-\ep}{4} \frac{{\rm TadP}(\mUV^2)}{\mUV^2}\,,
    \label{eq:uv-ctm-from-factor-integrated}\\
 \label{eq:uv-vert-integrated}
    \mathcal{R}_{\rm UV}\bare_{\textrm{vertex}}^{(1)}
    &=  -\ii  \nfinal  \mathcal{T}_1 (1-\ep)^3
            \frac{{\rm TadP}(\mUV^2)}{\mUV^2}\,,\\
    \label{eq:uv-self-energy-integrated}
   \mathcal{R}_{\rm UV}\bare_{\textrm{self-energy}}^{(1)}
   &= \ii  (\nfinal-1)
   \mathcal{T}_1 (1-\ep)^2 \frac{{\rm TadP}(\mUV^2)}{\mUV^2}\,,
   \end{align}
where the contributions from vertex and self-energy corrections combine as
\begin{align}
    \mathcal{R}_\text{UV}\bare^{(1)}
    &= \mathcal{R}_{\rm UV}\bare_{\textrm{vertex}}^{(1)} + \mathcal{R}_{\rm UV}\bare_{\textrm{self-energy}}^{(1)}\,.
\end{align}
The integrated two-loop counterterms are
\begin{align}
    \label{eq:form-factor-integrated2l}
    &Y_\text{IR}
    = \int \ddmsbar{l}\ddmsbar{k}
    \frac{\mathcal{R}_{\rm IR}\mathcal{M}^{(1)}(k)}{l^2(l+k)^2}
   = -\ii\left[  \mathcal{T}_1 \left( \frac{8}{3} \frac{1}{3 \ep-2}-\frac{1}{\ep}+\frac{1}{3} \right) + \mathcal{T}_2 \frac{1}{2}\frac{1}{2-3\ep}  \right] {\rm Tri}(s_{12})\,, \\
   \label{eq:uv-ctm-from-factor-integrated2l}
     &Y_\text{UV\&IR}
     =\int \ddmsbar{l}\ddmsbar{k} 
    \frac{\mathcal{R}_{\rm UV}\mathcal{R}_{\rm IR} \mathcal{M}^{(1)}(k)}{l^2(l+k)^2}
    = \ii\left(\ep+
    \frac{3}{2}
    \frac{1}{\ep-2}
    +\frac{1}{2} \right) \mathcal{T}_2\frac{{\rm Tad2P}(\mUV^2)}{\mUV^2}\,,\\
    \label{eq:uv-vert-integrated2l}
    &Y^\text{v}_\text{UV}
    = \int \ddmsbar{l}\ddmsbar{k} 
    \frac{\mathcal{R}_{\rm UV}\mathcal{M}_{\textrm{vertex}}^{(1)}(k)}{l^2(l+k)^2}
    =  -2\ii  \frac{(2\ep-1)(\ep-1)^3}{2-\ep} \nfinal\mathcal{T}_1 \frac{{\rm Tad2P}(\mUV^2)}{\mUV^2}\,, \\
    &Y^\text{se}_\text{UV}
    =\int \ddmsbar{l}\ddmsbar{k}  \frac{\mathcal{R}_{\rm UV}\mathcal{M}_{\textrm{self-energy}}^{(1)} (k)}{l^2(l+k)^2}
    = 2\ii \frac{(2\ep-1) (\ep-1)}{2-\ep}  (\nfinal-1)\mathcal{T}_1\frac{{\rm Tad2P}(\mUV^2)}{\mUV^2}\,,
    \label{eq:uv-self-energy-integrated2l}
\end{align}
such that $Y_\text{UV} = Y_\text{UV}^\text{v} +  Y_\text{UV}^\text{se}$.
We used that for $\nfinal$ external gauge bosons, we have $\nfinal$ diagrams with a vertex correction and $\nfinal-1$ diagrams with a self-energy insertion.
The master integrals ${\rm Bub},{\rm TadP},{\rm Tri}$ and ${\rm Tad2P}$ are presented in section~\ref{app:masters}.
We define the modified tree amplitudes $\mathcal{T}_1$ and $\mathcal{T}_2$ as
\begin{align}
    \label{eq:T1-def}
    \mathcal{T}_1
    &=  C_F \bare^{(0)}
    = C_F \bar{v}(p_2)\left[
    \widetilde{\mathcal{M}}^{(0)} \right]u(p_1)   \,, 
    \\
    \label{eq:T2-def}
    \mathcal{T}_2
    &= C_F \bar{v}(p_2)\gamma^{\mu} \gamma^{\nu}\left[
    \mathbf{P}_1 \widetilde{\mathcal{M}}^{(0)}\mathbf{P}_1\right] \gamma_{\nu}\gamma_{\mu} u(p_1)\,.
\end{align}
Finally, combining all contributions yields the final expressions
\begin{align}
\label{eq:M-singular-integrated}
  R^{(1)}_\text{singular}
  &= \left(\mathcal{T}_1 -\frac{\mathcal{T}_2}{4}\right) m_{\mu}
    -\left( l_{\mu}^{2}+\nfinal +4 l_{\mu} +10-\frac{\pi^{2}}{6}\right) \mathcal{T}_1 
    +\frac{\mathcal{T}_2}{4} \left(l_{\mu} +3\right)\,,\\
\label{eq:2-nf-fin-remainder-integrated}
        R^{(2,\Nf)}_\text{singular}
    &=  \frac{4}{9} T_F \Nf \left[
    \left(-\frac{3 \mathcal{T}_1}{2}+\frac{3 \mathcal{T}_2}{8}\right) m_{\mu}^{2}\right.  \nonumber\\
    &+\left(\left(3 \nfinal -\frac{7}{2}\right) \mathcal{T}_1 
     +\frac{\mathcal{T}_2}{8}\right) m_{\mu}
 \nonumber  \\
   &\left.
   + \left(l_{\mu}^{3}+11 l_{\mu}^{2}+\left(\pi^{2}+\frac{133}{3}\right) l_{\mu} +\zeta_3+\frac{23 \pi^{2}}{12}+2 \nfinal +\frac{5291}{72}\right) \mathcal{T}_1
   \right. 
  \nonumber \\ 
   &\left.
   -\frac{3}{8} \left(l_{\mu}^{2}+\frac{19}{3} l_{\mu} +\frac{25}{2}\right) \mathcal{T}_2\right]\,,
\end{align}
with the abbreviations $l_\mu = \ln (-\frac{\mu^2}{s_{12}+\ii \epsilon})$ and $m_{\mu}=\ln \! \left(\frac{\mu^{2}}{\mUV^{2}}\right)$.
\par
One important remark is in order.
The finite remainders in eqs.~\eqref{eq:M-singular-integrated} and \eqref{eq:2-nf-fin-remainder-integrated} feature $\mathcal{T}_1$ and $\mathcal{T}_2$, which contain gamma matrices (and may contain $\gamma^5$ depending on the external gauge bosons).
In particular, $\mathcal{T}_2$ contains two pairs of repeated Lorentz indices, which we retain in order to preserve the underlying tree-level structure intact.
One could naively believe that we need to treat $\mathcal{T}_1$ and $\mathcal{T}_2$ as $d$-dimensional objects when we evaluate eq.~\eqref{eq:2-nf-fin-remainder-integrated}.
However, it is not necessary to explicitly consider the evanescent contributions of such objects in dimensional regularisation, since only the order $O(\ep^0)$ will contribute to $R^{(1)}$ and $R^{(2,\Nf)}$. 

\subsection{Catani IR poles}
\label{app:catani}
For completeness we report the expressions for $\mathbf{Z}^{(1)}$ and $\mathbf{Z}^{(2,\Nf)}$
\begin{align}
    \label{eq:Z-1}
&\mathbf{Z}^{(1)}=-C_F \left(\frac{2 }{\ep^2}+
   \frac{ 2  l_\mu+3 }{\ep} \right)\,,
    \\
  &\mathbf{Z}^{(2,\Nf)}=  \Nf C_F T_F \left(-\frac{2 }{\ep^{3}}
  -\frac{4 }{9\ep^2} \left(3 l_{\mu} +2\right)
 + \frac{1 }{27\ep}\left(9 \pi^{2}+60 l_{\mu} +65\right)
  \right) \,,
    \label{eq:Z-2-Nf}
\end{align} 
where $l_\mu = \ln (-\frac{\mu^2}{s_{12}+\ii \epsilon})$.
We follow the same conventions as ref.~\cite{Chawdhry:2020for}.
$\mathbf{Z}^{(2,\Nf)}$ is the contribution proportional to $\Nf$ of $\mathbf{Z}^{(2)}$ in eq.~\eqref{eq:Z-2L}.

\subsection{Master integrals}
\label{app:masters}
The integrated IR and UV counterterms in~\crefrange{eq:form-factor-integrated}{eq:uv-self-energy-integrated} and \crefrange{eq:form-factor-integrated2l}{eq:uv-self-energy-integrated2l} are expressed in terms of the following master integrals.
At one loop we need
\begin{align}
    \label{eq:bub-master-integral}
    \text{Bub}(p^2)
    &=\int \ddmsbar{k}
    \frac{1}{(k+p)^2+\ii \epsilon}
    \frac{1}{k^2+\ii \epsilon}
    =\ii e^{\ep \gamma_E}\frac{\Gamma(\ep)\Gamma(1-\ep)^2}{\Gamma(2-2\ep)}
    \left(-\frac{\mu^2}{p^2+\ii \epsilon}\right)^\ep\,, \\
    \label{eq:tadpole-1L}
    \frac{{\rm TadP}(\mUV^2)}{\mUV^2}
    &= \frac{1}{\mUV^2}\int \ddmsbar{k}
    \frac{1}{k^2-\mUV^2+\ii \epsilon}
    = -\ii e^{\ep \gamma_E}\Gamma(\ep-1)\left(\frac{\mu^2}{\mUV^2-\ii \epsilon}\right)^{\ep}\,.
\end{align}
The relevant two-loop integrals are
\begin{align}
    \frac{{\rm Tad2P}(\mUV^2)}{\mUV^2}
    &= \frac{1}{\mUV^2}\int \ddmsbar{k} \ddmsbar{l}
    \frac{1}{l^2+\ii \epsilon}
    \frac{1}{(l+k)^2+\ii \epsilon}
    \frac{1}{k^2-\mUV^2+\ii \epsilon}
    \nonumber\\
    &= e^{2\ep\gamma_E} 
\frac{\Gamma (\ep) \Gamma (1-\ep)^2  \Gamma (2 \ep-1)}{\Gamma
   (2-\ep)} \left( \frac{\mu^2}{\mUV^2-\ii \epsilon}\right)^{2 \ep}\,,
   \label{eq:tadp2-master-integral}
   \\
    {\rm Tri}(p^2)
    &= \int \ddmsbar{k} \ddmsbar{l}
    \frac{1}{l^2+\ii \epsilon}
    \frac{1}{(l+k)^2+\ii \epsilon}
    \frac{1}{k^2+\ii \epsilon}
    \frac{1}{(k+p)^2+\ii \epsilon}\nonumber\\
 &= -e^{2\ep\gamma_E}  
 \frac{\Gamma(2\ep)\Gamma(1-2\ep)^2}{\Gamma(2-3e)}
 \frac{\Gamma(\ep)\Gamma(1-\ep)^2}{\Gamma(2-2e)}
 \left(-\frac{\mu^2}{p^2+\ii \epsilon}\right)^{2\ep}\,.
 \label{eq:tri-master-integral} 
\end{align}
They were calculated in ref.~\cite{Anastasiou:1999bn}.
The measure is defined as
\begin{align}
    \ddmsbar{k}
    = \mu^{2\epsilon} e^{\ep\gamma_E}\frac{\dd^d k}{\pi^{d/2}}\,.
\end{align}
and the additional factors of $e^{\ep\gamma_E}$ remove the $\gamma_E$-dependence in the results.


%% file: sections/phase_space_param.tex
\section{Two- and three-body phase space parametrisation}
\label{app:ps_parametrisation}
In this section we describe the phase space parametrisation we used for evaluating the virtual corrections for the $2\to2$ and $2\to3$ processes (cf.~\cite{Hilgart:1992xu,Berends:1994pv,Denner:1999gp,Roth:1999kk,Dittmaier:2002ap,Kersevan:2004yh}).
\par
The momentum fractions $x_1$ and $x_2$ are parameterised using the $\tau$ and the rapidity $\eta$, defined as
\begin{align}
    \tau = x_1 x_2, \qquad \eta = \frac{1}{2} \ln \frac{x_1}{x_2} \qquad \text{or} \qquad
    x_1 = \sqrt{\tau} \exp \eta, \qquad x_2 = \sqrt{\tau} \exp-\eta.
\end{align}
The integral measure and boundaries transform as
\begin{align}
    \int\limits_0^1 \dd x_1 \int\limits_0^1 \dd x_2 \longrightarrow \int\limits_0^1 \dd \tau \int\limits_{\frac{1}{2}\ln\tau}^{-\frac{1}{2}\ln\tau} \dd \eta.
\end{align}
We sample $\tau$ and $\eta$ linearly from the hypercube.
The incoming momenta are taken along the $z$-axis and expressed in the center-of-mass frame 
\begin{align}
    p_1 = \frac{\hat{E}_\text{CM}}{2}\begin{pmatrix}
        1, & 0, & 0, & 1
    \end{pmatrix}, \qquad
    p_2 = \frac{\hat{E}_\text{CM}}{2}\begin{pmatrix}
        1, & 0, & 0, & -1
    \end{pmatrix},
\end{align}
where $\hat{E}_\text{CM}=\sqrt{\tau} E_\text{CM} =\sqrt{\tau s}$.
\par
The two-body phase space we decompose as
\begin{align}
    \int \dd\Pi_2(q_1,q_2) = \frac{\lambda^{\frac{1}{2}}(s,s_1,s_2)}{8s} \int\limits_{-1}^1\dd \cos\theta \int\limits_0^{2\pi} \dd \varphi,
\end{align}
where $s_1=q_1^2$, $s_2 =q_2^2$, $s=p^2$ and $p=q_1+q_2$ is the total incoming momentum. 
In the rest frame of $p$ the momenta $q_1$ and $q_2$ are back-to-back along a direction determined by the angles $\varphi$ and $\cos\theta$.
The outgoing momenta read
\begin{align}
    q_1 = B(\vec{v})R(\varphi,\cos\theta)
    \begin{pmatrix}
        \frac{s + s_1 - s_2}{2\sqrt{s}}\\
        0\\
        0\\
        \frac{ \lambda^{1/2}(s,s_1,s_2)}{2\sqrt{s}}
    \end{pmatrix}
    \qquad 
    \text{and}
    \qquad
    q_2 = p-q_1
\end{align}
where $R(\varphi,\cos\theta)$ is a rotation matrix and $B(\vec{v})$ a boost with velocity $\vec{v}=\vec{p}/p^0$.
The three-body phase space we decompose as
\begin{align}
    \int \Pi_3 = \int \dd \Pi_2(q_1,q_{32}) \int\limits_{s_{32,\text{min}}}^{s_{32,\text{max}}} \dd s_{32}  \int \dd \Pi(q_3,q_2)
\end{align}
where the bounds are
\begin{align}
    s_{32,\text{min}} = (\sqrt{s_2}+\sqrt{s_3})^2 \qquad \text{and} \qquad
    s_{32,\text{max}} = (\hat{E}_\text{CM}-\sqrt{s_1})^2
\end{align}
for $s_i=q_i^2$.
\par
The invariant $s_{32}$ and the angle $\varphi$ are sampled linearly from the unit hypercube.
We further express the $\cos\theta$ of the two-body phase space building blocks as $t$-channel invariants.
For two incoming momenta $p_a$ and $p_b$ with $p=p_a+p_b$ the invariant reads
\begin{align}
\label{eq:t_channel}
    t = (q_1-p_1)^2 = s_1 + s_a - \frac{(s+s_1-s_2)(s+s_a-s_b)-\lambda^{\frac{1}{2}}(s,s_1,s_2)\lambda^{\frac{1}{2}}(s,s_a,s_b)\cos\theta}{2 s}
\end{align}
where $s_a=p_a^2$ and $s_b=p_b^2$.
The integral transforms as
\begin{align}
    \int\limits_{-1}^1 \dd \cos\theta = \frac{2 s}{\lambda^{\frac{1}{2}}(s,s_1,s_2)\lambda^{\frac{1}{2}}(s,s_a,s_b)}\int\limits_{-t_\text{max}}^{-t_\text{min}} \dd |t|
\end{align}
where the bounds of $t$ follow from the bounds of $\cos\theta$ and eq.~\eqref{eq:t_channel}.
We use importance sampling to map the invariant $t$ from the hypercube variable $r$ to explicitly account for the pure $t$-channel resonance structure $1/t^2$ using the change of variables
\begin{align}
    \phi(r) &= \left[(\phi_\text{max}-m^2)^{1-\nu}r+ (\phi_\text{min}-m^2)^{1-\nu}(1-r)\right]^\frac{1}{1-\nu} + m^2 \\
    \frac{\dd \phi}{\dd r} &= \frac{\left[(\phi_\text{max}-m^2)^{1-\nu}- (\phi_\text{min}-m^2)^{1-\nu}\right]}{1-\nu} (\phi-m^2)^\nu,
\end{align}
for $\phi=|t|$, $\phi_\text{min}=-t_\text{max}$ and $\phi_\text{max}=-t_\text{min}$.
We set $m^2=0$ if $\phi_\text{min}\neq0$ and otherwise  to a small parameter $m^2=-E_\text{CM}^2/10^2$.
In the latter case, the phase space is subject to a transverse momentum and/or rapidity cut in order for the cross section to be finite.
Instead of $\nu=2$, we find $\nu=1.5$ to result in better performance, which may be related to numerator effects and the interference contribution with the $u$-channel.
Note that we do not address the pure $u$-channel resonance $1/u^2$, which is typically accounted for using multi-channelling.
However, for indistinguishable particles in the final state we can effectively map the $u$-channel onto the $t$-channel.

%% file: sections/ps_points.tex
\section{Finite remainders at fixed phase-space points}\label{app:fixed_ps_results}
Since our code allows the phase-space integration to be disabled, we can compute finite remainders directly at fixed phase-space points by performing only the loop integration.
In the following appendix, we will leverage this functionality to validate our approach against existing amplitude calculations and to facilitate comparisons with future results.
Note that these results are produced specifically for validation purposes and do not represent intermediate results, as our approach is ultimately designed to also perform the phase-space integration.
\par
We define the tree-level, one-loop and two-loop $\Nf$ contributions to the helicity summed (but not averaged) finite remainders of the squared matrix element as
\begin{align}
\label{eq:fin_rem_def_start}
    F^{(0)} &= \sum_h \left|R_h^{(0)}\right|^2\,,\\
    F^{(1)}_{\overline{\text{MS}}} &= \sum_h 2\Re\left[\left(R_h^{(0)}\right)^*R_h^{(1)}\right]\,,\\
    F^{(2,\Nf)}_{\overline{\text{MS}}} &= \sum_h 2\Re\left[\left(R_h^{(0)}\right)^*R_h^{(2,\Nf)}\right]\,,
\end{align}
where the finite remainder of the amplitude $R_h$ is given in the $\overline{\text{MS}}$ scheme according to eqs.~\eqref{eq:Z-tree} to \eqref{eq:Z-2L}.
In tables~\ref{tab:ps_2to2} and \ref{tab:ps_2to3}, we present their values for six different processes at three random phase-space points generated with \textsc{RAMBO} \cite{Kleiss:1985gy}.
The phase-space points are provided in listings~\ref{lst:1} to \ref{lst:5}.
As in section~\ref{sec:results}, we give the two-loop $\Nf$ contributions in the $\overline{\text{MS}}$ scheme and the one-loop contributions follow the BLHA normalisation convention \cite{Binoth:2010xt,Alioli:2013nda}, which relates to the $\overline{\text{MS}}$ scheme via
\begin{equation}
    \label{eq:fin_rem_def_end}
    F^{(1)}_{\text{BLHA}} = F^{(1)}_{\overline{\text{MS}}}\,- \frac{\pi^2}{6}C_F F^{(0)}\,.\\
\end{equation}
\par
We validate our numerically integrated results with exact benchmark values obtained from analytic loop integration.
The one-loop reference results were generated with \textsc{MadGraph} \cite{Hirschi:2011pa,Alwall:2014hca}.
The reference results for the two-loop $\Nf$ parts were generated from the analytic calculations of refs.~\cite{Anastasiou:2002zn, Caola:2020dfu} for $\gamma\gamma$, refs.~\cite{Caola:2014iua, Gehrmann:2015ora} for $\gamma^*\gamma^*$ and $ZZ$ and ref.~\cite{Abreu:2023bdp} for $\gamma\gamma\gamma$ production.
To the best of our knowledge, no benchmark results exist for the corresponding contributions to the processes $q\bar{q}\to\gamma^*\gamma^*\gamma^*$ and $q\bar{q}\to Z\gamma_1^*\gamma_2^*$ and analytic computations are currently not possible since the master integrals are not available. 
\par
In table~\ref{tab:ps_points_ffs} we present for completeness and reproducibility the directly numerically integrated results using $10^9$ Monte~Carlo samples in the form factor subtraction scheme (FFS),
\begin{align}
\label{eq:fin_rem_ffs_1l}
    F^{(1)}_{\text{FFS}} &= \sum_h 2\Re\left[\left(R_h^{(0)}\right)^*M_{\text{finite},h}^{(1)}\right]\,,\\
\label{eq:fin_rem_ffs_2lnf}
    F^{(2,\Nf)}_{\text{FFS}} &= \sum_h 2\Re\left[\left(R_h^{(0)}\right)^*M_{\text{finite},h}^{(2,\Nf)}\right]\,,
\end{align}
which, after performing the scheme transformation described in eqs.~\eqref{eq:scheme_change_1l} and \eqref{eq:scheme_change_2l}, were used to obtain the results shown in tables~\ref{tab:ps_2to2} and \ref{tab:ps_2to3}.
\par

We highlight that the process $q\bar{q}\to Z\gamma_1^*\gamma_2^*$ receives a parity-sensitive contribution from the imaginary part of the absorptive amplitude according to eq.~\eqref{eq:re_summary}.
While this contribution is non-zero at fixed phase-space points, it vanishes in the phase-space integrated virtual corrections presented in tables~\ref{tab:new_results} and \ref{tab:2}.

\par
The Monte Carlo integration was performed with Vegas, where the parameters \texttt{nstart} and \texttt{nincrease} were set to $10^5$.
The colour factors $C_F$ and $\Nf C_F T_F$ are removed from the presented one- and two-loop contributions, respectively, and the electromagnetic charge of the quark is factored out.
The renormalisation and factorisation scales are set equal and fixed at $\mu^2 = s$.
\par
In the tables, $\Delta\,[\%]$ denotes the relative Monte Carlo error, and $\Delta\,[\sigma]$ the deviation between the numerical and reference results in units of the Monte Carlo error.
The relative error remains below one percent for all results, except for the $\Im\absorptive F_{\text{FFS}}$ contributions, which are more than an order of magnitude smaller than $F_{\text{FFS}}$ itself.
Moreover, for all processes, the benchmark results, where available, are consistent with the numerical estimates within $2\sigma$.
\par
\begin{figure}[H]
    \centering
    \includegraphics[width=\linewidth]{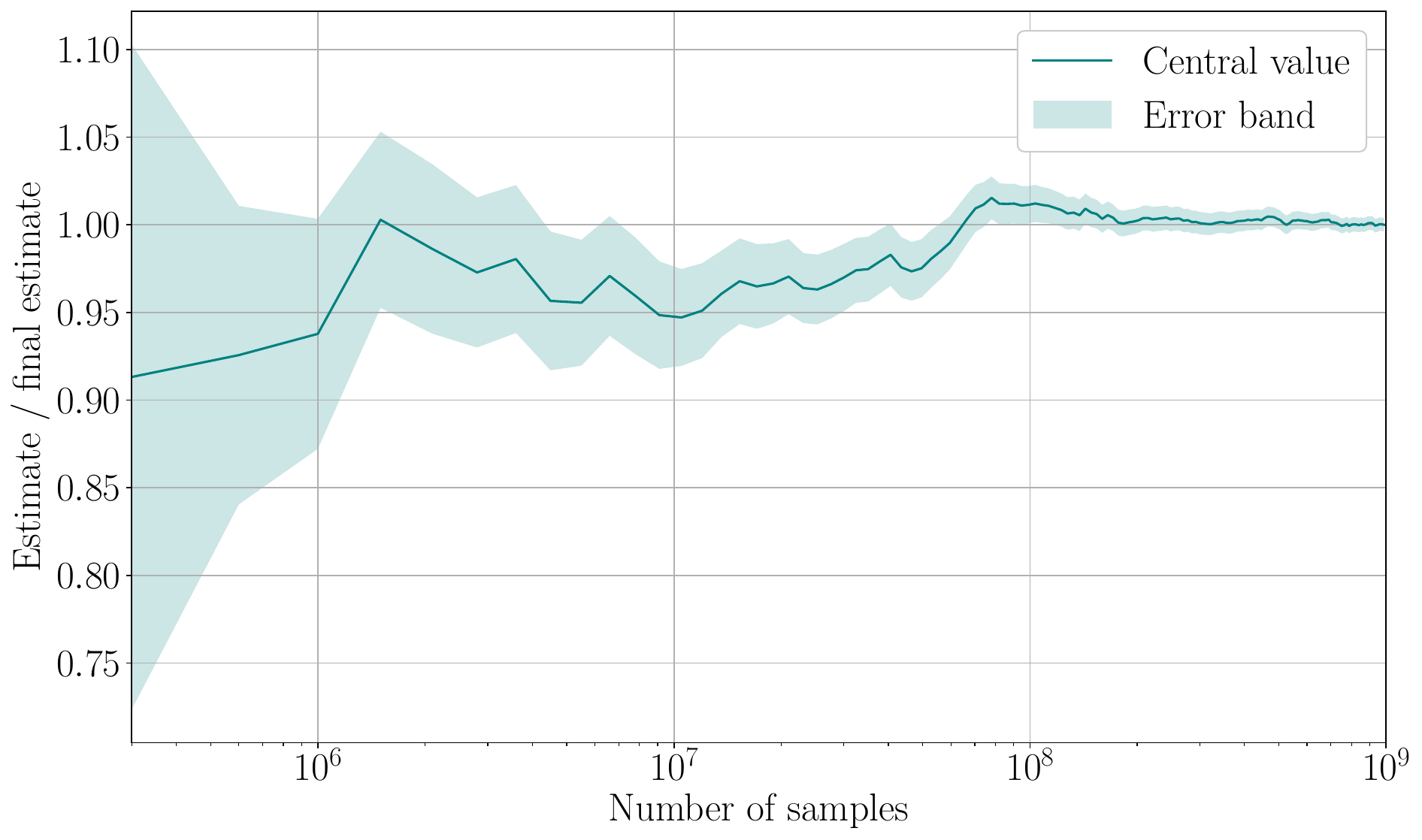}
    \caption{Convergence of the Monte Carlo integration for the contribution $F^{(2,\Nf)}_\text{FFS}$ to the process $q\bar{q}\to\gamma^*\gamma^*\gamma^*$ at the first phase-space point in listing~\ref{lst:4}.}
    \label{fig:convergence_1}
\end{figure}
\begin{figure}[H]
    \centering
    \includegraphics[width=\linewidth]{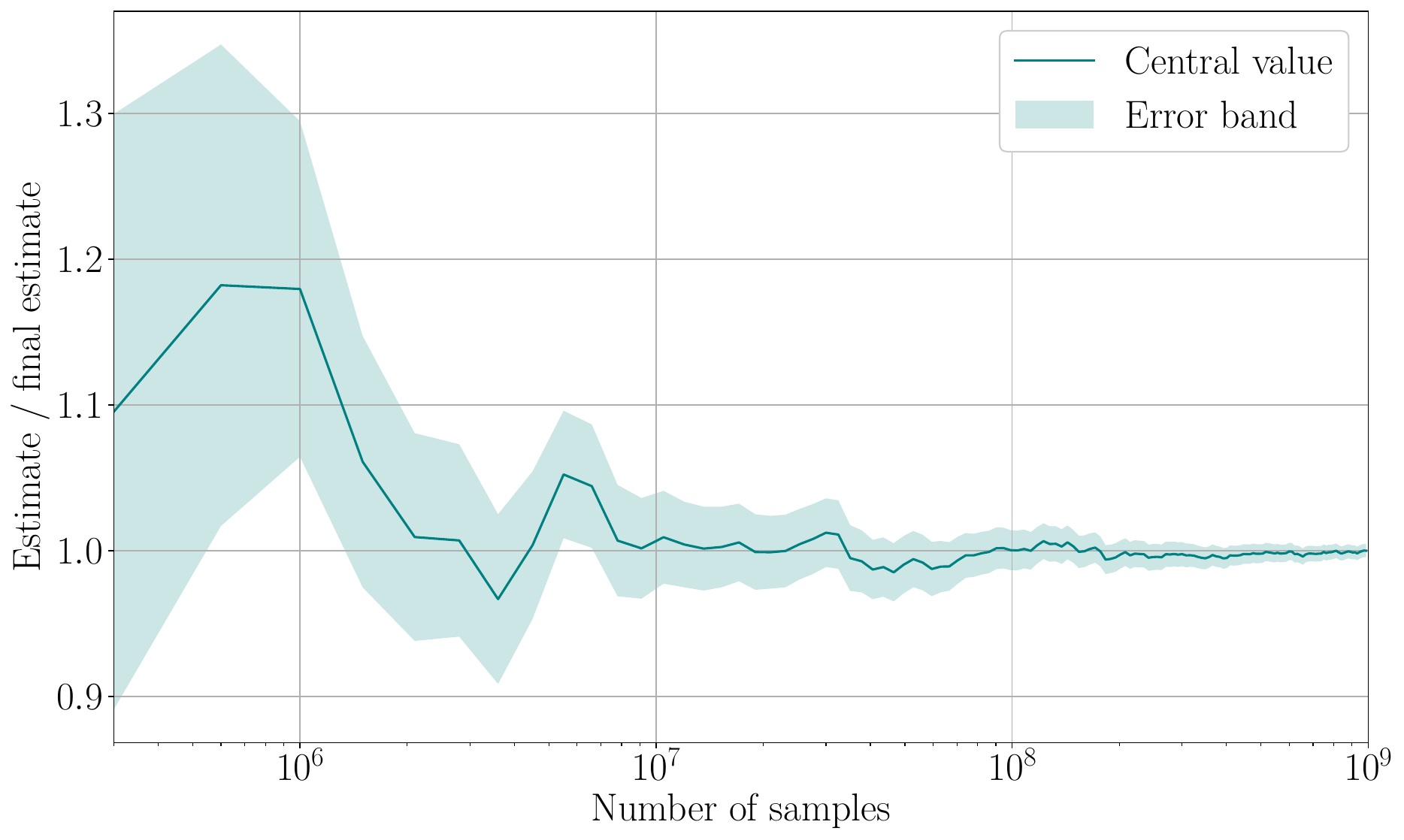}
    \caption{Convergence of the Monte Carlo integration for the contribution $F^{(2,\Nf)}_\text{FFS}$ to the process $q\bar{q}\to Z\gamma_1^*\gamma_2^*$ at the first phase-space point in listing~\ref{lst:5}.}
    \label{fig:convergence_2}
\end{figure}

Finally, figures~\ref{fig:convergence_1} and \ref{fig:convergence_2} illustrate the convergence of the Monte~Carlo integration for the two-loop $\Nf$-contributions to the $2\to3$ processes with massive final-state particles.
The first iteration is not shown, since its error would obscure the visibility of the error band.
As the final estimate remains within approximately $2\sigma$ throughout the iterations, we consider the result to be consistent and reliable.
\par

\def\treeone{F^{(0)}}
\def\fulloneloop{F^{(1)}_{\text{BLHA}}}
\def\fulltwoloopnf{F^{(2,\Nf)}_{\overline{\text{MS}}}}

\def\oneloopone{F^{(1)}_{\text{FFS}}}
\def\twoloopnfone{F^{(2,\Nf)}_{\text{FFS}}}
\def\onelooponedispersive{\Re\dispersive F^{(1)}_{\text{FFS}}}
\def\twoloopnfonedispersive{\Re\dispersive F^{(2,\Nf)}_{\text{FFS}}}
\def\onelooponeabsorptive{\Im\absorptive F^{(1)}_{\text{FFS}}}
\def\twoloopnfoneabsorptive{\Im\absorptive F^{(2,\Nf)}_{\text{FFS}}}

\vfill
\input{tables/ps_point_mom_2to2}
\newpage
\def\arraystretch{1.4}%
\input{tables/ps_points_tab_2to2}
\newpage
\input{tables/ps_point_mom_2to3}
\newpage
\input{tables/ps_points_tab_2to3}
\newpage
\def\arraystretch{1.25}%
\input{tables/ps_points_ffs}
\newpage

%% file: tables/ps_point_mom_2to2.tex
\begin{lstlisting}[language=Python,frame=single,caption={Phase space points for $q(p_1)\bar{q}(p_2)\to\gamma(q_1)\gamma(q_2)$},commentstyle=\color{teal},captionpos=b,label={lst:1}]
#PSP1
0.5000000000000000E+03  0.0000000000000000E+00  0.0000000000000000E+00  0.5000000000000000E+03
0.5000000000000000E+03  0.0000000000000000E+00  0.0000000000000000E+00 -0.5000000000000000E+03
0.4999999999999999E+03  0.1109242844438328E+03  0.4448307894881214E+03 -0.1995529299308788E+03
0.4999999999999999E+03 -0.1109242844438328E+03 -0.4448307894881214E+03  0.1995529299308787E+03
#PSP2
0.5000000000000000E+03  0.0000000000000000E+00  0.0000000000000000E+00  0.5000000000000000E+03
0.5000000000000000E+03  0.0000000000000000E+00  0.0000000000000000E+00 -0.5000000000000000E+03
0.5000000000000000E+03 -0.2819155058093908E+03 -0.3907909666011574E+03  0.1334393795218208E+03
0.5000000000000002E+03  0.2819155058093908E+03  0.3907909666011575E+03 -0.1334393795218208E+03
#PSP3
0.5000000000000000E+03  0.0000000000000000E+00  0.0000000000000000E+00  0.5000000000000000E+03
0.5000000000000000E+03  0.0000000000000000E+00  0.0000000000000000E+00 -0.5000000000000000E+03
0.4999999999999990E+03 -0.2406210290937675E+03  0.2804720059911282E+03 -0.3368040590805997E+03
0.5000000000000012E+03  0.2406210290937674E+03 -0.2804720059911282E+03  0.3368040590805997E+03
\end{lstlisting}

\begin{lstlisting}[language=Python,frame=single,caption={Phase space points for $q(p_1)\bar{q}(p_2)\to\gamma^*(q_1)\gamma^*(q_2)$ and $q(p_1)\bar{q}(p_2)\to Z(q_1)Z(q_2)$},commentstyle=\color{teal},captionpos=b,label={lst:2}]
#PSP1
0.5000000000000000E+03  0.0000000000000000E+00  0.0000000000000000E+00  0.5000000000000000E+03
0.5000000000000000E+03  0.0000000000000000E+00  0.0000000000000000E+00 -0.5000000000000000E+03
0.4999999999999998E+03  0.1090639744303390E+03  0.4373705369731355E+03 -0.1962062298314933E+03
0.4999999999999998E+03 -0.1090639744303390E+03 -0.4373705369731355E+03  0.1962062298314931E+03
#PSP2
0.5000000000000000E+03  0.0000000000000000E+00  0.0000000000000000E+00  0.5000000000000000E+03
0.5000000000000000E+03  0.0000000000000000E+00  0.0000000000000000E+00 -0.5000000000000000E+03
0.4999999999999999E+03 -0.2771875038119387E+03 -0.3842370153902377E+03  0.1312014690844130E+03
0.5000000000000002E+03  0.2771875038119388E+03  0.3842370153902378E+03 -0.1312014690844130E+03
#PSP3
0.5000000000000000E+03  0.0000000000000000E+00  0.0000000000000000E+00  0.5000000000000000E+03
0.5000000000000000E+03  0.0000000000000000E+00  0.0000000000000000E+00 -0.5000000000000000E+03
0.4999999999999990E+03 -0.2365855763331326E+03  0.2757682129140208E+03 -0.3311555217306899E+03
0.5000000000000011E+03  0.2365855763331325E+03 -0.2757682129140208E+03  0.3311555217306899E+03
\end{lstlisting}

%% file: tables/ps_points_tab_2to2.tex
\begin{table}[H]
\centering
\resizebox{\columnwidth}{!}{%
\begin{tabular}{lllrclrr}
\hline\hline
Process & PSP & Fin. Rem. & Exp. & Reference & \ \ \ \ \  Result $[\text{GeV}^0]$ & $\mathtt{\Delta \ [\sigma]}$ & $\Delta \ [\%]$ \\
\hline\hline\multirow{9}{*}{$q\bar{q}\to\gamma\gamma$}
& \multirow{3}{*}{1}
& $\treeone$
& {\ttlf$\texttt{10}^{\texttt{1}}$}
& 
& {\ttlf\texttt{~2.2063}}
& 
& {\ttlf\texttt{}}
\\
\cline{3-3}\cline{4-4}\cline{5-5}\cline{6-6}\cline{7-7}\cline{8-8}& & $\fulloneloop$
& {\ttlf$\texttt{10}^{\texttt{1}}$}
& {\ttlf\texttt{ 6.9439}}
& {\ttlf\texttt{~6.9422~$\pmtt$~0.0040}}
& {\ttlf\texttt{0.416}}
& {\ttlf\texttt{0.058}}
\\
\cline{3-3}\cline{4-4}\cline{5-5}\cline{6-6}\cline{7-7}\cline{8-8}& & $\fulltwoloopnf$
& {\ttlf$\texttt{10}^{\texttt{2}}$}
& {\ttlf\texttt{-6.1876}}
& {\ttlf\texttt{-6.2091~$\pmtt$~0.0176}}
& {\ttlf\texttt{{\color{black}1.221}}}
& {\ttlf\texttt{0.284}}
\\
[\doublerulesep]\ccline{2-2}\ccline{3-3}\ccline{4-4}\ccline{5-5}\ccline{6-6}\ccline{7-7}\ccline{8-8}& \multirow{3}{*}{2}
& $\treeone$
& {\ttlf$\texttt{10}^{\texttt{1}}$}
& 
& {\ttlf\texttt{~1.8454}}
& 
& {\ttlf\texttt{}}
\\
\cline{3-3}\cline{4-4}\cline{5-5}\cline{6-6}\cline{7-7}\cline{8-8}& & $\fulloneloop$
& {\ttlf$\texttt{10}^{\texttt{1}}$}
& {\ttlf\texttt{ 6.2117}}
& {\ttlf\texttt{~6.2098~$\pmtt$~0.0030}}
& {\ttlf\texttt{0.610}}
& {\ttlf\texttt{0.049}}
\\
\cline{3-3}\cline{4-4}\cline{5-5}\cline{6-6}\cline{7-7}\cline{8-8}& & $\fulltwoloopnf$
& {\ttlf$\texttt{10}^{\texttt{2}}$}
& {\ttlf\texttt{-5.4951}}
& {\ttlf\texttt{-5.4978~$\pmtt$~0.0138}}
& {\ttlf\texttt{0.199}}
& {\ttlf\texttt{0.251}}
\\
[\doublerulesep]\ccline{2-2}\ccline{3-3}\ccline{4-4}\ccline{5-5}\ccline{6-6}\ccline{7-7}\ccline{8-8}& \multirow{3}{*}{3}
& $\treeone$
& {\ttlf$\texttt{10}^{\texttt{1}}$}
& 
& {\ttlf\texttt{~4.2581}}
& 
& {\ttlf\texttt{}}
\\
\cline{3-3}\cline{4-4}\cline{5-5}\cline{6-6}\cline{7-7}\cline{8-8}& & $\fulloneloop$
& {\ttlf$\texttt{10}^{\texttt{2}}$}
& {\ttlf\texttt{ 1.3443}}
& {\ttlf\texttt{~1.3430~$\pmtt$~0.0012}}
& {\ttlf\texttt{{\color{black}1.052}}}
& {\ttlf\texttt{0.088}}
\\
\cline{3-3}\cline{4-4}\cline{5-5}\cline{6-6}\cline{7-7}\cline{8-8}& & $\fulltwoloopnf$
& {\ttlf$\texttt{10}^{\texttt{3}}$}
& {\ttlf\texttt{-1.1737}}
& {\ttlf\texttt{-1.1698~$\pmtt$~0.0059}}
& {\ttlf\texttt{0.668}}
& {\ttlf\texttt{0.503}}
\\
\hline\hline\multirow{9}{*}{$q\bar{q}\to\gamma^*\gamma^*$}
& \multirow{3}{*}{1}
& $\treeone$
& {\ttlf$\texttt{10}^{\texttt{1}}$}
& 
& {\ttlf\texttt{~2.3362}}
& 
& {\ttlf\texttt{}}
\\
\cline{3-3}\cline{4-4}\cline{5-5}\cline{6-6}\cline{7-7}\cline{8-8}& & $\fulloneloop$
& {\ttlf$\texttt{10}^{\texttt{1}}$}
& {\ttlf\texttt{ 7.0935}}
& {\ttlf\texttt{~7.0927~$\pmtt$~0.0035}}
& {\ttlf\texttt{0.211}}
& {\ttlf\texttt{0.049}}
\\
\cline{3-3}\cline{4-4}\cline{5-5}\cline{6-6}\cline{7-7}\cline{8-8}& & $\fulltwoloopnf$
& {\ttlf$\texttt{10}^{\texttt{2}}$}
& {\ttlf\texttt{-6.3009}}
& {\ttlf\texttt{-6.3031~$\pmtt$~0.0129}}
& {\ttlf\texttt{0.168}}
& {\ttlf\texttt{0.205}}
\\
[\doublerulesep]\ccline{2-2}\ccline{3-3}\ccline{4-4}\ccline{5-5}\ccline{6-6}\ccline{7-7}\ccline{8-8}& \multirow{3}{*}{2}
& $\treeone$
& {\ttlf$\texttt{10}^{\texttt{1}}$}
& 
& {\ttlf\texttt{~1.9633}}
& 
& {\ttlf\texttt{}}
\\
\cline{3-3}\cline{4-4}\cline{5-5}\cline{6-6}\cline{7-7}\cline{8-8}& & $\fulloneloop$
& {\ttlf$\texttt{10}^{\texttt{1}}$}
& {\ttlf\texttt{ 6.3054}}
& {\ttlf\texttt{~6.3055~$\pmtt$~0.0025}}
& {\ttlf\texttt{0.031}}
& {\ttlf\texttt{0.040}}
\\
\cline{3-3}\cline{4-4}\cline{5-5}\cline{6-6}\cline{7-7}\cline{8-8}& & $\fulltwoloopnf$
& {\ttlf$\texttt{10}^{\texttt{2}}$}
& {\ttlf\texttt{-5.6127}}
& {\ttlf\texttt{-5.6045~$\pmtt$~0.0095}}
& {\ttlf\texttt{0.853}}
& {\ttlf\texttt{0.170}}
\\
[\doublerulesep]\ccline{2-2}\ccline{3-3}\ccline{4-4}\ccline{5-5}\ccline{6-6}\ccline{7-7}\ccline{8-8}& \multirow{3}{*}{3}
& $\treeone$
& {\ttlf$\texttt{10}^{\texttt{1}}$}
& 
& {\ttlf\texttt{~4.4559}}
& 
& {\ttlf\texttt{}}
\\
\cline{3-3}\cline{4-4}\cline{5-5}\cline{6-6}\cline{7-7}\cline{8-8}& & $\fulloneloop$
& {\ttlf$\texttt{10}^{\texttt{2}}$}
& {\ttlf\texttt{ 1.4308}}
& {\ttlf\texttt{~1.4306~$\pmtt$~0.0010}}
& {\ttlf\texttt{0.171}}
& {\ttlf\texttt{0.073}}
\\
\cline{3-3}\cline{4-4}\cline{5-5}\cline{6-6}\cline{7-7}\cline{8-8}& & $\fulltwoloopnf$
& {\ttlf$\texttt{10}^{\texttt{3}}$}
& {\ttlf\texttt{-1.1964}}
& {\ttlf\texttt{-1.1913~$\pmtt$~0.0036}}
& {\ttlf\texttt{{\color{black}1.401}}}
& {\ttlf\texttt{0.306}}
\\
\hline\hline\multirow{9}{*}{$q\bar{q}\to ZZ$}
& \multirow{3}{*}{1}
& $\treeone$
& {\ttlf$\texttt{10}^{\texttt{1}}$}
& 
& {\ttlf\texttt{~1.2878}}
& 
& {\ttlf\texttt{}}
\\
\cline{3-3}\cline{4-4}\cline{5-5}\cline{6-6}\cline{7-7}\cline{8-8}& & $\fulloneloop$
& {\ttlf$\texttt{10}^{\texttt{1}}$}
& {\ttlf\texttt{ 3.9099}}
& {\ttlf\texttt{~3.9096~$\pmtt$~0.0019}}
& {\ttlf\texttt{0.113}}
& {\ttlf\texttt{0.049}}
\\
\cline{3-3}\cline{4-4}\cline{5-5}\cline{6-6}\cline{7-7}\cline{8-8}& & $\fulltwoloopnf$
& {\ttlf$\texttt{10}^{\texttt{2}}$}
& {\ttlf\texttt{-3.4730}}
& {\ttlf\texttt{-3.4777~$\pmtt$~0.0068}}
& {\ttlf\texttt{0.697}}
& {\ttlf\texttt{0.194}}
\\
[\doublerulesep]\ccline{2-2}\ccline{3-3}\ccline{4-4}\ccline{5-5}\ccline{6-6}\ccline{7-7}\ccline{8-8}& \multirow{3}{*}{2}
& $\treeone$
& {\ttlf$\texttt{10}^{\texttt{1}}$}
& 
& {\ttlf\texttt{~1.0822}}
& 
& {\ttlf\texttt{}}
\\
\cline{3-3}\cline{4-4}\cline{5-5}\cline{6-6}\cline{7-7}\cline{8-8}& & $\fulloneloop$
& {\ttlf$\texttt{10}^{\texttt{1}}$}
& {\ttlf\texttt{ 3.4755}}
& {\ttlf\texttt{~3.4756~$\pmtt$~0.0014}}
& {\ttlf\texttt{0.112}}
& {\ttlf\texttt{0.040}}
\\
\cline{3-3}\cline{4-4}\cline{5-5}\cline{6-6}\cline{7-7}\cline{8-8}& & $\fulltwoloopnf$
& {\ttlf$\texttt{10}^{\texttt{2}}$}
& {\ttlf\texttt{-3.0937}}
& {\ttlf\texttt{-3.0876~$\pmtt$~0.0055}}
& {\ttlf\texttt{{\color{black}1.103}}}
& {\ttlf\texttt{0.178}}
\\
[\doublerulesep]\ccline{2-2}\ccline{3-3}\ccline{4-4}\ccline{5-5}\ccline{6-6}\ccline{7-7}\ccline{8-8}& \multirow{3}{*}{3}
& $\treeone$
& {\ttlf$\texttt{10}^{\texttt{1}}$}
& 
& {\ttlf\texttt{~2.4561}}
& 
& {\ttlf\texttt{}}
\\
\cline{3-3}\cline{4-4}\cline{5-5}\cline{6-6}\cline{7-7}\cline{8-8}& & $\fulloneloop$
& {\ttlf$\texttt{10}^{\texttt{1}}$}
& {\ttlf\texttt{ 7.8863}}
& {\ttlf\texttt{~7.8867~$\pmtt$~0.0057}}
& {\ttlf\texttt{0.066}}
& {\ttlf\texttt{0.073}}
\\
\cline{3-3}\cline{4-4}\cline{5-5}\cline{6-6}\cline{7-7}\cline{8-8}& & $\fulltwoloopnf$
& {\ttlf$\texttt{10}^{\texttt{2}}$}
& {\ttlf\texttt{-6.5946}}
& {\ttlf\texttt{-6.6070~$\pmtt$~0.0214}}
& {\ttlf\texttt{0.577}}
& {\ttlf\texttt{0.325}}
\\
\cline{3-3}\cline{4-4}\cline{5-5}\cline{6-6}\cline{7-7}\cline{8-8}\hline\hline
\end{tabular}%
}%
\caption{\label{tab:ps_2to2}Finite remainders of squared matrix elements for three $2\to2$ processes evaluated at the phase-space points in listings~\ref{lst:1} and \ref{lst:2} obtained using $10^9$ Monte~Carlo samples. The shown contributions are defined in eqs.~\eqref{eq:fin_rem_def_start} to \eqref{eq:fin_rem_def_end}.}%
\end{table}

%% file: tables/ps_point_mom_2to3.tex
\newpage
\begin{lstlisting}[language=Python,frame=single,caption={Phase space points for $q(p_1)\bar{q}(p_2)\to\gamma(q_1)\gamma(q_2)\gamma(q_3)$},commentstyle=\color{teal},captionpos=b,label={lst:3}]
#PSP1
0.5000000000000000E+03  0.0000000000000000E+00  0.0000000000000000E+00  0.5000000000000000E+03
0.5000000000000000E+03  0.0000000000000000E+00  0.0000000000000000E+00 -0.5000000000000000E+03
0.4585787878854403E+03  0.1694532203096799E+03  0.3796536620781987E+03 -0.1935024746502525E+03
0.3640666207368177E+03 -0.1832986929319185E+02 -0.3477043013193672E+03  0.1063496077587081E+03
0.1773545913777421E+03 -0.1511233510164880E+03 -0.3194936075883155E+02  0.8715286689154436E+02
#PSP2
0.5000000000000000E+03  0.0000000000000000E+00  0.0000000000000000E+00  0.5000000000000000E+03
0.5000000000000000E+03  0.0000000000000000E+00  0.0000000000000000E+00 -0.5000000000000000E+03
0.4951533920773834E+03  0.1867229157692120E+03  0.3196835780850242E+03 -0.3288066974913060E+03
0.1026779138750091E+03 -0.7062042730772224E+02 -0.4345595295696615E+02  0.6055649756384902E+02
0.4021686940476072E+03 -0.1161024884614897E+03 -0.2762276251280580E+03  0.2682501999274569E+03
#PSP3
0.5000000000000000E+03  0.0000000000000000E+00  0.0000000000000000E+00  0.5000000000000000E+03
0.5000000000000000E+03  0.0000000000000000E+00  0.0000000000000000E+00 -0.5000000000000000E+03
0.2014347360150499E+03  0.1460392683472161E+03 -0.1127399318336771E+03 -0.8085909190808647E+02
0.4577010828828537E+03 -0.3806771896263719E+03 -0.7377601450610842E+02  0.2431712529348376E+03
0.3408641811020967E+03  0.2346379212791557E+03  0.1865159463397859E+03 -0.1623121610267509E+03
\end{lstlisting}

\begin{lstlisting}[language=Python,frame=single,caption={Phase space points for $q(p_1)\bar{q}(p_2)\to\gamma^*(q_1)\gamma^*(q_2)\gamma^*(q_3)$},commentstyle=\color{teal},captionpos=b,label={lst:4}]
#PSP1
0.5000000000000000E+03  0.0000000000000000E+00  0.0000000000000000E+00  0.5000000000000000E+03
0.5000000000000000E+03  0.0000000000000000E+00  0.0000000000000000E+00 -0.5000000000000000E+03
0.4477393491286647E+03  0.1619802634433327E+03  0.3629107790826664E+03 -0.1849689357540892E+03
0.3597595401382783E+03 -0.1752151450156574E+02 -0.3323703983032166E+03  0.1016595462179398E+03
0.1925011107330571E+03 -0.1444587489417669E+03 -0.3054038077944974E+02  0.8330938953614938E+02
#PSP2
0.5000000000000000E+03  0.0000000000000000E+00  0.0000000000000000E+00  0.5000000000000000E+03
0.5000000000000000E+03  0.0000000000000000E+00  0.0000000000000000E+00 -0.5000000000000000E+03
0.4763584456437345E+03  0.1763133172889205E+03  0.3018615679964441E+03 -0.3104760834666963E+03
0.1330984765879482E+03 -0.6668341566803356E+02 -0.4103333107364272E+02  0.5718053900827706E+02
0.3905430777683173E+03 -0.1096299016208870E+03 -0.2608282369228013E+03  0.2532955444584192E+03
#PSP3
0.5000000000000000E+03  0.0000000000000000E+00  0.0000000000000000E+00  0.5000000000000000E+03
0.5000000000000000E+03  0.0000000000000000E+00  0.0000000000000000E+00 -0.5000000000000000E+03
0.2133710236616980E+03  0.1398546405276659E+03 -0.1079655000888193E+03 -0.7743478422058600E+02
0.4477027246606839E+03 -0.3645558630552619E+03 -0.7065166858946638E+02  0.2328731754873389E+03
0.3389262516776180E+03  0.2247012225275959E+03  0.1786171686782860E+03 -0.1554383912667527E+03
\end{lstlisting}

\begin{lstlisting}[language=Python,frame=single,caption={Phase space points for $q(p_1)\bar{q}(p_2)\to Z(q_1)\gamma_1^*(q_2)\gamma_2^*(q_3)$},commentstyle=\color{teal},captionpos=b,label={lst:5}]
#PSP1
0.5000000000000000E+03  0.0000000000000000E+00  0.0000000000000000E+00  0.5000000000000000E+03
0.5000000000000000E+03  0.0000000000000000E+00  0.0000000000000000E+00 -0.5000000000000000E+03
0.4600427531794949E+03  0.1666212369563253E+03  0.3733087082963783E+03 -0.1902685686433855E+03
0.3585404168236821E+03 -0.1802353174107795E+02 -0.3418933005521544E+03  0.1045722421928131E+03
0.1814168299968238E+03 -0.1485977052152473E+03 -0.3141540774422388E+02  0.8569632645057239E+02
#PSP2
0.5000000000000000E+03  0.0000000000000000E+00  0.0000000000000000E+00  0.5000000000000000E+03
0.5000000000000000E+03  0.0000000000000000E+00  0.0000000000000000E+00 -0.5000000000000000E+03
0.4968913830841454E+03  0.1841960107343594E+03  0.3153573279314480E+03 -0.3243569849535531E+03
0.1032440625066204E+03 -0.6966472718600879E+02 -0.4286786731215435E+02  0.5973699173106596E+02
0.3998645544092341E+03 -0.1145312835483505E+03 -0.2724894606192936E+03  0.2646199932224872E+03
#PSP3
0.5000000000000000E+03  0.0000000000000000E+00  0.0000000000000000E+00  0.5000000000000000E+03
0.5000000000000000E+03  0.0000000000000000E+00  0.0000000000000000E+00 -0.5000000000000000E+03
0.2166619178935838E+03  0.1424891993643399E+03 -0.1099993365152814E+03 -0.7889348801664580E+02
0.4470224549143230E+03 -0.3714233067585944E+03 -0.7198259316305118E+02  0.2372600022668417E+03
0.3363156271920932E+03  0.2289341073942544E+03  0.1819819296783329E+03 -0.1583665142501957E+03
\end{lstlisting}

%% file: tables/ps_points_tab_2to3.tex
\begin{table}[H]
\centering
\resizebox{\columnwidth}{!}{%
\begin{tabular}{lllrclrr}
\hline\hline
Process & PSP & Fin. Rem. & Exp. & Reference & \ \ \ \ \  Result $[\text{GeV}^{-2}]$ & $\mathtt{\Delta \ [\sigma]}$ & $\Delta \ [\%]$ \\
\hline\hline\multirow{9}{*}{$q\bar{q}\to\gamma\gamma\gamma$}
& \multirow{3}{*}{1}
& $\treeone$
& {\ttlf$\texttt{10}^{\texttt{-3}}$}
& 
& {\ttlf\texttt{~3.9166}}
& 
& {\ttlf\texttt{}}
\\
\cline{3-3}\cline{4-4}\cline{5-5}\cline{6-6}\cline{7-7}\cline{8-8}& & $\fulloneloop$
& {\ttlf$\texttt{10}^{\texttt{-2}}$}
& {\ttlf\texttt{ 1.2394}}
& {\ttlf\texttt{~1.2374~$\pmtt$~0.0017}}
& {\ttlf\texttt{{\color{black}1.112}}}
& {\ttlf\texttt{0.140}}
\\
\cline{3-3}\cline{4-4}\cline{5-5}\cline{6-6}\cline{7-7}\cline{8-8}& & $\fulltwoloopnf$
& {\ttlf$\texttt{10}^{\texttt{-1}}$}
& {\ttlf\texttt{-1.3680}}
& {\ttlf\texttt{-1.3675~$\pmtt$~0.0052}}
& {\ttlf\texttt{0.104}}
& {\ttlf\texttt{0.382}}
\\
[\doublerulesep]\ccline{2-2}\ccline{3-3}\ccline{4-4}\ccline{5-5}\ccline{6-6}\ccline{7-7}\ccline{8-8}& \multirow{3}{*}{2}
& $\treeone$
& {\ttlf$\texttt{10}^{\texttt{-2}}$}
& 
& {\ttlf\texttt{~2.6286}}
& 
& {\ttlf\texttt{}}
\\
\cline{3-3}\cline{4-4}\cline{5-5}\cline{6-6}\cline{7-7}\cline{8-8}& & $\fulloneloop$
& {\ttlf$\texttt{10}^{\texttt{-2}}$}
& {\ttlf\texttt{ 8.7289}}
& {\ttlf\texttt{~8.7325~$\pmtt$~0.0198}}
& {\ttlf\texttt{0.179}}
& {\ttlf\texttt{0.227}}
\\
\cline{3-3}\cline{4-4}\cline{5-5}\cline{6-6}\cline{7-7}\cline{8-8}& & $\fulltwoloopnf$
& {\ttlf$\texttt{10}^{\texttt{-1}}$}
& {\ttlf\texttt{-9.0831}}
& {\ttlf\texttt{-9.1160~$\pmtt$~0.0572}}
& {\ttlf\texttt{0.576}}
& {\ttlf\texttt{0.627}}
\\
[\doublerulesep]\ccline{2-2}\ccline{3-3}\ccline{4-4}\ccline{5-5}\ccline{6-6}\ccline{7-7}\ccline{8-8}& \multirow{3}{*}{3}
& $\treeone$
& {\ttlf$\texttt{10}^{\texttt{-3}}$}
& 
& {\ttlf\texttt{~3.8137}}
& 
& {\ttlf\texttt{}}
\\
\cline{3-3}\cline{4-4}\cline{5-5}\cline{6-6}\cline{7-7}\cline{8-8}& & $\fulloneloop$
& {\ttlf$\texttt{10}^{\texttt{-3}}$}
& {\ttlf\texttt{ 9.3301}}
& {\ttlf\texttt{~9.3019~$\pmtt$~0.0168}}
& {\ttlf\texttt{{\color{black}1.679}}}
& {\ttlf\texttt{0.180}}
\\
\cline{3-3}\cline{4-4}\cline{5-5}\cline{6-6}\cline{7-7}\cline{8-8}& & $\fulltwoloopnf$
& {\ttlf$\texttt{10}^{\texttt{-1}}$}
& {\ttlf\texttt{-1.1603}}
& {\ttlf\texttt{-1.1558~$\pmtt$~0.0051}}
& {\ttlf\texttt{0.884}}
& {\ttlf\texttt{0.442}}
\\
\hline\hline\multirow{9}{*}{$q\bar{q}\to\gamma^*\gamma^*\gamma^*$}
& \multirow{3}{*}{1}
& $\treeone$
& {\ttlf$\texttt{10}^{\texttt{-3}}$}
& 
& {\ttlf\texttt{~4.0277}}
& 
& {\ttlf\texttt{}}
\\
\cline{3-3}\cline{4-4}\cline{5-5}\cline{6-6}\cline{7-7}\cline{8-8}& & $\fulloneloop$
& {\ttlf$\texttt{10}^{\texttt{-2}}$}
& {\ttlf\texttt{ 2.1319}}
& {\ttlf\texttt{~2.1315~$\pmtt$~0.0015}}
& {\ttlf\texttt{0.272}}
& {\ttlf\texttt{0.068}}
\\
\cline{3-3}\cline{4-4}\cline{5-5}\cline{6-6}\cline{7-7}\cline{8-8}& & $\fulltwoloopnf$
& {\ttlf$\texttt{10}^{\texttt{-1}}$}
& 
& {\ttlf\texttt{-1.8543~$\pmtt$~0.0057}}
& 
& {\ttlf\texttt{0.307}}
\\
[\doublerulesep]\ccline{2-2}\ccline{3-3}\ccline{4-4}\ccline{5-5}\ccline{6-6}\ccline{7-7}\ccline{8-8}& \multirow{3}{*}{2}
& $\treeone$
& {\ttlf$\texttt{10}^{\texttt{-2}}$}
& 
& {\ttlf\texttt{~1.6447}}
& 
& {\ttlf\texttt{}}
\\
\cline{3-3}\cline{4-4}\cline{5-5}\cline{6-6}\cline{7-7}\cline{8-8}& & $\fulloneloop$
& {\ttlf$\texttt{10}^{\texttt{-1}}$}
& {\ttlf\texttt{ 1.0541}}
& {\ttlf\texttt{~1.0529~$\pmtt$~0.0017}}
& {\ttlf\texttt{0.702}}
& {\ttlf\texttt{0.162}}
\\
\cline{3-3}\cline{4-4}\cline{5-5}\cline{6-6}\cline{7-7}\cline{8-8}& & $\fulltwoloopnf$
& {\ttlf$\texttt{10}^{\texttt{-1}}$}
& 
& {\ttlf\texttt{-8.6105~$\pmtt$~0.0651}}
& 
& {\ttlf\texttt{0.756}}
\\
[\doublerulesep]\ccline{2-2}\ccline{3-3}\ccline{4-4}\ccline{5-5}\ccline{6-6}\ccline{7-7}\ccline{8-8}& \multirow{3}{*}{3}
& $\treeone$
& {\ttlf$\texttt{10}^{\texttt{-3}}$}
& 
& {\ttlf\texttt{~4.1521}}
& 
& {\ttlf\texttt{}}
\\
\cline{3-3}\cline{4-4}\cline{5-5}\cline{6-6}\cline{7-7}\cline{8-8}& & $\fulloneloop$
& {\ttlf$\texttt{10}^{\texttt{-2}}$}
& {\ttlf\texttt{ 2.0747}}
& {\ttlf\texttt{~2.0731~$\pmtt$~0.0016}}
& {\ttlf\texttt{0.973}}
& {\ttlf\texttt{0.078}}
\\
\cline{3-3}\cline{4-4}\cline{5-5}\cline{6-6}\cline{7-7}\cline{8-8}& & $\fulltwoloopnf$
& {\ttlf$\texttt{10}^{\texttt{-1}}$}
& 
& {\ttlf\texttt{-1.7871~$\pmtt$~0.0058}}
& 
& {\ttlf\texttt{0.322}}
\\
\hline\hline\multirow{9}{*}{$q\bar{q}\to Z\gamma_1^*\gamma_2^*$}
& \multirow{3}{*}{1}
& $\treeone$
& {\ttlf$\texttt{10}^{\texttt{-3}}$}
& 
& {\ttlf\texttt{~2.1867}}
& 
& {\ttlf\texttt{}}
\\
\cline{3-3}\cline{4-4}\cline{5-5}\cline{6-6}\cline{7-7}\cline{8-8}& & $\fulloneloop$
& {\ttlf$\texttt{10}^{\texttt{-3}}$}
& {\ttlf\texttt{ 8.8148}}
& {\ttlf\texttt{~8.8178~$\pmtt$~0.0088}}
& {\ttlf\texttt{0.343}}
& {\ttlf\texttt{0.099}}
\\
\cline{3-3}\cline{4-4}\cline{5-5}\cline{6-6}\cline{7-7}\cline{8-8}& & $\fulltwoloopnf$
& {\ttlf$\texttt{10}^{\texttt{-2}}$}
& 
& {\ttlf\texttt{-8.6776~$\pmtt$~0.0286}}
& 
& {\ttlf\texttt{0.330}}
\\
[\doublerulesep]\ccline{2-2}\ccline{3-3}\ccline{4-4}\ccline{5-5}\ccline{6-6}\ccline{7-7}\ccline{8-8}& \multirow{3}{*}{2}
& $\treeone$
& {\ttlf$\texttt{10}^{\texttt{-2}}$}
& 
& {\ttlf\texttt{~1.4476}}
& 
& {\ttlf\texttt{}}
\\
\cline{3-3}\cline{4-4}\cline{5-5}\cline{6-6}\cline{7-7}\cline{8-8}& & $\fulloneloop$
& {\ttlf$\texttt{10}^{\texttt{-2}}$}
& {\ttlf\texttt{ 6.4174}}
& {\ttlf\texttt{~6.4005~$\pmtt$~0.0139}}
& {\ttlf\texttt{{\color{black}1.220}}}
& {\ttlf\texttt{0.217}}
\\
\cline{3-3}\cline{4-4}\cline{5-5}\cline{6-6}\cline{7-7}\cline{8-8}& & $\fulltwoloopnf$
& {\ttlf$\texttt{10}^{\texttt{-1}}$}
& 
& {\ttlf\texttt{-5.7577~$\pmtt$~0.0424}}
& 
& {\ttlf\texttt{0.737}}
\\
[\doublerulesep]\ccline{2-2}\ccline{3-3}\ccline{4-4}\ccline{5-5}\ccline{6-6}\ccline{7-7}\ccline{8-8}& \multirow{3}{*}{3}
& $\treeone$
& {\ttlf$\texttt{10}^{\texttt{-3}}$}
& 
& {\ttlf\texttt{~2.0477}}
& 
& {\ttlf\texttt{}}
\\
\cline{3-3}\cline{4-4}\cline{5-5}\cline{6-6}\cline{7-7}\cline{8-8}& & $\fulloneloop$
& {\ttlf$\texttt{10}^{\texttt{-3}}$}
& {\ttlf\texttt{ 8.9921}}
& {\ttlf\texttt{~8.9706~$\pmtt$~0.0130}}
& {\ttlf\texttt{{\color{black}1.655}}}
& {\ttlf\texttt{0.145}}
\\
\cline{3-3}\cline{4-4}\cline{5-5}\cline{6-6}\cline{7-7}\cline{8-8}& & $\fulltwoloopnf$
& {\ttlf$\texttt{10}^{\texttt{-2}}$}
& 
& {\ttlf\texttt{-8.4459~$\pmtt$~0.0534}}
& 
& {\ttlf\texttt{0.632}}
\\
\cline{3-3}\cline{4-4}\cline{5-5}\cline{6-6}\cline{7-7}\cline{8-8}\hline\hline
\end{tabular}%
}%
\caption{\label{tab:ps_2to3}Finite remainders of squared matrix elements for three $2\to3$ processes evaluated at the phase-space points in listings~\ref{lst:3} to \ref{lst:5} obtained using $10^9$ Monte~Carlo samples. The shown contributions are defined in eqs.~\eqref{eq:fin_rem_def_start} to \eqref{eq:fin_rem_def_end}.}%
\end{table}

%% file: tables/ps_points_ffs.tex
\begin{table}[H]
\centering
\resizebox*{!}{0.99\textheight}{%
\begin{tabular}{p{3.8cm}p{1.5cm}p{4.1cm}p{2.1cm}p{5cm}r}
\hline\hline
Process & PSP & Fin. Rem. & Exp. & \ \ \ \ \  Result $[\text{GeV}^{4-2N}]$ & $\Delta \ [\%]$ \\
\hline\hline\multirow{6}{*}{$q\bar{q}\to\gamma\gamma$}
& \multirow{2}{*}{1}
& $\oneloopone$
& {\ttlf$\texttt{10}^{\texttt{1}}$}
& {\ttlf\texttt{~3.1050~$\pmtt$~0.0040}}
& {\ttlf\texttt{0.129}}
\\
\cline{3-3}\cline{4-4}\cline{5-5}\cline{6-6}& & $\twoloopnfone$
& {\ttlf$\texttt{10}^{\texttt{2}}$}
& {\ttlf\texttt{-3.4238~$\pmtt$~0.0176}}
& {\ttlf\texttt{0.515}}
\\
[\doublerulesep]\ccline{2-2}\ccline{3-3}\ccline{4-4}\ccline{5-5}\ccline{6-6}& \multirow{2}{*}{2}
& $\oneloopone$
& {\ttlf$\texttt{10}^{\texttt{1}}$}
& {\ttlf\texttt{~3.0003~$\pmtt$~0.0030}}
& {\ttlf\texttt{0.101}}
\\
\cline{3-3}\cline{4-4}\cline{5-5}\cline{6-6}& & $\twoloopnfone$
& {\ttlf$\texttt{10}^{\texttt{2}}$}
& {\ttlf\texttt{-3.1860~$\pmtt$~0.0138}}
& {\ttlf\texttt{0.433}}
\\
[\doublerulesep]\ccline{2-2}\ccline{3-3}\ccline{4-4}\ccline{5-5}\ccline{6-6}& \multirow{2}{*}{3}
& $\oneloopone$
& {\ttlf$\texttt{10}^{\texttt{1}}$}
& {\ttlf\texttt{~6.0247~$\pmtt$~0.0118}}
& {\ttlf\texttt{0.195}}
\\
\cline{3-3}\cline{4-4}\cline{5-5}\cline{6-6}& & $\twoloopnfone$
& {\ttlf$\texttt{10}^{\texttt{2}}$}
& {\ttlf\texttt{-6.3238~$\pmtt$~0.0588}}
& {\ttlf\texttt{0.930}}
\\
[\doublerulesep]\ccline{1-1}\ccline{2-2}\ccline{3-3}\ccline{4-4}\ccline{5-5}\ccline{6-6}\multirow{6}{*}{$q\bar{q}\to\gamma^*\gamma^*$}
& \multirow{2}{*}{1}
& $\oneloopone$
& {\ttlf$\texttt{10}^{\texttt{1}}$}
& {\ttlf\texttt{~3.0295~$\pmtt$~0.0035}}
& {\ttlf\texttt{0.114}}
\\
\cline{3-3}\cline{4-4}\cline{5-5}\cline{6-6}& & $\twoloopnfone$
& {\ttlf$\texttt{10}^{\texttt{2}}$}
& {\ttlf\texttt{-3.3422~$\pmtt$~0.0129}}
& {\ttlf\texttt{0.386}}
\\
[\doublerulesep]\ccline{2-2}\ccline{3-3}\ccline{4-4}\ccline{5-5}\ccline{6-6}& \multirow{2}{*}{2}
& $\oneloopone$
& {\ttlf$\texttt{10}^{\texttt{1}}$}
& {\ttlf\texttt{~2.8910~$\pmtt$~0.0025}}
& {\ttlf\texttt{0.087}}
\\
\cline{3-3}\cline{4-4}\cline{5-5}\cline{6-6}& & $\twoloopnfone$
& {\ttlf$\texttt{10}^{\texttt{2}}$}
& {\ttlf\texttt{-3.1317~$\pmtt$~0.0095}}
& {\ttlf\texttt{0.305}}
\\
[\doublerulesep]\ccline{2-2}\ccline{3-3}\ccline{4-4}\ccline{5-5}\ccline{6-6}& \multirow{2}{*}{3}
& $\oneloopone$
& {\ttlf$\texttt{10}^{\texttt{1}}$}
& {\ttlf\texttt{~6.5562~$\pmtt$~0.0105}}
& {\ttlf\texttt{0.160}}
\\
\cline{3-3}\cline{4-4}\cline{5-5}\cline{6-6}& & $\twoloopnfone$
& {\ttlf$\texttt{10}^{\texttt{2}}$}
& {\ttlf\texttt{-6.3004~$\pmtt$~0.0364}}
& {\ttlf\texttt{0.578}}
\\
[\doublerulesep]\ccline{1-1}\ccline{2-2}\ccline{3-3}\ccline{4-4}\ccline{5-5}\ccline{6-6}\multirow{6}{*}{$q\bar{q}\to ZZ$}
& \multirow{2}{*}{1}
& $\oneloopone$
& {\ttlf$\texttt{10}^{\texttt{1}}$}
& {\ttlf\texttt{~1.6700~$\pmtt$~0.0019}}
& {\ttlf\texttt{0.114}}
\\
\cline{3-3}\cline{4-4}\cline{5-5}\cline{6-6}& & $\twoloopnfone$
& {\ttlf$\texttt{10}^{\texttt{2}}$}
& {\ttlf\texttt{-1.8457~$\pmtt$~0.0068}}
& {\ttlf\texttt{0.366}}
\\
[\doublerulesep]\ccline{2-2}\ccline{3-3}\ccline{4-4}\ccline{5-5}\ccline{6-6}& \multirow{2}{*}{2}
& $\oneloopone$
& {\ttlf$\texttt{10}^{\texttt{1}}$}
& {\ttlf\texttt{~1.5935~$\pmtt$~0.0014}}
& {\ttlf\texttt{0.086}}
\\
\cline{3-3}\cline{4-4}\cline{5-5}\cline{6-6}& & $\twoloopnfone$
& {\ttlf$\texttt{10}^{\texttt{2}}$}
& {\ttlf\texttt{-1.7245~$\pmtt$~0.0055}}
& {\ttlf\texttt{0.319}}
\\
[\doublerulesep]\ccline{2-2}\ccline{3-3}\ccline{4-4}\ccline{5-5}\ccline{6-6}& \multirow{2}{*}{3}
& $\oneloopone$
& {\ttlf$\texttt{10}^{\texttt{1}}$}
& {\ttlf\texttt{~3.6149~$\pmtt$~0.0057}}
& {\ttlf\texttt{0.158}}
\\
\cline{3-3}\cline{4-4}\cline{5-5}\cline{6-6}& & $\twoloopnfone$
& {\ttlf$\texttt{10}^{\texttt{2}}$}
& {\ttlf\texttt{-3.5132~$\pmtt$~0.0214}}
& {\ttlf\texttt{0.611}}
\\
[\doublerulesep]\ccline{1-1}\ccline{2-2}\ccline{3-3}\ccline{4-4}\ccline{5-5}\ccline{6-6}\multirow{6}{*}{$q\bar{q}\to\gamma\gamma\gamma$}
& \multirow{2}{*}{1}
& $\oneloopone$
& {\ttlf$\texttt{10}^{\texttt{-2}}$}
& {\ttlf\texttt{~1.3396~$\pmtt$~0.0017}}
& {\ttlf\texttt{0.129}}
\\
\cline{3-3}\cline{4-4}\cline{5-5}\cline{6-6}& & $\twoloopnfone$
& {\ttlf$\texttt{10}^{\texttt{-2}}$}
& {\ttlf\texttt{-9.7767~$\pmtt$~0.0523}}
& {\ttlf\texttt{0.535}}
\\
[\doublerulesep]\ccline{2-2}\ccline{3-3}\ccline{4-4}\ccline{5-5}\ccline{6-6}& \multirow{2}{*}{2}
& $\oneloopone$
& {\ttlf$\texttt{10}^{\texttt{-2}}$}
& {\ttlf\texttt{~9.4180~$\pmtt$~0.0198}}
& {\ttlf\texttt{0.211}}
\\
\cline{3-3}\cline{4-4}\cline{5-5}\cline{6-6}& & $\twoloopnfone$
& {\ttlf$\texttt{10}^{\texttt{-1}}$}
& {\ttlf\texttt{-6.5189~$\pmtt$~0.0571}}
& {\ttlf\texttt{0.877}}
\\
[\doublerulesep]\ccline{2-2}\ccline{3-3}\ccline{4-4}\ccline{5-5}\ccline{6-6}& \multirow{2}{*}{3}
& $\oneloopone$
& {\ttlf$\texttt{10}^{\texttt{-2}}$}
& {\ttlf\texttt{~1.0297~$\pmtt$~0.0017}}
& {\ttlf\texttt{0.163}}
\\
\cline{3-3}\cline{4-4}\cline{5-5}\cline{6-6}& & $\twoloopnfone$
& {\ttlf$\texttt{10}^{\texttt{-2}}$}
& {\ttlf\texttt{-7.6400~$\pmtt$~0.0511}}
& {\ttlf\texttt{0.669}}
\\
[\doublerulesep]\ccline{1-1}\ccline{2-2}\ccline{3-3}\ccline{4-4}\ccline{5-5}\ccline{6-6}\multirow{6}{*}{$q\bar{q}\to\gamma^*\gamma^*\gamma^*$}
& \multirow{2}{*}{1}
& $\oneloopone$
& {\ttlf$\texttt{10}^{\texttt{-2}}$}
& {\ttlf\texttt{~2.2366~$\pmtt$~0.0015}}
& {\ttlf\texttt{0.065}}
\\
\cline{3-3}\cline{4-4}\cline{5-5}\cline{6-6}& & $\twoloopnfone$
& {\ttlf$\texttt{10}^{\texttt{-1}}$}
& {\ttlf\texttt{-1.4917~$\pmtt$~0.0057}}
& {\ttlf\texttt{0.382}}
\\
[\doublerulesep]\ccline{2-2}\ccline{3-3}\ccline{4-4}\ccline{5-5}\ccline{6-6}& \multirow{2}{*}{2}
& $\oneloopone$
& {\ttlf$\texttt{10}^{\texttt{-1}}$}
& {\ttlf\texttt{~1.0958~$\pmtt$~0.0017}}
& {\ttlf\texttt{0.156}}
\\
\cline{3-3}\cline{4-4}\cline{5-5}\cline{6-6}& & $\twoloopnfone$
& {\ttlf$\texttt{10}^{\texttt{-1}}$}
& {\ttlf\texttt{-7.2107~$\pmtt$~0.0651}}
& {\ttlf\texttt{0.903}}
\\
[\doublerulesep]\ccline{2-2}\ccline{3-3}\ccline{4-4}\ccline{5-5}\ccline{6-6}& \multirow{2}{*}{3}
& $\oneloopone$
& {\ttlf$\texttt{10}^{\texttt{-2}}$}
& {\ttlf\texttt{~2.1814~$\pmtt$~0.0016}}
& {\ttlf\texttt{0.074}}
\\
\cline{3-3}\cline{4-4}\cline{5-5}\cline{6-6}& & $\twoloopnfone$
& {\ttlf$\texttt{10}^{\texttt{-1}}$}
& {\ttlf\texttt{-1.4077~$\pmtt$~0.0058}}
& {\ttlf\texttt{0.409}}
\\
[\doublerulesep]\ccline{1-1}\ccline{2-2}\ccline{3-3}\ccline{4-4}\ccline{5-5}\ccline{6-6}\multirow{12}{*}{$q\bar{q}\to Z\gamma_1^*\gamma_2^*$}
& \multirow{4}{*}{1}
& $\onelooponedispersive$
& {\ttlf$\texttt{10}^{\texttt{-3}}$}
& {\ttlf\texttt{~9.2691~$\pmtt$~0.0086}}
& {\ttlf\texttt{0.093}}
\\
\cline{3-3}\cline{4-4}\cline{5-5}\cline{6-6}& & $\onelooponeabsorptive$
& {\ttlf$\texttt{10}^{\texttt{-3}}$}
& {\ttlf\texttt{-0.1190~$\pmtt$~0.0017}}
& {\ttlf\texttt{{\color{black}1.422}}}
\\
\cline{3-3}\cline{4-4}\cline{5-5}\cline{6-6}& & $\twoloopnfonedispersive$
& {\ttlf$\texttt{10}^{\texttt{-2}}$}
& {\ttlf\texttt{-6.3862~$\pmtt$~0.0285}}
& {\ttlf\texttt{0.446}}
\\
\cline{3-3}\cline{4-4}\cline{5-5}\cline{6-6}& & $\twoloopnfoneabsorptive$
& {\ttlf$\texttt{10}^{\texttt{-2}}$}
& {\ttlf\texttt{~0.2000~$\pmtt$~0.0024}}
& {\ttlf\texttt{{\color{black}1.192}}}
\\
[\doublerulesep]\ccline{2-2}\ccline{3-3}\ccline{4-4}\ccline{5-5}\ccline{6-6}& \multirow{4}{*}{2}
& $\onelooponedispersive$
& {\ttlf$\texttt{10}^{\texttt{-2}}$}
& {\ttlf\texttt{~6.6869~$\pmtt$~0.0138}}
& {\ttlf\texttt{0.206}}
\\
\cline{3-3}\cline{4-4}\cline{5-5}\cline{6-6}& & $\onelooponeabsorptive$
& {\ttlf$\texttt{10}^{\texttt{-2}}$}
& {\ttlf\texttt{-0.0911~$\pmtt$~0.0015}}
& {\ttlf\texttt{{\color{black}1.630}}}
\\
\cline{3-3}\cline{4-4}\cline{5-5}\cline{6-6}& & $\twoloopnfonedispersive$
& {\ttlf$\texttt{10}^{\texttt{-1}}$}
& {\ttlf\texttt{-4.3195~$\pmtt$~0.0423}}
& {\ttlf\texttt{0.980}}
\\
\cline{3-3}\cline{4-4}\cline{5-5}\cline{6-6}& & $\twoloopnfoneabsorptive$
& {\ttlf$\texttt{10}^{\texttt{-1}}$}
& {\ttlf\texttt{~0.0787~$\pmtt$~0.0024}}
& {\ttlf\texttt{{\color{black}3.012}}}
\\
[\doublerulesep]\ccline{2-2}\ccline{3-3}\ccline{4-4}\ccline{5-5}\ccline{6-6}& \multirow{4}{*}{3}
& $\onelooponedispersive$
& {\ttlf$\texttt{10}^{\texttt{-3}}$}
& {\ttlf\texttt{~9.3092~$\pmtt$~0.0129}}
& {\ttlf\texttt{0.138}}
\\
\cline{3-3}\cline{4-4}\cline{5-5}\cline{6-6}& & $\onelooponeabsorptive$
& {\ttlf$\texttt{10}^{\texttt{-3}}$}
& {\ttlf\texttt{-0.1954~$\pmtt$~0.0019}}
& {\ttlf\texttt{0.979}}
\\
\cline{3-3}\cline{4-4}\cline{5-5}\cline{6-6}& & $\twoloopnfonedispersive$
& {\ttlf$\texttt{10}^{\texttt{-2}}$}
& {\ttlf\texttt{-6.3487~$\pmtt$~0.0533}}
& {\ttlf\texttt{0.840}}
\\
\cline{3-3}\cline{4-4}\cline{5-5}\cline{6-6}& & $\twoloopnfoneabsorptive$
& {\ttlf$\texttt{10}^{\texttt{-2}}$}
& {\ttlf\texttt{~0.1705~$\pmtt$~0.0026}}
& {\ttlf\texttt{{\color{black}1.533}}}
\\
\cline{3-3}\cline{4-4}\cline{5-5}\cline{6-6}\hline\hline
\end{tabular}%
}%
\caption{\label{tab:ps_points_ffs}Finite remainders in the factor subtraction scheme for six processes evaluated at the phase-space points in listings~\ref{lst:1} to \ref{lst:5} obtained using $10^9$ Monte~Carlo samples. The shown contributions are defined in eqs.~\eqref{eq:fin_rem_ffs_1l} and \eqref{eq:fin_rem_ffs_2lnf}.}%
\end{table}